\pdfoutput=1 
\documentclass[11pt,
       a4paper,
       onecolumn,
       oneside,
       notitlepage]{article}
\usepackage{sectsty}
\sectionfont{\fontsize{12}{15}\selectfont}
\subsectionfont{\fontsize{11}{15}\selectfont}
\usepackage[utf8]{inputenc}
\usepackage[T1]{fontenc}
\usepackage{lmodern}
\usepackage[british]{babel}
\usepackage[numbers, square, sort&compress]{natbib}
\let\BibOrigCite\cite
\renewcommand\cite[1]{\textsuperscript{\BibOrigCite{#1}}}
\let\BibOrigCitep\citep
\renewcommand\citep[1]{\textsuperscript{\BibOrigCitep{#1}}}

\renewcommand\citet[1]{\citeauthor{#1}\textsuperscript{[\citenum{#1}]}}

\usepackage[labeled,resetlabels]{multibib}
\newcites{S}{Supplementary Information References}
\makeatletter
\let\mb@OrigCiteS\citeS
\DeclareRobustCommand{\citeS}{\mb@OrigCiteS}
\let\mb@OrigCitepS\citepS
\DeclareRobustCommand{\citepS}{\mb@OrigCitepS}
\let\mb@OrigCitetS\citetS
\DeclareRobustCommand{\citetS}{\mb@OrigCitetS}
\let\mb@OrigCitealtS\citealtS
\DeclareRobustCommand{\citealtS}{\mb@OrigCitealtS}
\let\mb@OrigCitealpS\citealpS
\DeclareRobustCommand{\citealpS}{\mb@OrigCitealpS}
\makeatother
\usepackage{nicefrac}
\usepackage{amsthm}
\usepackage{xparse}
\usepackage{mathtools}
\usepackage{xcolor}
\usepackage{graphicx}
\usepackage{epstopdf}
\DeclareGraphicsExtensions{.pdf,.eps,.png,.jpg}
\usepackage{caption}
\usepackage{subcaption}
\DeclareCaptionFont{tenpt}{\fontsize{10pt}{12pt}\selectfont}
\usepackage[affil-it]{authblk}
\usepackage{amsmath, amssymb}
\usepackage[adobe-helvetica,scaled]{mathdesign}   

\usepackage[eulergreek]{sansmath}                 
\sansmath                                         
\usepackage{verbatim}                             
\usepackage{siunitx}                              
\usepackage[version=3]{mhchem}                    
\usepackage[left=2cm, top=2cm, right=2cm, bottom=2cm, columnsep=0.5cm]{geometry} 
\usepackage[T1]{fontenc}
\usepackage[pdftex]{hyperref}
\usepackage[symbol]{footmisc}
\usepackage{booktabs}
\usepackage{mathtools}
\usepackage{tikz}
\usepackage{pgfplots}
\pgfplotsset{compat=1.18}
\usepackage{xspace}

\definecolor{lime}{HTML}{A6CE39}
\DeclareRobustCommand{\orcidicon}{%
	\begin{tikzpicture}
	\draw[lime, fill=lime] (0,0) 
	circle [radius=0.16] 
	node[white] {{\fontfamily{qag}\selectfont \tiny ID}};
	\draw[white, fill=white] (-0.0625,0.095) 
	circle [radius=0.007];
	\end{tikzpicture}
	\hspace{-2mm}
}
\foreach \x in {A, ..., Z}{%
	\expandafter\xdef\csname orcid\x\endcsname{\noexpand\href{https://orcid.org/\csname orcidauthor\x\endcsname}{\noexpand\orcidicon}}
}

\DeclarePairedDelimiter\abs{\lvert}{\rvert}   
\newcommand{\defeq}{\vcentcolon=}             

\usepackage[activate={true,nocompatibility},final,
    tracking=true,kerning=true,spacing=true,factor=1100,
    expansion=alltext,stretch=5,shrink=5]{microtype}
\microtypecontext{spacing=nonfrench}
\hypersetup{colorlinks,
    linktocpage=true,
    urlcolor=gray,   
    citecolor=black, 
    linkcolor=black, 
    pdfauthor={S.R.G. Balestra, A. Rivas-Blanco, S. Hamad, A.R. Ru\'iz-Salvador},
    pdftitle={A topology-tuned pressure valve across the isoreticular RHO zeolite family},
    pdfsubject={arXiv preprint},
    pdfkeywords={zeolite, RHO, isoreticular, phase transition, Landau theory, OPES, molecular dynamics}
}
\makeatletter
\newcommand*{\email}[1]{%
    \normalsize\href{mailto:#1}{#1}\par
}
\newcommand*{\doi}[1]{DOI: \href{http://dx.doi.org/#1}{#1}}

\def\vs{\emph{vs.\ }}

\newcommand{\centric}{\textit{centric}\xspace}
\newcommand{\acentric}{\textit{acentric}\xspace}
\newcommand{\centtoacent}{\textit{centric}$\to$\textit{acentric}\xspace}
\newcommand{\centacent}{\textit{centric}$\leftrightarrow$\textit{acentric}\xspace}

\newbox\abstract@box
\renewenvironment{abstract}
  {\global\setbox\abstract@box=\vbox\bgroup
     \hsize=\textwidth\linewidth=\textwidth
    \small
    \begin{center}%
    {\bfseries \abstractname\vspace{-.5em}\vspace{\z@}}%
    \end{center}%
    \quotation}
  {\endquotation\egroup}
  \expandafter\def\expandafter\@maketitle\expandafter{\@maketitle
  \ifvoid\abstract@box\else\unvbox\abstract@box\if@twocolumn\vskip1.5em\fi\fi}
\makeatother
\begin{document}
\title{\textbf{{{\huge A topology-tuned pressure valve across the isoreticular RHO zeolite family}}}}
\author[1,*]{Salvador R.G. Balestra\, \orcidA{}}
\author[2,3]{Antonio Rivas-Blanco\ \orcidD{}}
\author[2,3]{Said Hamad\ \orcidC{}}
\author[2,3]{A. Rabdel Ruíz-Salvador\ \orcidB{}}
\affil[1]{Departamento de Física Atómica, Molecular y Nuclear,
Área de Física Teórica, 
Universidad de Sevilla, ES-41012 Sevilla, Spain}
\affil[2]{Centro de Nanociencia y Tecnologías Sostenibles (CNATS),
Universidad Pablo de Olavide, ES-41013 Sevilla, Spain}
\affil[3]{Departamento de Sistemas Físicos, Químicos y Naturales,
Universidad Pablo de Olavide, ES-41013 Sevilla, Spain}

\affil[*]{E-Mail: \email{srodriguez9@us.es}}
\date{}
\begin{abstract}
\emph{The isoreticular index of the eight-member embedded RHO zeolite hierarchy operates as a \textbf{phenomenological design knob} that tunes the mechanical critical pressure of the framework valve from $\sim 0.94$~\si{\giga\Pa} for parent RHO down to a predicted $\lesssim 0.03$~\si{\giga\Pa} for the largest member PST-28, more than an order of magnitude lower across a single family of synthesizable nanoporous solids.
The molecular-valve effect that defines zeolite RHO (a reversible \centtoacent phase transition triggered by water, gas pressure, or mechanical loading) is shown here to be not a peculiarity of the smallest member but a generic property of the whole hierarchy, with a critical pressure that decays exponentially with the isoreticular order $k$.
Combining lattice dynamics, full elastic-constant tensors, and finite-temperature free-energy reconstructions within a classical core-shell description of the pure-silica frameworks (independently validated against r$^2$SCAN+rVV10 density-functional theory on $G_1$ and $G_2$), we find that the entire family is well described by an effective mean-field Landau picture in which the framework distortion couples quadratically to the volumetric strain.
We emphasise that $p_c$ is a mechanical (hydrostatic) critical pressure of the bare pure-silica framework, used as a proxy for the intrinsic framework softness and for the cation- and dehydration-driven response of the real aluminosilicates; it is not a gas-adsorption pressure. On this scale the exponential extrapolation places $p_c(G_6$-$G_8) \lesssim 0.1$~\si{\giga\Pa} (model-dependent band $0.03$-$0.15$~\si{\giga\Pa}), identifying the higher-order members as the softest, most stimuli-responsive frameworks of the hierarchy; whether this intrinsic softness translates into guest-driven switching at low gas activities will depend on the Al distribution, extra-framework cations and adsorbed molecules, and remains to be tested experimentally.}
\end{abstract}
\maketitle


\section{Introduction}
\label{sec:Introduction}
Zeolite RHO is the archetypal molecular valve: its reversible \centacent{} transition reshapes the eight-membered-ring windows that control access to the $\alpha$-cages. The embedded isoreticular RHO hierarchy $(G_k)_{k=1}^{8}$ raises a broader question: how does this framework flexibility change when the same local window motif is placed in progressively larger frameworks?\cite{Guo2015, Shin2016, Lee2018} 
In other words, isoreticular expansion offers a controlled way to test whether molecular-valve behaviour is a fixed property of a single framework or a tunable response governed by framework size, connectivity and softness.

Isoreticular chemistry turns topology into a design variable by preserving the connectivity of a framework while modifying the structural motifs from which it is built. 
This idea has shaped the chemistry of nanoporous solids, most prominently in metal-organic frameworks,\cite{Rosi2003, Fan2021} but also in supertetrahedral chalcogenides\cite{Zhang2020a} and in increasingly complex zeolite families, including UTL-type, embedded RHO-type and proposed FAU-type extensions.\cite{Zheng2002, US4247416A, Guo2015, Shin2016, ilkov2016, Min2017, Cho2017, Lee2018, DelgadoFriedrichs2020} 
For zeolites, however, topology alone does not determine function: cooperative framework flexibility controls how pores open, close and respond to external stimuli. 
It underpins chiral separation in STW frameworks,\cite{BuenoPerez2018} enhances diffusion in Ge-containing topologies,\cite{GutierrezSevillano2016} enables stepped adsorption under weak host-guest coupling,\cite{Min2018Langmuir} and defines a broader class of responsive behaviours.\cite{Ghojavand2023} 
Isoreticular zeolite families therefore provide a direct way to test whether flexibility is mainly controlled by a local structural motif or by the larger framework in which that motif is embedded.

\begin{figure*}[htp]
\includegraphics[width=\textwidth]{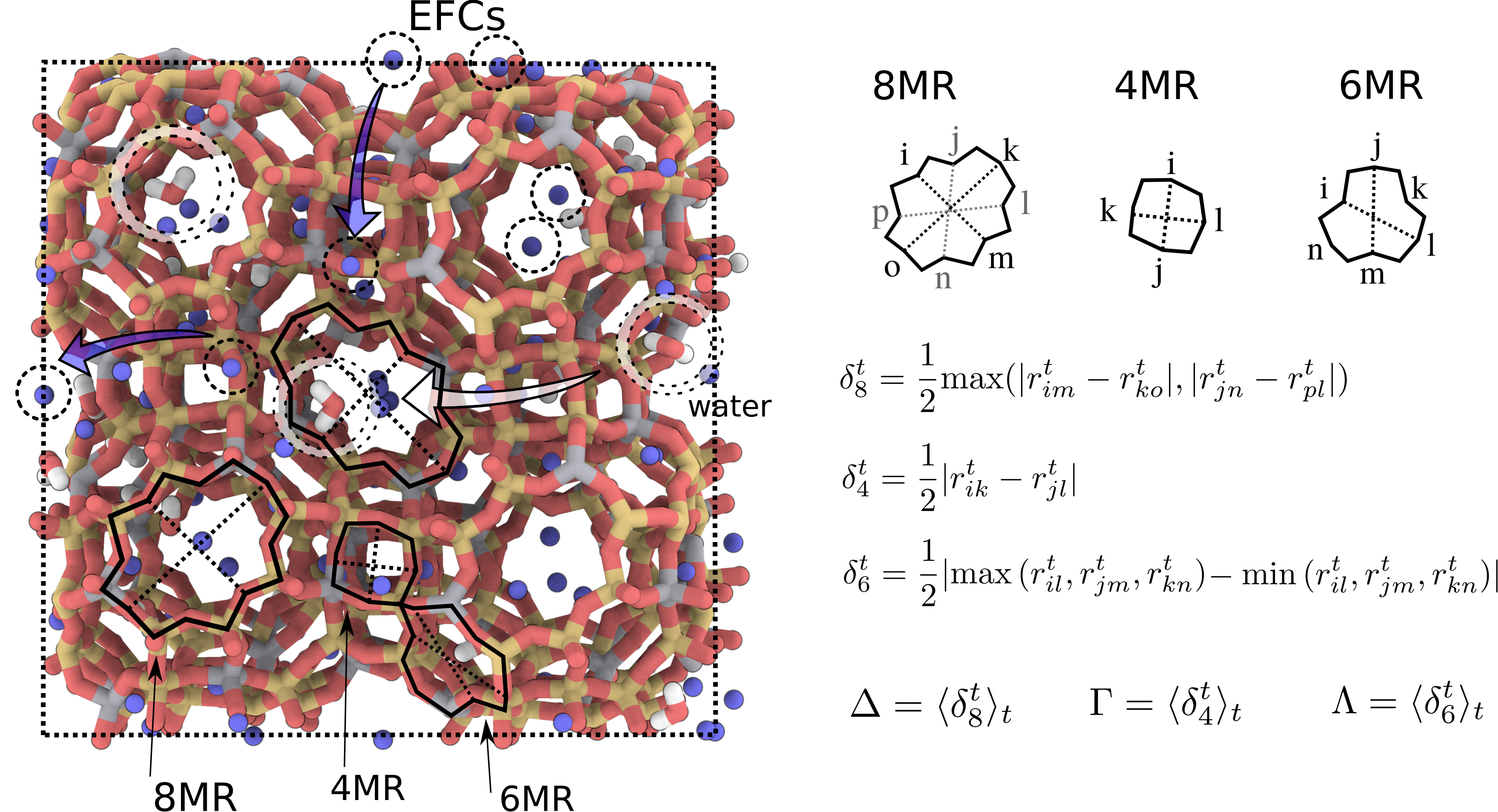}
\caption{\label{fig1:panel}
(Left) Snapshot of the $2\times 2\times 2$ supercell of a low hydrated Na-RHO structure with $\nicefrac{\text{Si}}{\text{Al}} = 4.5$. 
The 4-, 6-, and 8-membered rings are highlighted. Colour code for Si, Al, O, Na and H atoms: gold, grey, salmon, purple and white, respectively.
(Right) Schematic view of the rings of the structure and the order parameter definitions, following \citet{Balestra2015}: $\Delta \equiv \langle \delta_8 \rangle$ for the 8MR (D8R) distortion, $\Gamma \equiv \langle \delta_4 \rangle$ for the 4MR distortion and $\Lambda \equiv \langle \delta_6 \rangle$ for the 6MR distortion.
Abbreviations: EFC, extra-framework cation; D8R, double eight-membered ring; 8MR, 6MR and 4MR denote eight-, six- and four-membered rings, respectively.
}
\end{figure*}

RHO is a particularly suitable system to address this question because its flexibility is expressed as a crystallographic switch. 
Zeolite RHO is a synthetic aluminosilicate with $3.6 < \nicefrac{\text{Si}}{\text{Al}} < 5$,\cite{Chatelain1995, ROBSON1973} and the aluminosilicate counterpart of the rare beryllium-phosphate zeolite pahasapaite. 
Hydrated RHO adopts the \centric{} phase ($\mathrm{Im}\bar{3}\mathrm{m}$, \# 229), whereas partial or complete dehydration lowers the symmetry to the \acentric{} phase ($\mathrm{I}\bar{4}3\mathrm{m}$, \# 217), distorting the double eight-membered-ring (D8R) windows that gate the $\alpha$-cages into elliptical apertures (\textbf{Figure} \ref{fig1:panel}).

The aperture can be tuned by relative humidity, temperature, guest pressure or hydrostatic load,\cite{Parise1983, Parise1984a, Parise1984b, Corbin1990, Parise1991, Reisner2000, Lee2001} providing the structural basis for the nanovalve behaviour exploited in \ce{CO2}/\ce{CH4}/\ce{N2} separations.\cite{Lee2001, Palomino2012, Lozinska2012} 
Diffraction, adsorption and simulation studies have further shown that this gating response is controlled by the coupled motion of extra-framework cations, guest molecules and framework oxygen atoms at the window.\cite{Lozinska2012, Lozinska2014, Lozinska2016, Balestra2013, Grand2020, Lozinska2021, Clatworthy2023, Ghojavand2025}

Previous theoretical work has clarified the RHO molecular-valve transition from two complementary perspectives. Rigid-unit-mode analyses identified low-energy collective rotations of nearly rigid tetrahedra as natural pathways for the symmetry-lowering distortion in RHO-type frameworks,\cite{Bieniok1998, Dove2019, Hammonds1997, Hammonds1998} whereas atomistic simulations, ranging from classical force fields based on the Sanders-Leslie-Catlow core-shell potential\cite{Sanders1984, Catlow1986, Higgins1997} and de Leeuw-Parker water models\cite{deLeeuw1998, deLeeuw2004, Higgins2001, Lewis2002} to \emph{ab initio} molecular dynamics,\cite{Coudert2017} reproduced the main features of cation-mediated window gating.\cite{AlmoraBarrios2001, Maurin2004, White2004, Balestra2015}

These studies established the microscopic ingredients of the parent RHO valve, but they also exposed the limitation that motivates the present work: following the \centtoacent transition while treating Al disorder, extra-framework cations and adsorbed guests explicitly remains demanding for RHO and becomes prohibitive for the larger members of the isoreticular family.

Paulingite (PAU), one of the most structurally complex minerals known,\cite{Kent1966, Krivovichev2013} provides the first natural bridge between RHO and the larger embedded members of the family. 
Its synthetic analogue, (Na,H)-ECR-18, exhibits a closely related \centacent transition under dehydration and \ce{CO2} pressure,\cite{Bieniok1996, Bieniok1997, Greenaway2015} together with high-pressure volume shrinkage.\cite{Gatta2015} 
A topological connection between RHO, PAU and a then-unknown intermediate member with $a\simeq 25$~\si[mode=text]{\angstrom} was proposed by \citeauthor{Vaughan1999}\cite{Vaughan1999} and later placed on a crystallochemical basis by \citet{Shevchenko2008}. 
In this construction, stacking bipyramidal complexes generates a hierarchy of ideal \centric frameworks with cell parameters
\begin{equation}
\nicefrac{a_k}{\text{\AA}} \simeq 15 + 10(k-1),
\end{equation}
where $k$ is the number of layers.

This hierarchy was subsequently realised experimentally through ZSM-25,\cite{US4247416A, Guo2015} PST-20 and PST-25,\cite{Guo2015} PST-26 and PST-28,\cite{Shin2016} and PST-29,\cite{Lee2018} the predicted $a\simeq 25$~\si[mode=text]{\angstrom} member. 
We denote the eight-member family as $(G_k)_{k=1}^{8}$, corresponding to RHO, PST-29 (PWN), PAU, ZSM-25 (MWF), PST-20, PST-25, PST-26 and PST-28. 
The framework size grows rapidly with $k$, with T-atom counts
\begin{equation}
\label{eq:Nk-counts}
\left(N_k\right)_{k=1}^{8} = (48, 240, 672, 1440, 2640, 4368, 6720, 9792),
\end{equation}
generated by
\begin{equation}
N_k= 8k(1+3k+2k^2),
\end{equation}
for $k>0$. 
Combined with an approximately constant framework density of $\rho\sim 15$ T-atoms per \SI[mode=text]{1000}{\angstrom\cubed}, this expression is consistent, to leading order in $k$, with the empirical lattice-parameter law $a_k\simeq 14.9 + 10.76\,(k-1)$~\si[mode=text]{\angstrom} reported by \citeauthor{Shevchenko2008}.\cite{Shevchenko2008}

Adsorption measurements show that this structural hierarchy is not merely geometric. 
\ce{CO2} adsorption across the RHO isoreticular series exhibits trapdoor behaviour,\cite{Min2018} although analogous selectivity can also occur in comparatively rigid frameworks such as CHA, KFI and LTA.\cite{DeBaerdemaeker2013, Shang2012, Remy2013, Liu2010} 
For the RHO family, several experimental observations point to a direct role of framework flexibility. 
Cell-volume changes upon dehydration have been reported across members of the series,\cite{Min2017b, Min2018} K-ZSM-25 ($G_4$) achieves \ce{N2}/\ce{CH4} selectivity above $30$ through temperature-regulated cation gating,\cite{Zhao2021} electric-field studies further support the field-tunable character of trapdoor ZSM-25,\cite{Chen2024} and \ce{CO2} uptake induces framework expansion in Li-ZSM-25.\cite{Zhao2018} 
Most directly, the \centtoacent  transition with a $\sim 4.4\%$ cell-parameter contraction has been resolved in dehydrated Na,H-ECR-18 ($G_3$).\cite{Greenaway2015} 
Recent in-situ TEM and synchrotron measurements of breathing RHO nanocrystals complete the experimental picture for the parent framework.\cite{Clatworthy2023, Ghojavand2025} 
The available literature, summarised quantitatively in \textbf{Tables} \ref{tab:SI-validation} and \ref{tab:SI-expdata} of the Supporting Information, establishes a strong experimental basis for flexibility-mediated gating in the lower and intermediate members of the RHO family. 
It does not, however, determine whether the same soft-mode mechanism persists in $G_6$-$G_8$, which lack pure phases and direct flexibility measurements.

Atomistic modelling provides the natural route to address this question. 
We build on previous work on crystallographically reliable modelling of zeolite flexibility\cite{AlmoraBarrios2001, RabdelRuizSalvador2007, Balestra2015} and seek the general rules that govern the pressure-induced \centtoacent  response across the RHO isoreticular hierarchy. 
The isoreticular construction itself gives rise to scaling behaviours, which we characterise quantitatively and connect with an effective Landau description of structural phase transitions.

Realistic modelling that simultaneously includes Al disorder, extra-framework cations and adsorbed molecules remains out of reach for the larger members of the family because of their unit-cell size. 
We therefore use external hydrostatic pressure on the pure-silica frameworks as a mechanical proxy for framework softness, guided by the structural analogy with the dehydration-driven transition previously reported for zeolite RHO.\cite{Balestra2015} 
The Sanders-Leslie-Catlow (SLC) core-shell potential used here was benchmarked against dispersion-corrected DFT at the r$^2$SCAN+rVV10 level for $G_1$ and $G_2$ (Supporting Information, \textbf{Section} \ref{si:slc-validation} and \textbf{Figure} \ref{fig:SI-slc-dft-richard}). 
For both materials, SLC and DFT agree on the existence, approximate location and qualitative phenomenology of the \centtoacent transition, with $p_c$ within $\sim 20\%$ between the two methods; DFT gives a sharper first-order character, especially for $G_1$.

This $G_1$-$G_2$ benchmark, together with the crystallographic and adsorption data summarised above, supports the use of SLC as a semiquantitative model for trends across the family. Neural-network-based interatomic potentials\cite{Erlebach2022, Brugnoli2024} provide the natural route towards fully ab initio-quality simulations of the larger frameworks. Nevertheless, the benchmarks shown in the Supporting Information indicate that current foundation models will require zeolite-specific training or targeted fine-tuning to reproduce the \centtoacent  transitions reliably.

Using this framework, we combine pressure-dependent structural optimisation, lattice dynamics, elastic-constant analysis and finite-temperature free-energy reconstruction to examine how isoreticular order controls framework softness in the RHO family. 
Rather than treating the molecular-valve transition as a simple yes-or-no property, we follow how its critical pressure, distortion amplitude, elastic anomaly and free-energy landscape change from $G_1$ to $G_5$, and use these trends to predict the response of $G_6$-$G_8$.

\section{Methodology}
\label{sec:Methodology}

\subsection{Static optimisation and pressure scans}

Energy minimisations, pressure-dependent structural relaxations, phonon calculations and elastic-constant calculations were performed with LAMMPS and GULP.\cite{Plimpton1995, Gale1997, Gale2003}
For the pure-silica zeolite frameworks, we used the Sanders-Leslie-Catlow (SLC) classical core-shell potential.\cite{Sanders1984, Catlow1986}
This potential was used for all members of the RHO isoreticular family considered in the pressure scans, phonon calculations, elastic-constant calculations and finite-temperature free-energy simulations.
The pressure-scan, phonon and elastic-constant calculations were carried out on pure-silica frameworks, without extra-framework cations or guest molecules.
Because the SLC potential is a classical model fitted to silica polymorphs, the absolute transition pressures should be regarded as semiquantitative; the trends across the family are the main quantity of interest in the present work.

The pressure-dependent structural optimisations were first carried out with LAMMPS.
At each imposed hydrostatic pressure, both the atomic positions and the simulation cell were relaxed.
The cell volume and shape were optimised anisotropically using the Polak-Ribiere conjugate-gradient algorithm,\cite{polak1969note} followed by relaxation of the atomic positions with the FIRE damped-dynamics minimiser.\cite{Bitzek2006, Gunol2020}
This procedure provided the structural pressure scans used to follow the cell volume, the equivalent cubic-cell parameter and the D8R distortion across the transition.

\subsection{Phonons and elastic constants}

Phonon frequencies and elastic constants require tighter convergence than the structural pressure scans.
Therefore, selected structures close to the critical pressure were further refined with GULP.
The GULP optimisations used Broyden-Fletcher-Goldfarb-Shanno (BFGS) minimisation followed by rational-function optimisation (RFO), until convergence to a well-defined local minimum was achieved.
The refined structures were then used to compute the $\Gamma$-point phonon frequencies and elastic constants.
When the parent \centric{} structure developed an imaginary phonon mode, the structure was displaced along the corresponding eigenvector and reoptimised in the lower-symmetry branch.
This procedure was used to follow the \centtoacent{} pathway and to identify the onset of the soft-mode instability.
The detailed convergence thresholds used in the LAMMPS and GULP optimisations are reported in the Supporting Information.

\subsection{Hydrated aluminosilicate models}

For the chemically realistic tests on hydrated and dehydrated aluminosilicates, representative Ca-aluminosilicate models were used for the first members of the family.
The framework was described with the same SLC model, and the water-framework interactions were described with the de Leeuw-Parker water potentials and related parametrisations previously used for zeolite-water simulations.\cite{deLeeuw1998, deLeeuw2004, Higgins2001, Lewis2002}
The divalent-cation distributions were taken from the corresponding experimental structural models, and the hydrated structures used water loadings obtained from rigid-framework adsorption calculations with RASPA at \SI[mode=text]{1}{\bar}.\cite{Dubbeldam2016}
The resulting hydrated and dehydrated aluminosilicate structures were then fully relaxed to compare the stability of the \centric{} and \acentric{} forms in chemically realistic models.

\subsection{Order parameter definition}
The equivalent cubic-cell parameter was calculated from the instantaneous volume.
For unit-cell simulations, we used
\begin{equation}
a=V^{1/3},
\end{equation}
whereas for $2\times2\times2$ supercell simulations we used the corresponding primitive-cell value,
\begin{equation}
a=(V/8)^{1/3}.
\end{equation}
In the finite-temperature simulations, this supercell convention was used for $G_1$, whereas $G_2$-$G_5$ were treated with their crystallographic unit cells.

The D8R distortion was quantified from the two diagonals of each eight-membered ring. 
For a given 8MR at time $t$, we define
\begin{equation}
r_{i,m}^{t} =
\left|
\mathbf{r}_{i}^{t} - \mathbf{r}_{m}^{t}
\right|,
\qquad
r_{k,o}^{t} =
\left|
\mathbf{r}_{k}^{t} - \mathbf{r}_{o}^{t}
\right|,
\end{equation}
where $(\mathrm{O}_{i},\mathrm{O}_{m})$ and $(\mathrm{O}_{k},\mathrm{O}_{o})$ are the two pairs of oxygen atoms located on opposite sides of the ring. 
The instantaneous distortion of that 8MR is then
\begin{equation}
 \label{eq:delta_t}
 \delta_{\mathrm{D8R}}^{t} =
 \frac{1}{2}\,
 \max\!\Bigl(
   \left| r_{i,m}^{t} - r_{k,o}^{t} \right|,\;
   \left| r_{j,n}^{t} - r_{l,p}^{t} \right|
 \Bigr),
\end{equation}
where $(\mathrm{O}_{j}, \mathrm{O}_{n})$ and $(\mathrm{O}_{l}, \mathrm{O}_{p})$ are the second orthogonal pair of opposite-oxygen diagonals of the ring; the $\max$ over the two pairs ensures a non-negative measure that is rotation-invariant within the ring plane (consistent with the convention used by the White Rabbit code and with \textbf{Figure} \ref{fig1:panel}).
This quantity is zero for an undistorted 8MR and becomes finite when the ring becomes elliptical. 
The order parameter reported in the paper is the average over the relevant D8R-related 8MR windows and, for molecular dynamics simulations, over the trajectory,
\begin{equation}
 \label{eq:Delta}
 \Delta =
 \left\langle
 \delta_{\mathrm{D8R}}^{t}
 \right\rangle_{t,\,\mathrm{NPT}} .
\end{equation}

\subsection{Finite-temperature OPES/free-energy simulations}
Finite-temperature molecular dynamics simulations were performed in the isothermal-isobaric (NPT) ensemble.
The time step was \SI[mode=text]{0.2}{\femto\second}.
A Nosé-Hoover chain thermostat and an anisotropic Parrinello-Rahman-type barostat were used, with damping times of 10 and 100 time steps, respectively.
All finite-temperature simulations were carried out in $P1$, without imposing spatial symmetry on the atomic positions.

To sample the finite-temperature free-energy landscape spanned by the D8R distortion and the cell parameter, we used the OPES exploration protocol (PLUMED action \texttt{OPES\_METAD\_EXPLORE}) applied to the two structural collective variables defined below.\cite{Invernizzi2020, Invernizzi2020b} This protocol places an adaptive bias on the chosen collective variables at a fixed reference state point $(T, p)$; it is distinct from the OPES-Expanded variants used to sample temperature- or pressure-expanded ensembles, which are not employed here.
The biased collective variables were the pair $(\Delta,a)$.
Here $\Delta$ is the D8R distortion defined in \textbf{Equation} \ref{eq:delta_t}, averaged on the fly over the six 8MRs that form the three orthogonal D8Rs of the cubic (super)cell, and $a$ is the equivalent cubic-cell parameter defined above.
The OPES bias was updated every 500 MD steps at the reference temperature $T=\SI[mode=text]{298.15}{\kelvin}$, with a barrier parameter of \SI[mode=text]{0.2}{\kilo\joule\per\mole} per framework T-atom, of the order of the elastic strain energy of one D8R at the transition.
Four independent walkers were combined in parallel for each pressure point.

Because the simulations were performed in $P1$, the three D8Rs aligned with the cubic-cell axes were treated as independent observables rather than symmetry-locked replicas.
Each 8MR also admits two equivalent polarisations of the elliptical distortion.
In larger supercells, local domains with opposite polarisation could therefore coexist along the same crystallographic direction.
The averaged order parameter used here captures the dominant D8R distortion but does not resolve this intra-axis heterogeneity, which is left for future simulations on larger supercells.

One biased NPT simulation was run at each nominal hydrostatic pressure around the transition region of each member.
The pressures were \SI[mode=text]{1.0}{\giga\Pa}, \SI[mode=text]{1.2}{\giga\Pa}, \SI[mode=text]{1.225}{\giga\Pa} and \SI[mode=text]{1.3}{\giga\Pa} for $G_1$; \SI[mode=text]{0.5}{\giga\Pa}, \SI[mode=text]{0.75}{\giga\Pa}, \SI[mode=text]{0.82}{\giga\Pa}, \SI[mode=text]{0.9}{\giga\Pa} and \SI[mode=text]{1.0}{\giga\Pa} for $G_2$; \SI[mode=text]{0.1}{\giga\Pa}, \SI[mode=text]{0.4}{\giga\Pa}, \SI[mode=text]{0.5}{\giga\Pa}, \SI[mode=text]{0.7}{\giga\Pa} and \SI[mode=text]{1.0}{\giga\Pa} for $G_3$; \SI[mode=text]{0.2}{\giga\Pa}, \SI[mode=text]{0.5}{\giga\Pa}, \SI[mode=text]{0.6}{\giga\Pa} and \SI[mode=text]{0.8}{\giga\Pa} for $G_4$; and \SI[mode=text]{0.3}{\giga\Pa}, \SI[mode=text]{0.4}{\giga\Pa}, \SI[mode=text]{0.5}{\giga\Pa} and \SI[mode=text]{0.7}{\giga\Pa} for $G_5$.
Each run was extended for $1$-$2\times10^{7}$ MD steps after equilibration.

The two-dimensional free-energy surface $\Delta g^{*}(\Delta,a)$ at each target $(T,p)$ was reconstructed by reweighting the biased trajectories and combining the four walker histories.
The maps were evaluated on a $200\times200$ grid spanning the relevant $(\Delta,a)$ window for each member and pressure.
The resulting free-energy surfaces were shifted so that the minimum of each map is zero. Free energies are reported in units of $100\,k_B T$ per framework T-atom.


\section{\label{sec:Results}Results}

\subsection{Hydration-controlled distortions in realistic aluminosilicate models}
\begin{figure*}[!hpt!]
  \centering
  \begin{subfigure}{\textwidth}
  \centering
  \caption{\label{fig:RHO_Ca}(Left) Dehydrated (\acentric form, $\Delta\simeq 1.6$~\si[mode=text]{\angstrom}, $a\simeq 14.2$~\si[mode=text]{\angstrom}) and (Right) hydrated (\centric form, $a\simeq 15$~\si[mode=text]{\angstrom}) Ca-RHO with 10 Al per unit cell ($\nicefrac{\text{Si}}{\text{Al}}=3.8$) and 60 water molecules per unit cell.}
  \includegraphics[width=0.8\textwidth]{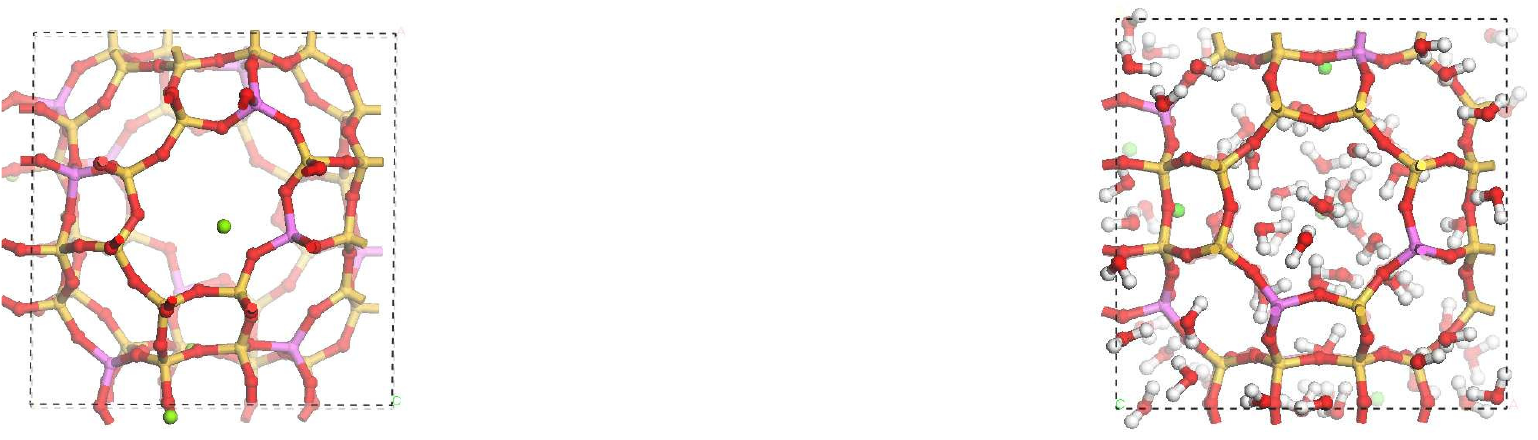}
  \end{subfigure}
  \begin{subfigure}{\textwidth}
  \centering
  \caption{\label{fig:PST-29_Ca}(Left) Dehydrated (\acentric form, $\Delta\simeq 1.9$~\si[mode=text]{\angstrom}, $a = 24$~\si[mode=text]{\angstrom}) and (Right) hydrated (\centric form, $a\simeq 25.1$~\si[mode=text]{\angstrom}) Ca-PST-29 with $\nicefrac{\text{Si}}{\text{Al}}=4.45$ and 166 water molecules per unit cell.}
  \includegraphics[width=0.9\textwidth]{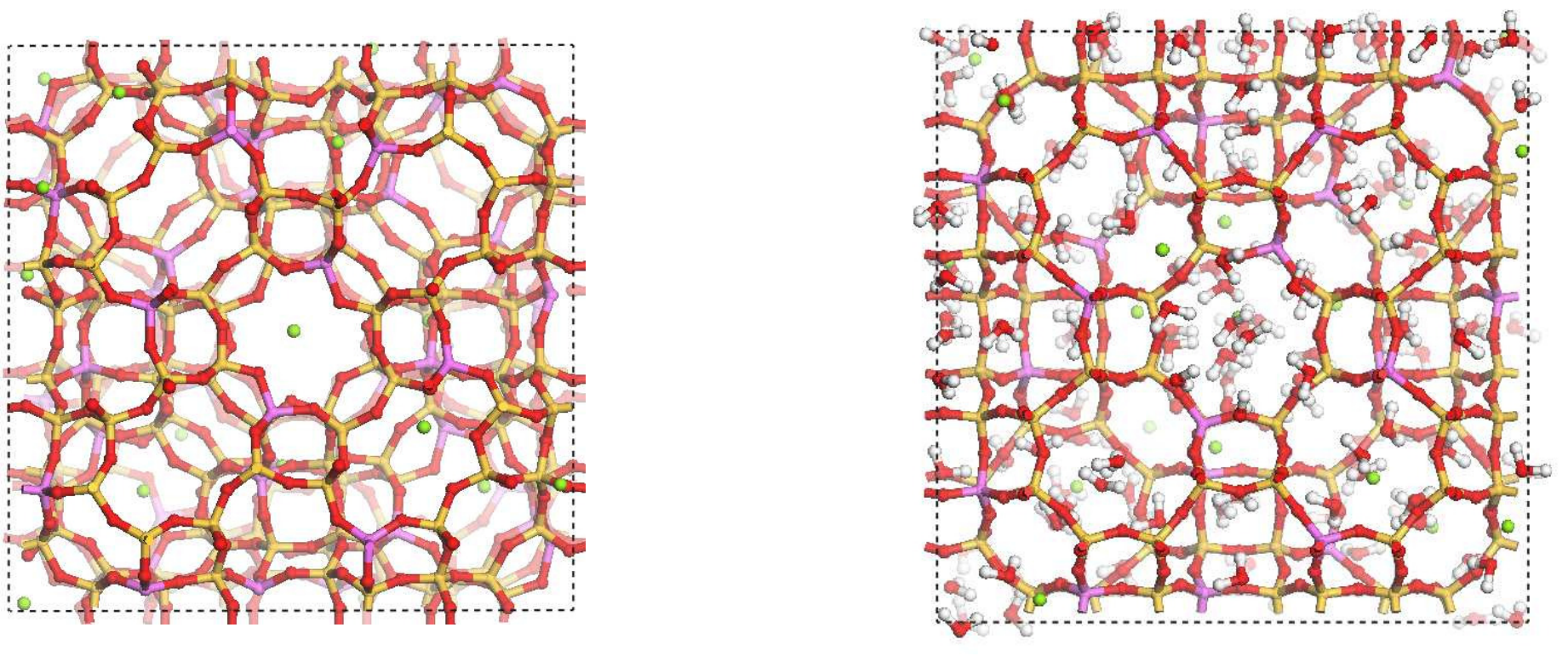}
  \end{subfigure}
  \begin{subfigure}{\textwidth}
  \centering
  \caption{\label{fig:PAU_Ca}(Left) Dehydrated (\acentric form, $\Delta\simeq 1.92$~\si[mode=text]{\angstrom}, $a = 33.6$~\si[mode=text]{\angstrom}) and (Right) hydrated (\centric form, $a = 34.412$~\si[mode=text]{\angstrom}) Ca-PAU with $\nicefrac{\text{Si}}{\text{Al}}=3.2$ and 288 water molecules per unit cell.}
  \includegraphics[width=\textwidth]{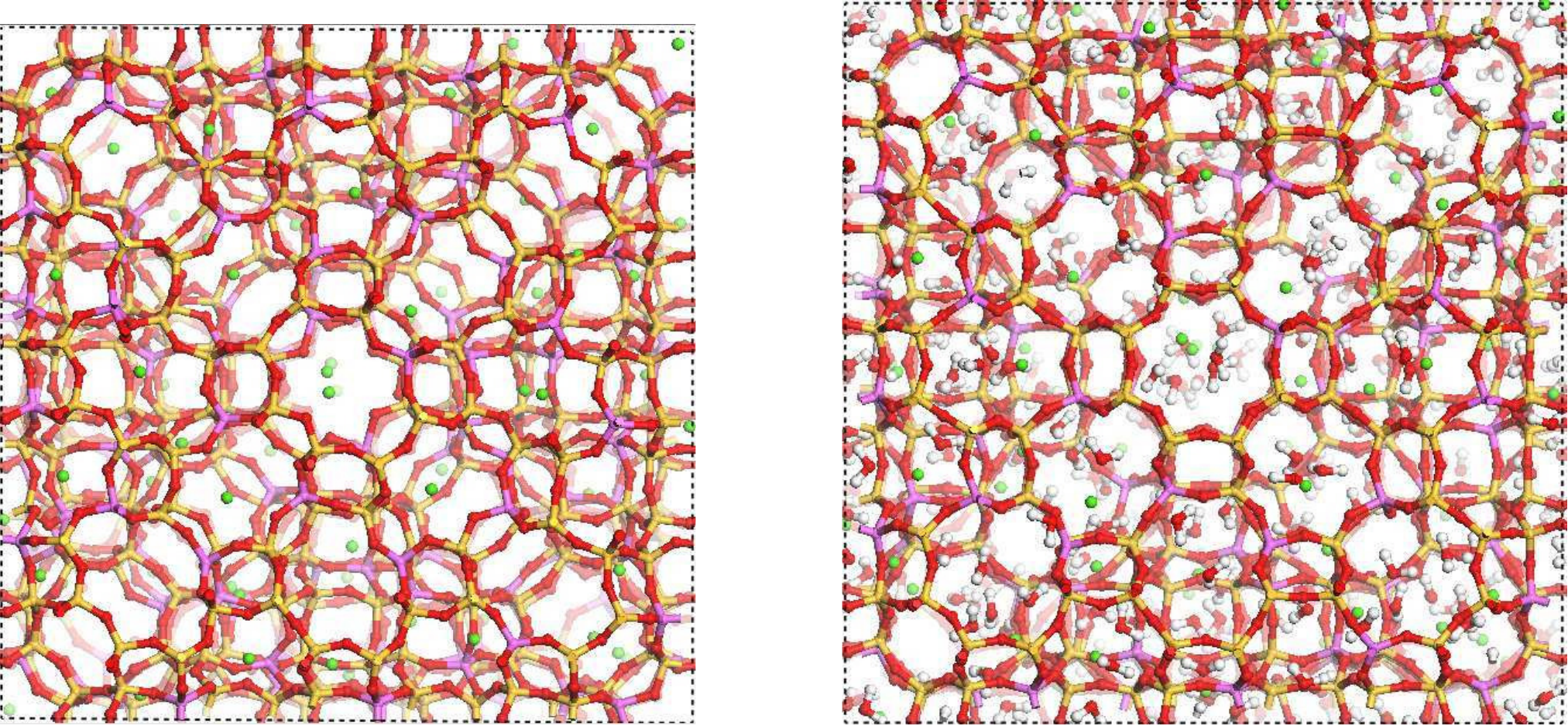}
  \end{subfigure}
  \caption{\label{fig:G1-3_hydrated_dehydrated}
  Aluminosilicate $\text{Ca-}(G_k)_{k=1}^{3}$ models optimised at \SI[mode=text]{0}{\giga\Pa} in their hydrated (right) and fully dehydrated (left) states. Reported cell parameters are the equivalent cubic edges $a = V^{1/3}$ of the optimised cell. Each model uses one representative Si/Al ordering compatible with the experimental Si/Al ratio.
  }
\end{figure*}

We first tested whether the hydration-driven \centacent{} response known for RHO is retained when realistic aluminosilicate chemistry is introduced in the first larger members of the isoreticular family. 
For this purpose, we performed fully flexible energy optimisations of representative \ce{Ca}-aluminosilicate models of $G_1$, $G_2$ and $G_3$, comparing hydrated and fully dehydrated states in each case. 
\textbf{Figure} \ref{fig:G1-3_hydrated_dehydrated} collects the optimised structures for $G_1$ (Ca-RHO, $\nicefrac{\text{Si}}{\text{Al}} = 3.8$), $G_2$ (Ca-PST-29, $\nicefrac{\text{Si}}{\text{Al}} = 4.45$) and $G_3$ (Ca-PAU, $\nicefrac{\text{Si}}{\text{Al}} = 3.2$), and reveals a common hydration-controlled structural response in the three models.
Each model includes the experimentally reported divalent-cation distribution, while the hydrated structures carry water loadings obtained by rigid-framework adsorption with RASPA\cite{Dubbeldam2016} at \SI[mode=text]{1}{\bar}.

In the hydrated state, all three frameworks remain in the symmetric \centric{} phase and the D8R windows are essentially undistorted ($\Delta\simeq 0$).
This is consistent with the \ce{Ca^{2+}} cations and adsorbed water molecules stabilising the high-symmetry, open configuration through their interaction with the 8-ring windows. 
After complete dehydration, the same frameworks relax into distorted \acentric{} structures, with substantial D8R distortions of $\Delta\simeq 1.6$, $1.9$ and $1.92$~\si[mode=text]{\angstrom} for $G_1$, $G_2$ and $G_3$, respectively.  These distortions are accompanied by cell-parameter contractions of about $5\%$, $4\%$ and $2.3\%$, which decrease with the isoreticular order.

The nearly constant value of $\Delta$ in the dehydrated structures suggests that the local D8R distortion is preserved from RHO to PAU, whereas the macroscopic strain becomes smaller as the same local distortion is embedded in larger frameworks.  This is the first indication that isoreticular expansion does not suppress the local switching motif, but changes how strongly it is expressed at the unit-cell level.  The structural assignments are also consistent with available experiments. 
For $G_1$, the dehydrated Ca-RHO cell parameter ($a\simeq\SI[mode=text]{14.2}{\angstrom}$) and \acentric{} symmetry agree with the experimental Ca-RHO structure of \citeauthor{Corbin1990}\cite{Corbin1990} within $\sim 1\%$. 
For $G_2$, the experimental reference is limited to the hydrated PST-29 structure reported by \citeauthor{Lee2018};\cite{Lee2018} to our knowledge, the corresponding dehydrated form has not yet been characterised experimentally. The Ca-aluminosilicate cell parameters bracket the SLC pure-silica value $a = 24.67$~\si[mode=text]{\angstrom} of \textbf{Table} \ref{tab:SI-structures} (dehydrated Ca-PST-29 $a \simeq 24$~\si[mode=text]{\angstrom}, hydrated Ca-PST-29 $a \simeq 25.1$~\si[mode=text]{\angstrom}), as expected if extra-framework cations and adsorbed water act as a chemical analogue of the mechanical compression and expansion that defines $p_c$ in the pure-silica framework.
For $G_3$, the dehydrated cell parameter ($a=\SI[mode=text]{33.6}{\angstrom}$) and \acentric{} symmetry reproduce the dehydrated paulingite reported by \citeauthor{Bieniok1997}.\cite{Bieniok1997} The cell parameters quoted in the main text and in the figure captions are equivalent cubic edges $a = V^{1/3}$ of the optimised cell. This convention is used because one Si/Al periodic ordering locally lowers the cubic symmetry and introduces small artefactual distortions of the metric tensor, which the statistical Al distribution of the real material averages out at the macroscopic scale.

These optimisations should be read as a chemically realistic qualitative test, not as a full thermodynamic description of the hydrated frameworks. 
The water loading was obtained from rigid-framework adsorption rather than from a fully relaxed framework-water equilibrium, the \ce{Si}/\ce{Al} distribution in each cell is one representative configuration rather than an ensemble average, and static minimisation does not give access to the elastic response or to the free-energy barriers along the transition pathway. Even with these limitations, the calculations show that realistic Ca-aluminosilicate models of $G_1$-$G_3$ follow the same hydration-controlled pattern: hydration stabilises the open \centric{} form, whereas dehydration drives the framework towards the distorted \acentric{} form. They therefore provide chemically explicit support for the idea that the local switching motif is retained in the lower members of the family, complementing the experimental record summarised in the Introduction (\textbf{Tables} \ref{tab:SI-validation} and \ref{tab:SI-expdata}).

Extending the same explicit aluminosilicate strategy beyond $G_3$ is not practical with the present level of theory. For example, a realistic $G_4$ model (MWF, $1440$ T-atoms) with $\nicefrac{\text{Si}}{\text{Al}} = 3$ would contain about $360$ Al atoms and a very large number of inequivalent \ce{Si}/\ce{Al} orderings satisfying L\"owenstein's rule. A rigorous treatment would require symmetry-aware configurational enumeration, for example with the SOD code,\cite{GrauCrespo2007} followed by water adsorption and framework relaxation for the relevant configurations. Even a single pressure point would then require many thousands of independent optimisations before any elastic or free-energy analysis could be attempted. 
For the larger members we therefore reduce the chemical complexity to the pure-silica frameworks and use external hydrostatic pressure as a mechanical proxy for framework softness.

The configurational cost of such an explicit aluminosilicate treatment is already prohibitive for the smaller members of the family.
For the representative compositions used above, the number of possible \ce{Si}/\ce{Al} arrangements before applying any energetic selection is $\binom{48}{10}=6.5\times10^{9}$ for $G_1$ ($\nicefrac{\mathrm{Si}}{\mathrm{Al}}=3.8$), $\binom{240}{44}=3.0\times10^{48}$ for $G_2$ ($\nicefrac{\mathrm{Si}}{\mathrm{Al}}\simeq4.45$), and $\binom{672}{160}=5.6\times10^{158}$ for $G_3$ ($\nicefrac{\mathrm{Si}}{\mathrm{Al}}=3.2$).
Löwenstein's rule reduces this space, but it does not make it tractable: a conservative lower bound is obtained by placing all Al atoms on one independent sublattice of the bipartite T-site graph, which gives $\binom{24}{10}=2.0\times10^{6}$, $\binom{120}{44}=1.3\times10^{33}$ and $\binom{336}{160}=4.2\times10^{99}$ configurations for $G_1$, $G_2$ and $G_3$, respectively.
A full symmetry-aware enumeration with SOD\cite{GrauCrespo2007} would reduce these numbers by crystallographic equivalence, but only by factors that are negligible on this logarithmic scale, and each remaining framework would still require charge-balancing cation placement, water loading and structural relaxation.
Thus, the difficulty is not only the cost of the force-field or DFT optimisation itself, but the combinatorial explosion of realistic aluminosilicate models before any relaxation can be performed.
For this reason, the pure-silica pressure scans used below should be read as the closest currently tractable atomistic route to the intrinsic framework response of the larger isoreticular members, while the chemically explicit aluminosilicate models of $G_1$-$G_3$ provide the validation of the local switching motif.

This mapping is motivated by previous work on RHO. \citeauthor{Balestra2015} showed that the electrostatic action of extra-framework cations on framework oxygen atoms can produce a structural effect similar to external compression, and that progressive screening of the cation charge, used as a proxy for hydration, follows the same \centtoacent{} pathway as hydrostatic compression of the bare framework,\cite{Balestra2015} with a structural transition at $p_c \simeq 1$~\si[mode=text]{\giga\Pa} producing an acentricity $\Delta\simeq 0.25$~\si[mode=text]{\angstrom} at $p_c$ and a cell contraction of about $5\%$ that match the Ca-aluminosilicate values reported above to within the uncertainty of the rigid-framework water adsorption step.\cite{Balestra2015} The dehydration of an EFC-loaded aluminosilicate and the hydrostatic compression of the bare pure-silica framework therefore access the same Landau-type free-energy landscape, and the pressure-driven response of the pure-silica family can be read as a semi-quantitative proxy for the dehydration-driven response of its aluminosilicate counterpart, with the quantitative match established explicitly for RHO ($G_1$).

\subsection{Soft mode and the critical pressure}

\begin{figure*}[!htbp]
 \centering
 \includegraphics[width=0.66\textwidth]{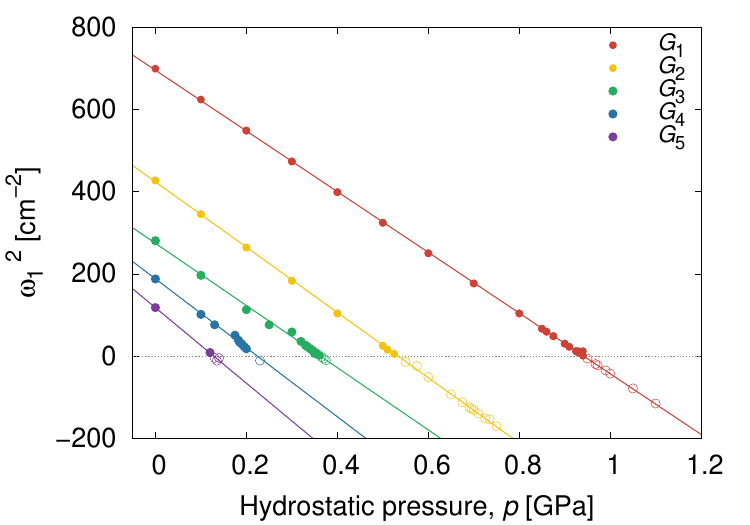}
 \caption{\label{fig:Phonons}Squared soft-mode frequency $\omega_1^{2}(p)$ for $G_1$ to $G_5$ on the parent-symmetry branch, coloured by member, together with the Cowley-Levanyuk linear fits $\omega_1^{2}(p) = \alpha(p_c - p)$ of \textbf{Equation} \ref{eq:cowley-levanyuk} (solid lines, vertical dashed line at each $p_c$). Filled circles correspond to RFO-converged Hessians with all real frequencies ($\omega_1^{2} > 0$, stable parent branch); open circles correspond to the structures relaxed with the GULP lower-symmetry option, where the parent-symmetry Hessian carries a negative-curvature direction and the soft mode is therefore reported as imaginary ($\omega_1^{2} < 0$, plotted on the negative ordinate with the conventional sign convention). The horizontal grey line is $\omega_1^{2} = 0$. The critical pressure of each member, obtained from the zero-crossing of its fit, is tabulated in \textbf{Table} \ref{tab:parameters}.}
\end{figure*}

To identify the microscopic driver of the transition and obtain a canonical critical pressure for each member of the family, we computed the lowest $\Gamma$-point phonon modes as a function of hydrostatic pressure for $G_1$-$G_5$. The detailed pressure evolution of the six lowest phonon branches per member, including the D8R-distortion-colour-coded transition from the parent (\centric{}) to the lower-symmetry (\acentric{}) branch, is reported in \textbf{Figure} \ref{fig:SI-phonon-modes} of the Supporting Information. In all members, only the lowest mode $\omega_1$ softens with increasing pressure and reaches $\omega_1 = 0$ at a topology-dependent critical pressure $p_c$, while the next five branches remain essentially flat in the range scanned. The same mode becomes formally imaginary above $p_c$ when the parent symmetry is retained, and is plotted with the conventional sign convention on the ordinate of \textbf{Figure} \ref{fig:Phonons}.

The pressure dependence of the soft-mode frequency on the parent-symmetry branch follows the classical Cochran soft-mode (Cowley-Levanyuk) linear law of a second-order structural transition,\cite{Cowley1976, LevanyukSigov1988}
\begin{equation}
   \label{eq:cowley-levanyuk}
   \omega^{2}(p) = \alpha\,(p_c - p),
\end{equation}
where $\alpha > 0$ is a member-dependent slope and $p_c$ is the critical pressure.
The transition $Im\bar{3}m \to I\bar{4}3m$ removes spatial inversion while preserving every other generator of the cubic parent group, so a discrete $\mathbb{Z}_2$ symmetry is broken. The corresponding order parameter $\Delta$ transforms as the one-dimensional irreducible representation $A_{2u}$ of $Im\bar{3}m$, which is the only odd-parity irrep that branches onto the identity $A_1$ of $I\bar{4}3m$ in the $O_h \to T_d$ reduction. Under the broken inversion centre $\Delta \to -\Delta$, so the Landau expansion of $\Delta$ contains only even powers by symmetry, and the broken-phase vacuum is the discrete two-point set $\{+\Delta_\text{eq},-\Delta_\text{eq}\}$ rather than a continuous manifold. The soft mode of \textbf{Equation} \ref{eq:cowley-levanyuk} is therefore a Cochran-type displacive instability, not a Goldstone mode: it freezes only at $p_c$ and re-stiffens on either side, in contrast to the gapless Nambu-Goldstone branch that would accompany a spontaneously broken continuous symmetry.
A linear regression of $\omega^{2}(p)$ on the parent branch, restricted to the monotonically decreasing portion below the transition, returns the $p_c(G_k)$ values listed in \textbf{Table} \ref{tab:parameters}.
The same procedure was used for $G_1$ through $G_4$, where dense phonon calculations were performed across the transition window with GULP.
For $G_1$ and $G_2$, the scan covers both sides of $p_c$ and yields a clean linear law with measured imaginary modes above $p_c$ ($\omega_1^{2} < 0$); for $G_3$ and $G_4$, the analysis covers the stable \centric{} branch up to and slightly past $p_c$, with three (one) imaginary points available for $G_3$ ($G_4$), and the linear extrapolation remains robust within the restricted upper window.
For $G_5$ (MWF, $2640$ T-atoms), the rational-function-optimisation phonon refinement requires the assembly and inversion of the full Hessian matrix, whose memory footprint scales as the square of the number of degrees of freedom and saturates the available memory on our hardware; a full pressure scan of the phonon spectrum was therefore prohibitive. The six parent-branch rows available around $p_c$, plus the exclusion of a spurious post-transition reappearance of a positive real mode at $p\simeq 0.16$~\si[mode=text]{\giga\Pa}, anchor the same linear extrapolation and return a $p_c$ with a broader uncertainty (full diagnostic in \textbf{Section} \ref{si:G5-phonon} of the Supporting Information).
The fitted slopes are remarkably uniform across the family, $\alpha\simeq 7.4$-$9.2\times 10^{2}~\si[mode=text]{\per\centi\meter\squared\per\giga\Pa}$, consistent with the topology-independent pre-transition bulk stiffness $K_0 \simeq 1/|\kappa_1|$ discussed in the Landau picture below.
For $G_6$-$G_8$ (T-atom counts of $4368$, $6720$ and $9792$), the same memory bottleneck applies even more severely and no direct phonon analysis is attempted in the present work; their critical pressures are obtained from the exponential scaling of $p_c$ with isoreticular order discussed below.

These canonical $p_c$ values are used as input for the structural fits of the next subsection and for the elastic and Landau analyses that follow.

\subsection{Pressure-driven structural response of the pure-silica family}
\begin{figure}[!htbp]
 \centering
 \begin{subfigure}{0.48\textwidth}
  \centering
  \caption{\label{fig:VolumeVsPressure}}
  \includegraphics[width=\textwidth]{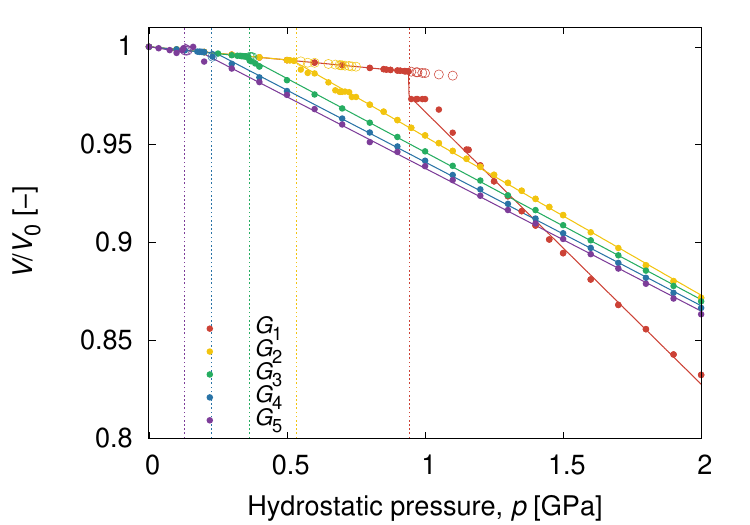}
 \end{subfigure}
 \begin{subfigure}{0.48\textwidth}
  \centering
  \caption{\label{fig:CellVsPressure}}
  \includegraphics[width=\textwidth]{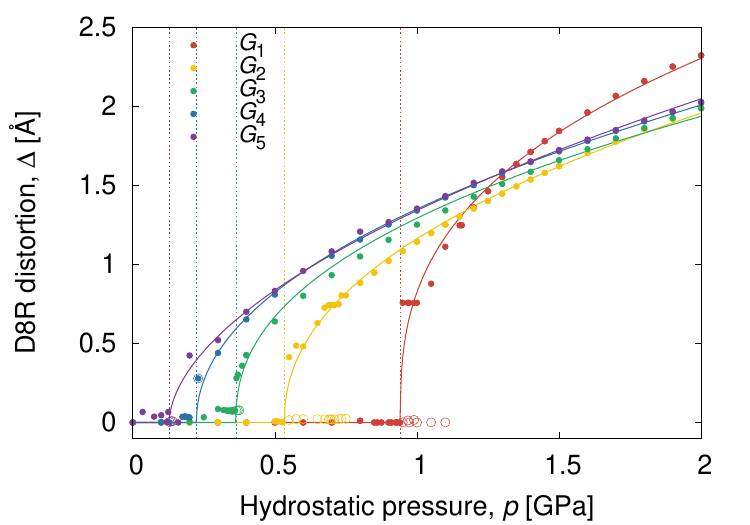}
 \end{subfigure}
 \caption{\label{fig:VolDelta}(a) Volume shrinkage, $v$, and (b) order parameter, $\Delta$ (D8R distortion) \vs applied hydrostatic pressure, $p$, for $G_1$ to $G_5$ structures. Solid lines are fitted functions \textbf{Equations} \ref{eq:kappa} and \ref{eq:fit}, for a) and b) panels. Filled and open circles follow the same convention as in \textbf{Figure} \ref{fig:Phonons}: filled circles are RFO-converged structures with all real frequencies on the parent branch, and open circles are structures relaxed with the GULP lower-symmetry option where the parent-symmetry Hessian carries an imaginary soft mode.}
\end{figure}

With the canonical $p_c(G_k)$ from the soft-mode analysis as input, we characterised the structural response of the pure-silica family under hydrostatic pressure between 0 and \SI[mode=text]{2}{\giga\Pa} (\textbf{Figure} \ref{fig:VolDelta}).
All five frameworks show the same basic response: the cell volume decreases linearly at low pressure, while the D8R distortion remains essentially zero up to $p_c$.
Above $p_c$, the framework enters the distorted \acentric{} branch, the volume decreases with a steeper slope, and the order parameter $\Delta$ grows continuously or weakly discontinuously depending on the member of the family.

We first fitted the relative volume, $v\defeq V/V_0$, where $V_0$ is the volume at \SI[mode=text]{0}{\giga\Pa}. 
Because $v(0)=1$ by definition, the equation of state was described with separate linear branches below and above the transition:
\begin{equation}
   v(p) =
   \begin{cases}
      1+\kappa_1 p, & p<p_c,\\
      v_c^+ + \kappa_2(p-p_c), & p\geq p_c,
   \end{cases}
 \label{eq:kappa}
\end{equation}
where $\kappa_1$ and $\kappa_2$ are compressibility-like slopes and $v_c^+$ is the fitted volume of the distorted branch at $p_c$. Each branch is fitted independently, with a gap of half-width $h = 0.06$~\si[mode=text]{\giga\Pa} excluded around $p_c$ to prevent the very-near-critical points (where the parent and broken branches mix) from biasing the linear slopes.
This form allows for a volume discontinuity at the transition and avoids forcing the two branches to meet at $v=1$.
The pre-transition slope is very similar across the family, with an average value $\kappa_1\simeq -0.014~\si[mode=text]{\per\giga\Pa}$ for $G_1$-$G_4$.
The post-transition slope is much larger in magnitude, especially for $G_1$ ($\kappa_2=-0.139~\si[mode=text]{\per\giga\Pa}$), and lies between $-0.073$ and $-0.081~\si[mode=text]{\per\giga\Pa}$ for $G_2$-$G_5$ (\textbf{Table} \ref{tab:parameters}).
This indicates that the distorted \acentric{} branch is mechanically softer than the initial \centric{} branch. The fitted intercepts $v_c^+$ above $p_c$ give a small apparent volume jump at the transition, $|\Delta v|<0.5\%$ for $G_2$-$G_5$, while $G_1$ stands apart with a much larger and negative apparent jump ($\Delta v = -1.23\%$), consistent with its singular (near-tricritical) position at the SLC level (\textbf{Section} \ref{si:landau-k-parametrisation}).

The pressure dependence of the D8R distortion gives the second descriptor of the transition. 
For all members, $\Delta\simeq 0$ below $p_c$ and increases above $p_c$ following a power-law-like behaviour in the reduced pressure. 
We fitted the data with the phenomenological expression
\begin{equation}
    \label{eq:fit}
    \Delta = \delta \left(\frac{p}{p_c}-1\right)^{\beta} H(p-p_c),
\end{equation}
where $H(p-p_c)$ is a Heaviside function centred at $p_c$. 
Here $\delta$ is an amplitude parameter that sets the scale of the distorted branch, while $\beta$ is an effective exponent describing how fast the distortion grows after the transition. 
Both parameters should be read as phenomenological descriptors of the pressure scan, not as exact critical quantities.

The fitted parameters show a clear dependence on isoreticular order.
The critical pressure $p_c$, fixed at the canonical soft-mode value (\textbf{Equation} \ref{eq:cowley-levanyuk}), decreases monotonically from $p_c=0.942~\si[mode=text]{\giga\Pa}$ for RHO to $p_c=0.128~\si[mode=text]{\giga\Pa}$ for $G_5$.
The amplitude $\delta$ decreases from $2.21~\si[mode=text]{\angstrom}$ in $G_1$ to $0.46~\si[mode=text]{\angstrom}$ in $G_5$, showing that the same D8R distortion becomes less strongly expressed at the unit-cell scale as the framework grows.
The effective exponent $\beta$ clusters between $0.43$ and $0.55$ for $G_2$-$G_5$, fully compatible with the mean-field value $1/2$ within the fit uncertainty, whereas RHO ($G_1$) returns a noticeably lower value, $\beta = 0.355 \pm 0.009$.
This singular position of $G_1$ in the family, together with the apparent negative volume jump at $p_c$ ($\Delta v = -1.23\%$, the only negative apparent jump of the family), is interpreted as the sextic Landau crossover of a near-tricritical continuous transition on the SLC scale in the dedicated subsection below; the DFT benchmark suggests that the true transition is weakly first-order, an interpretation that is fully consistent with the same Landau framework once the sign of $U_\text{eff}(G_1)$ is reversed (\textbf{Section} \ref{si:landau-k-parametrisation}).
Structurally, the singular position of $G_1$ is consistent with the fact that all 8MRs in RHO belong to D8Rs, so the distortion propagates through the whole framework rather than being diluted by larger intermediate cages.

\begin{table}[h!]
\caption{\label{tab:parameters}Phenomenological parameters of the pressure-driven transition for each $G_k$: soft-mode critical pressure $p_c$, Heaviside-power amplitude $\delta$ and effective exponent $\beta$ of \textbf{Equation} \ref{eq:fit}, pre- and post-transition compressibility-like slopes $\kappa_1$, $\kappa_2$ of \textbf{Equation} \ref{eq:kappa}, and volume jump at the transition $\Delta v$. Errors are one standard deviation from the non-linear fit.}
\centering
\small
\setlength{\tabcolsep}{6pt}
 \begin{tabular}{c c c c c c c}
 \toprule
    $k$ & $p_c$ [\si[mode=text]{\giga\Pa}] & $\delta$ [\AA] & $\beta$ &
    $\kappa_1$ [\si[mode=text]{\per\giga\Pa}] &
    $\kappa_2$ [\si[mode=text]{\per\giga\Pa}] & $\Delta v$ \\
 \midrule
 1 & $0.9418 \pm 0.0006$ & $2.210$ & $0.355 \pm 0.009$ & $-0.0135$ & $-0.1392$ & $-0.0123$ \\
 2 & $0.5338 \pm 0.0007$ & $1.236$ & $0.455 \pm 0.005$ & $-0.0134$ & $-0.0810$ & $-0.0010$ \\
 3 & $0.3627 \pm 0.0019$ & $1.014$ & $0.429 \pm 0.016$ & $-0.0141$ & $-0.0753$ & $-0.0008$ \\
 4 & $0.2241 \pm 0.0024$ & $0.744$ & $0.480 \pm 0.012$ & $-0.0139$ & $-0.0734$ & $+0.0012$ \\
 5 & $0.1282 \pm 0.0026$ & $0.464$ & $0.554 \pm 0.018$ & $-0.0217$ & $-0.0730$ & $+0.0043$ \\
 \bottomrule
 \end{tabular}
\end{table}

For $G_1$, the jump of $\Delta$ at $p_c$ is the largest of the family, of order $0.4$~\si[mode=text]{\angstrom} within a $\sim 100$-bar interval. 
This produces appreciable scatter in the \acentric{} branch of $\Delta(p)$ when the 48-T-atom unit cell is used. 
To reduce this finite-size noise, the $G_1$ parameters reported in \textbf{Table} \ref{tab:parameters} were obtained from a $2\times 2\times 2$ supercell (\textbf{Section} \ref{si:finite-size}). 
The unit-cell and supercell calculations give the same critical pressure and elastic constants within the fit uncertainty, so the $G_1$ values can be compared directly with those of $G_2$-$G_5$. 
A unified interpretation of these fitted parameters, together with the elastic signatures reported next, is given in the dedicated Landau picture subsection below.
The full finite-size comparison and the fitted volume parameters are reported in \textbf{Sections} \ref{si:finite-size} and \ref{si:volume-fits} of the Supporting Information.


\subsection{Mechanical softening and exponential $p_c(k)$ law}

\begin{figure*}[!t]
\centering
\begin{subfigure}{0.32\textwidth}
  \centering
  \caption{\label{fig:Mechanical:G1}$C_{ij}$ for Rho ($G_1$)}
  \includegraphics[width=\textwidth]{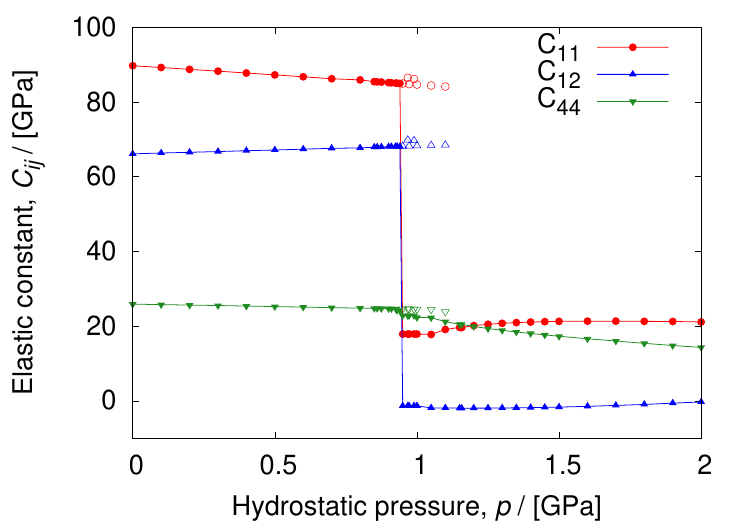}
\end{subfigure}\hfill
\begin{subfigure}{0.32\textwidth}
  \centering
  \caption{\label{fig:Mechanical:G2}$C_{ij}$ for PST-29 ($G_2$)}
  \includegraphics[width=\textwidth]{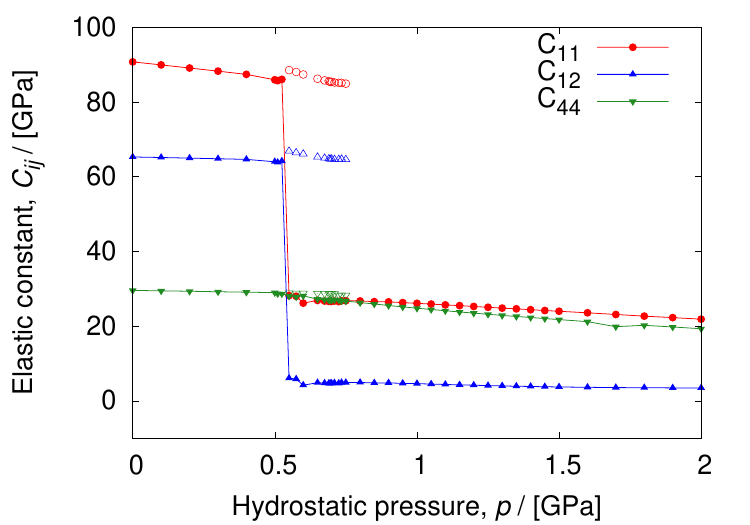}
\end{subfigure}\hfill
\begin{subfigure}{0.32\textwidth}
  \centering
  \caption{\label{fig:Mechanical:G3}$C_{ij}$ for Pau ($G_3$)}
  \includegraphics[width=\textwidth]{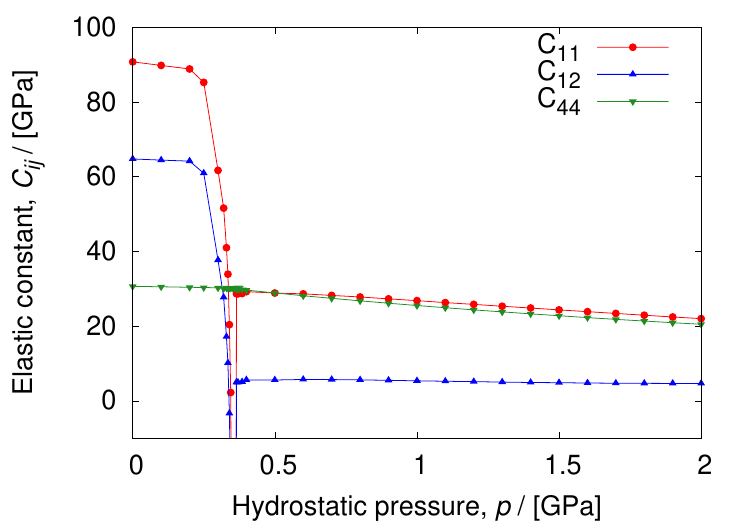}
\end{subfigure}

\vspace{0.4em}

\begin{subfigure}{0.48\textwidth}
  \centering
  \caption{\label{fig:Mechanical:kappa}$\kappa_T$ for $G_1$ to $G_5$}
  \includegraphics[width=\textwidth]{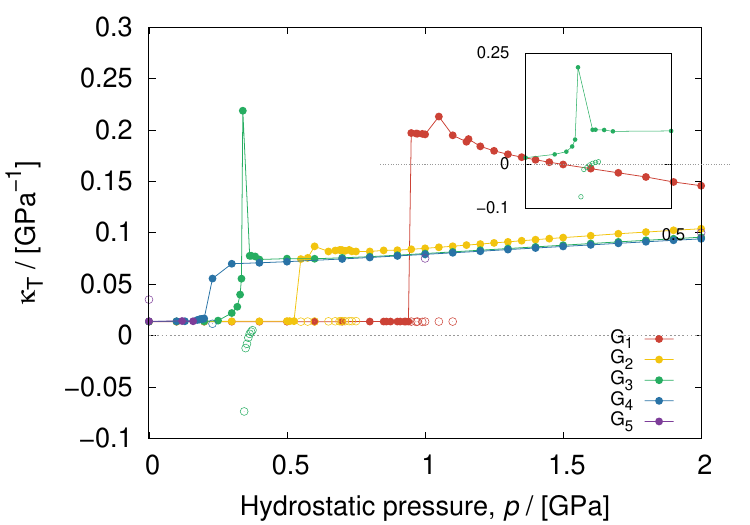}
\end{subfigure}\hfill
\begin{subfigure}{0.48\textwidth}
  \centering
  \caption{\label{fig:Mechanical:pc}$p_c(k\!-\!1) = 0.93\,e^{-0.49\,(k-1)}$ GPa}
  \includegraphics[width=\textwidth]{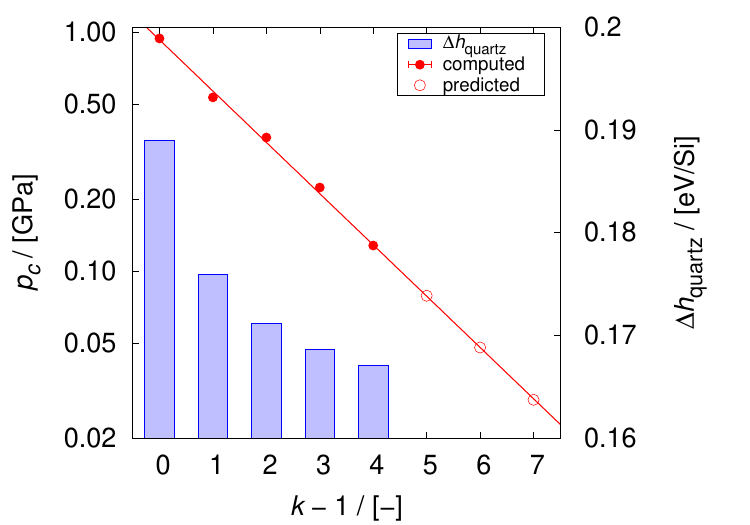}
\end{subfigure}
\caption{\label{fig:Mechanical}Pressure response of the cubic elastic constants and the soft-mode critical pressure across the RHO isoreticular family. Panels (a)-(c): $C_{11}$, $C_{12}$, $C_{44}$ \vs hydrostatic pressure for $G_1$-$G_3$. Panel (d): isothermal compressibility $\kappa_T = -V^{-1}(\partial V/\partial p)_T$ for $G_1$-$G_5$; filled circles connected by lines are phonon- and elastically-stable points (\mbox{$\omega_1 \geq 0$ AND $\kappa_T > 0$}), open circles are phonon- or elastically-unstable points (\mbox{$\omega_1 < 0$ OR $\kappa_T < 0$}). Dashed horizontal line at $\kappa_T = 0$ as a stability marker. Panel (e): soft-mode-derived $p_c$ \vs isoreticular order $k-1$, filled red circles for the computed members $G_1$-$G_5$ and open red circles for the predicted $G_6$-$G_8$, together with the exponential fit of \textbf{Equation} \ref{eq:pc-exponential} (red line); per-Si enthalpy relative to $\alpha$-quartz $\Delta h_\text{quartz}$ overlaid as blue bars on the secondary right axis.}
\end{figure*}

A $\Gamma$-point soft mode is expected to couple to homogeneous distortions of the unit cell, and therefore to the elastic response of the framework.
The pressure evolution of the cubic elastic constants $C_{11}$, $C_{12}$ and $C_{44}$, together with the isothermal compressibility $\kappa_T = -V^{-1}(\partial V/\partial p)_T$, is reported in \textbf{Figure} \ref{fig:Mechanical}.
Panels (a)-(c) show the elastic constants for $G_1$-$G_3$, while panel (d) gives $\kappa_T$ for the full $G_1$-$G_5$ series.

The elastic data follow the same trend as the phonons.
At the transition, $C_{11}$ drops from about $90$~\si{\giga\Pa} to $20$-$25$~\si{\giga\Pa}, $C_{12}$ decreases from about $65$~\si{\giga\Pa} to values close to zero, and $C_{44}$ shows a milder softening.
In the \centric{} phase, all five frameworks have a very similar compressibility, $\kappa_T\approx 0.013$-$0.015$~\si{\per\giga\Pa}, corresponding to a bulk modulus of about $K_T\approx 70$~\si{\giga\Pa}.
After the transition, $\kappa_T$ increases by roughly one order of magnitude and reaches $\kappa_T\approx 0.08$-$0.10$~\si{\per\giga\Pa} for $G_2$-$G_5$.
The response of $G_1$ is sharper, with a pronounced peak around $p_c$, consistent with the more first-order-like behaviour inferred from the pressure dependence of $\Delta$.
The bulk modulus obtained from the elastic constants, $K_T=(C_{11}+2C_{12})/3$, agrees with $1/\kappa_T$ on both sides of the transition for the members where both quantities were computed, providing an internal consistency check of the mechanical analysis.

A narrow negative excursion of $\kappa_T$ is observed for $G_3$ in the range $0.345 \lesssim p \lesssim 0.36$~\si{\giga\Pa} (open circles below the dashed $\kappa_T=0$ line in panel (d)).
We interpret this feature as the signature of a metastable \centric{} branch held slightly beyond its mechanical-stability limit, where $\omega_1\simeq 0$ and $\partial V/\partial p>0$.
It should therefore not be read as a stable negative-compressibility regime, but as a numerical marker of the instability of the parent branch close to $p_c$.
A denser pressure sampling near $p_c$ for $G_4$ does not show a comparable excursion, indicating that this feature is not a general requirement of the transition.

Finally, panel (e) summarises the critical pressures, taken from the Cowley-Levanyuk soft-mode fits (\textbf{Equation} \ref{eq:cowley-levanyuk} and \textbf{Table} \ref{tab:parameters}), as a function of isoreticular order.
A non-linear least-squares fit of the form
\begin{equation}
\label{eq:pc-exponential}
p_c(k)=p_0\,\exp[-\lambda(k-1)]
\end{equation}
gives $p_0 = 0.930 \pm 0.022$~\si{\giga\Pa} and $\lambda = 0.494 \pm 0.023$.
The fitted prefactor is statistically indistinguishable from the computed value for RHO, $p_c(G_1)=0.942$~\si{\giga\Pa}.
The exponential law captures the soft-mode-derived critical pressures of $G_1$-$G_5$ within a residual sum of squares roughly half that obtained when the structurally-fitted $p_c$ are used instead, supporting the canonical (Cowley-Levanyuk) definition of $p_c$ adopted in \textbf{Table} \ref{tab:parameters}.
The per-Si enthalpy relative to $\alpha$-quartz, $\Delta h_\text{quartz}$, plotted as blue bars on the right axis, decreases monotonically from about $0.19$ to $0.17$~\si{\electronvolt}/Si along the series, with consecutive gaps of order $\sim 0.005$~\si{\electronvolt}/Si, that is, a fraction of room-temperature $k_B T \simeq 0.026$~\si{\electronvolt}, for $k\gtrsim 5$.

The comparison with experiment is not equally direct across the series, because the available measurements do not probe the same observable in every member.
For $G_1$ (RHO), high-pressure single-crystal diffraction on Cd-RHO locates the \centtoacent{} transition at $p_c^\text{exp}\simeq 0.4$~\si[mode=text]{\giga\Pa}.\cite{Lee2001}
The pure-silica SLC value, $p_c=0.942$~\si[mode=text]{\giga\Pa}, is higher, as expected because the divalent \ce{Cd^{2+}} cations provide an additional electrostatic compression that must be supplied externally in the pure-silica model.
This is the same structural mapping discussed above for Ca-RHO and previously established by \citet{Balestra2015}.
Thus, for $G_1$ the calculation captures the pressure scale and the symmetry change, but should not be read as a quantitative reproduction of the experimental transition pressure.

For $G_3$ (PAU/ECR-18), dehydration-induced powder diffraction reports a cell-parameter contraction of $\Delta a/a=-4.4\%$, corresponding to $\Delta V/V\simeq -12.6\%$, between the hydrated $Im\bar{3}m$ phase and the fully dehydrated $I\bar{4}3m$ phase.\cite{Greenaway2015}
The pure-silica SLC pressure scan gives a smaller cumulative contraction over the same structural interval (from $p = 0$ to $\sim 2$~\si[mode=text]{\giga\Pa}, i.e. integrated across the transition rather than restricted to the discontinuity at $p_c$), $\Delta a/a\simeq -1.5\%$ and $\Delta V/V\simeq -4.5\%$; the discrete volume jump localised at $p_c$ from the piecewise-linear fit of \textbf{Equation} \ref{eq:kappa} is much smaller ($\Delta v = -0.08\%$, \textbf{Table} \ref{tab:parameters}), with the bulk of the contraction occurring smoothly above $p_c$ as $\Delta$ grows.
The sign and order of magnitude are therefore reproduced, but the magnitude is underestimated, consistently with the absence of the extra-framework cations and hydration effects present in the real aluminosilicate.
For $G_4$ (ZSM-25), the available evidence is not a direct crystallographic measurement of the \centtoacent{} transition.
Instead, K-ZSM-25 shows trapdoor \ce{N2}/\ce{CH4} selectivity through temperature-regulated cation gating,\cite{Zhao2021} which is consistent with a flexible framework response of the type predicted here, but does not provide a direct experimental value of $p_c$ or $\Delta$ for comparison.
The full set of available experimental observables for $G_1$-$G_5$ is summarised in \textbf{Tables} \ref{tab:SI-validation} and \ref{tab:SI-expdata} of the Supporting Information.

\subsection{Landau picture and effective scaling}

The previous sections identified four connected signatures of the same pressure-driven instability in the RHO isoreticular series: a soft $\Gamma$-point phonon that reaches zero at $p_c$ (\textbf{Figure} \ref{fig:Phonons}), a sharp softening of the cubic elastic constants $C_{11}$ and $C_{12}$ at the transition (\textbf{Figure} \ref{fig:Mechanical}), a piecewise-linear volume response with different slopes below and above $p_c$ (\textbf{Figure} \ref{fig:VolDelta} and \textbf{Equation} \ref{eq:kappa}), and an order-parameter growth $\Delta \propto (p/p_c - 1)^{\beta}$ with effective exponents close to the mean-field value $1/2$ for $G_2$-$G_5$ but a noticeably lower value, $\beta\simeq 0.36$, for $G_1$ accompanied by a finite negative volume jump at $p_c$ ($\Delta v = -1.23\%$, \textbf{Table} \ref{tab:parameters}).
We now show that these observations can be rationalised with a minimal Landau description in which the D8R distortion is coupled to the volume response of the framework.
The same description also explains why $G_2$-$G_5$ are controlled mainly by the quartic term, whereas $G_1$ requires the sixth-order term to capture its lower effective exponent and its proximity to the tricritical condition.

Let $\Delta$ denote the amplitude of the D8R distortion order parameter (\textbf{Equation} \ref{eq:delta_t}).
Following the discussion of \textbf{Equation} \ref{eq:cowley-levanyuk} above, $\Delta$ denotes the magnitude of the signed $A_{2u}$ amplitude; the two equivalent broken-phase variants $\pm\Delta_\text{eq}$ are not distinguished in what follows.
To describe the volume response, we separate the ordinary elastic compression of the parent \centric{} branch from the additional volume change induced by the D8R distortion.
We define the reduced volume $v \defeq V/V_0$, with $V_0$ the zero-pressure volume of the \centric{} phase, and introduce the residual volumetric strain
\begin{equation}
\label{eq:residual-strain}
e \defeq v - v_\mathrm{c}(p),
\qquad
v_\mathrm{c}(p)=1+\kappa_1 p,
\end{equation}
where $v_\mathrm{c}(p)$ is the linear volume response of the \centric{} branch below the transition.
Thus, $e$ measures the part of the volume change associated with the symmetry-breaking distortion, not the smooth compression already present before $p_c$.

The two variables $\Delta$ and $e$ capture the relevant physics of the transition.
$\Delta$ distinguishes the \centric{} and \acentric{} structures: it is zero in the high-symmetry phase and becomes finite when the D8R windows distort.
The residual strain $e$, in contrast, is fully symmetric ($A_{1g}$) and acts as a secondary elastic variable.
Group-theoretically, $\Delta^2$ transforms as $A_{2u}\otimes A_{2u}=A_{1g}$ and can therefore multiply $e$ in the totally symmetric scalar $e\,\Delta^2$, whereas $\Delta$ itself transforms as $A_{2u}$ and is forbidden from entering a linear coupling with $e$.
A minimal effective Landau functional must therefore contain the pressure-dependent harmonic term in $\Delta$, the quartic and sixth-order stabilising terms, the elastic cost of the residual strain, and the lowest-order strain-order-parameter coupling:

\begin{equation}
\label{eq:LandauF}
f(\Delta,e;p) =
\tfrac{1}{2}A\Bigl(1-\nicefrac{p}{p_c}\Bigr)\Delta^2
+\tfrac{1}{4}U\Delta^4
+\tfrac{1}{6}W\Delta^6
+\tfrac{1}{2}K_0 e^2
+G e\Delta^2 .
\end{equation}

Each term in \textbf{Equation} \ref{eq:LandauF} has a direct physical role.
The quadratic term in $\Delta$ is the pressure-dependent stiffness of the \centric{} phase: for $p<p_c$ it is positive and stabilises $\Delta=0$, whereas for $p>p_c$ it becomes negative and drives the symmetry-breaking distortion.
The quartic and sixth-order terms describe the nonlinear stiffness of the D8R distortion.
For most members, the quartic term is sufficient over the pressure range sampled here, but the sixth-order term is needed to describe the more abrupt response of $G_1$; in all cases $W>0$ is imposed by thermodynamic stability of the truncated functional, as required for the broken-phase probability density of $\Delta$ to be normalisable.
The term $K_0 e^2/2$ is the elastic cost of the residual volumetric strain, after subtracting the smooth compression of the parent \centric{} branch.
The coupling term $G e\Delta^2$ is the lowest-order symmetry-allowed coupling between the D8R distortion and the volume response (the linear coupling $\Delta\,e$ is forbidden by the $A_{2u}$ character of $\Delta$, as already noted above).
It expresses the fact that the structural distortion produces an additional volume change beyond the normal elastic compression.

At fixed $\Delta$, minimisation of \textbf{Equation} \ref{eq:LandauF} with respect to the residual strain $e$ gives
\begin{equation}
\label{eq:emin}
e_\text{min}(\Delta) = -\frac{G}{K_0}\Delta^2 .
\end{equation}
Since $e=v-v_\mathrm{c}(p)$, this result gives
\begin{equation}
\label{eq:vmin}
v(p)=v_\mathrm{c}(p)-\frac{G}{K_0}\Delta^2 .
\end{equation}
Thus, the volume follows the ordinary elastic compression of the parent \centric{} branch as long as $\Delta=0$, and develops an additional contraction once the D8R distortion appears.

Substituting \textbf{Equation} \ref{eq:emin} back into \textbf{Equation} \ref{eq:LandauF} gives an effective Landau free energy in $\Delta$ alone,
\begin{equation}
\label{eq:LandauEff}
f_\text{eff}(\Delta;p) =
\tfrac{1}{2}A\Bigl(1-\nicefrac{p}{p_c}\Bigr)\Delta^2
+\tfrac{1}{4}U_\text{eff}\Delta^4
+\tfrac{1}{6}W\Delta^6,
\end{equation}
with
\begin{equation}
U_\text{eff}=U-\frac{2G^2}{K_0}.
\end{equation}
The strain-order-parameter coupling therefore renormalises the quartic coefficient.
Physically, the framework can lower the cost of the D8R distortion by relaxing its volume, and this reduces the effective stiffness of the \acentric{} branch.
This is the characteristic behaviour of a co-elastic transition, in which a symmetry-breaking, non-strain order parameter is coupled to a fully symmetric (volumetric) strain through a term linear in the strain and quadratic in the order parameter ($G\,e\,\Delta^{2}$).\cite{CarpenterSalje1998, Salje2012, BruceCowley1981} Because $Im\bar{3}m \to I\bar{4}3m$ is a cubic-to-cubic transformation, the point group $\bar{4}3m$ admits only an isotropic (volumetric) spontaneous strain and forbids any deviatoric spontaneous strain; the transition therefore does not generate a ferroelastic (stress-switchable) spontaneous strain, and is co-elastic rather than (improper) ferroelastic.

The form of \textbf{Equation} \ref{eq:LandauEff} also anticipates the different effective exponents obtained from the pressure scans.
When $U_\text{eff}$ is sufficiently positive and the sixth-order term is small over the fitted pressure window, the quartic term controls the growth of the order parameter and gives the usual Landau behaviour $\Delta\propto(p/p_c-1)^{1/2}$, as observed for $G_2$-$G_5$.
For $G_1$, the strain coupling $-2G^2/K_0$ pushes $U_\text{eff}$ closest to zero in the family, so that the sixth-order term controls the curvature of the distorted basin and $G_1$ sits closest to the tricritical condition $U_\text{eff} = 0$ on the SLC scale. The small $\Delta v = -1.23\%$ returned by the piecewise-linear fit of \textbf{Equation} \ref{eq:kappa} is best read as an upper bound on a first-order discontinuity rather than evidence of one (\textbf{Section} \ref{si:landau-k-parametrisation}); the DFT benchmark of \textbf{Section} \ref{si:slc-validation} indicates that the true $G_1$ transition is more strongly first-order than its SLC counterpart, corresponding to $U_\text{eff}(G_1) < 0$ at the DFT level.
In this near-tricritical regime, the Heaviside-power fit of \textbf{Equation} \ref{eq:fit} cannot represent the discontinuity explicitly and absorbs it as an effective exponent below the mean-field value, $\beta\simeq 0.36$, that phenomenologically mimics the abrupt rise of $\Delta$ at $p_c$.
The quantitative extraction of the crossover parameter for $G_1$ is revisited in \textbf{Section} \ref{si:beta-G1-derivation} of the Supporting Information.

The effective model gives direct links with the simulation observables.
In the quartic-dominated regime, where the sixth-order term is small over the fitted pressure range, minimisation of \textbf{Equation} \ref{eq:LandauEff} gives
\begin{equation}
\label{eq:delta-amp}
\Delta^2(p) \simeq \frac{A}{U_\text{eff}}\left(\nicefrac{p}{p_c}-1\right),
\qquad p>p_c .
\end{equation}
This gives the usual Landau exponent $\beta=1/2$ and identifies the fitted amplitude of \textbf{Equation} \ref{eq:fit} as
\begin{equation}
\label{eq:delta-ratio}
\delta^2 \simeq \frac{A}{U_\text{eff}} .
\end{equation}
This limit describes $G_2$-$G_5$ well.
For $G_1$, the same expression gives only the initial growth of $\Delta$ close to $p_c$, because the sixth-order term becomes relevant over the pressure interval used in the fit.

The volume response follows from \textbf{Equation} \ref{eq:vmin}.
Using $v_\mathrm{c}(p)=1+\kappa_1p$, the distorted branch can be written as
\begin{equation}
\label{eq:v-landau}
v(p) \simeq 1+\kappa_1p-\frac{G}{K_0}\Delta^2(p).
\end{equation}
Thus, below the transition $\Delta=0$ and the volume follows the ordinary elastic compression of the parent \centric{} branch.
Above the transition, the D8R distortion produces an additional contraction proportional to $\Delta^2$.
In the quartic regime, this gives a second linear branch,
\begin{equation}
\label{eq:kappa2-landau}
\kappa_2-\kappa_1 \simeq -\frac{G}{K_0}\frac{\delta^2}{p_c},
\end{equation}
which is the Landau counterpart of the piecewise-linear behaviour fitted in \textbf{Equation} \ref{eq:kappa}.
The same strain-order-parameter coupling also accounts for the softening of the elastic constants at $p_c$: when the structure is allowed to relax along $\Delta$, part of the elastic response is transferred into the symmetry-breaking distortion.
This explains the sharp changes in $C_{11}$ and $C_{12}$ reported in \textbf{Figure} \ref{fig:Mechanical}(a)-(c), and the nearly topology-independent pre-transition compressibility, $\kappa_T \approx 1/K_0$, observed for $G_1$-$G_5$.

These relations determine combinations of Landau coefficients, rather than all coefficients separately.
The pre-transition slope gives $K_0\simeq 1/|\kappa_1|$; the order-parameter amplitude gives the ratio $A/U_\text{eff}$; and the change in volume slope gives the coupling combination $(G/K_0)\delta^2$.
The corresponding quantities are collected in \textbf{Table} \ref{tab:SI-landau} of the Supporting Information.
Two trends are relevant for the present discussion.
First, $K_0$ varies only weakly across the series, consistent with its interpretation as a local framework-stiffness parameter dominated by the SiO$_4$ network.
Second, the ratio $U_\text{eff}/A=1/\delta^2$ increases strongly from $G_1$ to $G_5$, showing that isoreticular order changes the nonlinear stiffness of the distorted branch while leaving the pre-transition elastic stiffness much less affected.

The role of the sixth-order term is most clearly seen by minimising \textbf{Equation} \ref{eq:LandauEff} with respect to $\Delta$.
For the distorted branch, $\Delta \neq 0$, one obtains
\begin{equation}
\label{eq:DeltaSq-sextic-body}
\Delta^{2}(p) =
\frac{U_\text{eff}}{2W}
\left[
\sqrt{1+\eta\left(\nicefrac{p}{p_c}-1\right)}-1
\right],
\qquad
\eta \equiv \frac{4WA}{U_\text{eff}^{2}} .
\end{equation}
This expression interpolates between two limiting behaviours.
When $\eta(p/p_c-1)\ll 1$, the quartic term controls the response and $\Delta\propto(p/p_c-1)^{1/2}$.
When $\eta(p/p_c-1)\gg 1$, the sixth-order term becomes dominant and the apparent exponent approaches the tricritical value $1/4$. Both limits describe a \emph{continuous} onset: within a sextic Landau functional with $W>0$, any $U_\text{eff} > 0$, including the limit $U_\text{eff}\to 0^+$, yields $\Delta(p_c) = 0$ and a continuous transition, whereas a genuine first-order discontinuity of $\Delta$ at $p_c$ would require $U_\text{eff} < 0$. The fitted $U_\text{eff}(G_1) > 0$ therefore places $G_1$, within the SLC description, just on the continuous (near-tricritical) side of the transition, the closest member of the family to tricriticality. The more sharply first-order character returned by the DFT benchmark of \textbf{Section} \ref{si:slc-validation} corresponds to $U_\text{eff}(G_1) < 0$ at the DFT level; the SLC and DFT descriptions agree on the singular position of $G_1$ and differ only in the sign of $U_\text{eff}(G_1)$ (see \textbf{Section} \ref{si:landau-k-parametrisation} for the detailed discussion).
The values fitted for $G_2$-$G_5$ lie close to the quartic regime over the pressure range sampled here, giving effective exponents near $1/2$.
RHO ($G_1$), in contrast, has the smallest $U_\text{eff}/A$ ratio of the family and therefore samples the crossover towards the sixth-order regime, which explains both its lower effective exponent ($\beta\simeq 0.36$) and the small first-order signature in the volume curve without requiring a different microscopic mechanism.
The detailed extraction of the crossover parameter for $G_1$ is given in \textbf{Section} \ref{si:beta-G1-derivation} of the Supporting Information.

Within this picture, the isoreticular index $k$ acts as a structural control variable for the transition.
It controls first the location of the instability, which follows the exponential law of \textbf{Equation} \ref{eq:pc-exponential} with $p_0 = 0.930 \pm 0.022$~\si[mode=text]{\giga\Pa} and $\lambda = 0.494 \pm 0.023$ from a non-linear least-squares fit with both parameters free.
Extrapolating the exponential law gives $p_c \simeq 0.079$, $0.048$ and $0.029$~\si{\giga\Pa} for $G_6$, $G_7$ and $G_8$ respectively. A formal comparison of three candidate functional forms (exponential, power law in $a_k$, power law in $N_T$; \textbf{Section} \ref{si:pc-models} of the Supporting Information) favours the exponential at $\Delta$AIC $> 6$, but yields a conservative across-model prediction band $p_c(G_6) \in [0.08, 0.15]$, $p_c(G_7) \in [0.05, 0.13]$ and $p_c(G_8) \in [0.03, 0.11]$~\si[mode=text]{\giga\Pa}, all well below \SI[mode=text]{0.2}{\giga\Pa} on the mechanical-pressure scale.
It also controls the nonlinear stiffness of the distorted branch through $U_\text{eff}(k)$, and therefore the magnitude of the volume softening above $p_c$.
Thus, increasing the isoreticular order preserves the same local symmetry-breaking distortion, but progressively lowers the mechanical scale and flattens the response.

A natural concern with the mean-field reading of the broken-phase exponents is that, for an isolated Landau-Ginzburg field in three dimensions, fluctuations of the order parameter would shift the effective exponent of a continuous transition away from $1/2$ towards the Ising-3D value $\beta\simeq 0.326$ once the reduced pressure $\nicefrac{p}{p_c} - 1$ enters the Ginzburg interval $|t| \lesssim t_G$.\cite{Ginzburg1960, LevanyukSigov1988} For the RHO family this regime is not reached for two reasons. First, the $T=0$ static minimisations on which the soft-mode and the $\Delta(p)$ fits are based contain no thermal fluctuations and therefore reproduce the mean-field result by construction. Second, the strong elastic coupling between $\Delta$ and the volumetric strain $e$ is of long range and effectively shrinks $t_G$ to a value below the present resolution. The combination of a long-range strain-mediated interaction and a short-range Landau-Ginzburg interaction has been shown, both for proper ferroelastic and for co-elastic transitions,\cite{Cowley1976, BruceCowley1981, LevanyukSigov1988} to suppress critical fluctuations and to recover mean-field exponents over essentially the whole reduced pressure window. The narrowest fit window probed in our scans, $\nicefrac{p}{p_c} - 1 \gtrsim 10^{-2}$ for $G_1$, lies orders of magnitude outside any plausible $t_G$ once the strain coupling has been included; the mean-field reading of $\beta\simeq 1/2$ for $G_2$-$G_5$ and the sextic-crossover reading of $\beta\simeq 0.36$ for $G_1$ are therefore consistent with the strain-coupled Landau picture and are not corrected by Ising-3D-like critical fluctuations.

The microscopic origin of the soft mode is naturally consistent with the rigid-unit mode (RUM) picture developed by \citet{Hammonds1997, Hammonds1998} and \citet{Bieniok1998}.
The eigenvector of the lowest $\Gamma$-point mode in \textbf{Figure} \ref{fig:Phonons} corresponds to a coordinated rotation of nearly rigid SiO$_4$ tetrahedra that opens the D8R windows along one diagonal while closing them along the orthogonal one.
This is consistent with the symmetry-allowed rotational degrees of freedom discussed by \citet{Dove2019} for embedded isoreticular topologies.
The Landau and RUM descriptions are therefore complementary: the RUM picture identifies the symmetry-allowed pathway of the distortion, while the Landau description captures the residual stiffness, the coupling to volume strain and the scaling of the instability with isoreticular order. This assignment is further supported by the ring-resolved finite-temperature histograms reported in \textbf{Section} \ref{si:8mr-histograms} of the Supporting Information. 
The pressure-induced redistribution is concentrated in the 8MR distortion distributions, both for D8R-paired and isolated 8MRs, whereas the 4MR and 6MR distortion distributions remain much less sensitive to the transition. 
This confirms that the soft mode is expressed primarily through the 8MR windows rather than through a uniform deformation of all ring types.


\subsection{Non-zero temperatures}
\begin{figure*}[!t]
 \centering
 \begin{subfigure}{\textwidth}
  \centering
  \includegraphics[width=\textwidth]{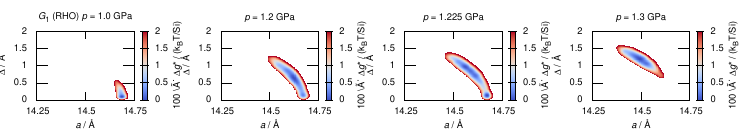}
 \end{subfigure}
 \begin{subfigure}{\textwidth}
  \centering
  \includegraphics[width=\textwidth]{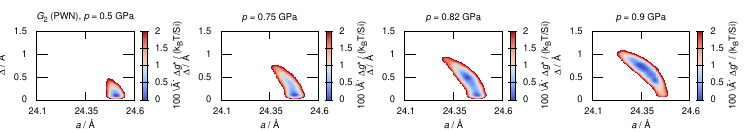}
 \end{subfigure}
 \begin{subfigure}{\textwidth}
  \centering
  \includegraphics[width=\textwidth]{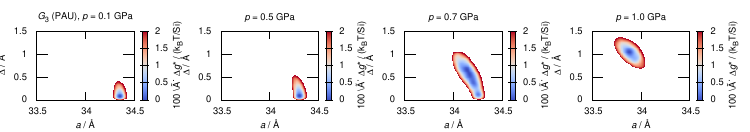}
 \end{subfigure}
 \begin{subfigure}{\textwidth}
  \centering
  \includegraphics[width=\textwidth]{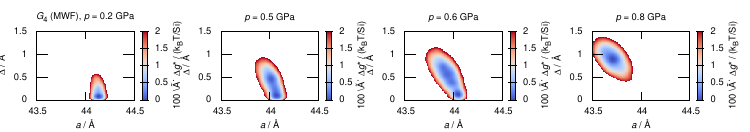}
 \end{subfigure}
 \begin{subfigure}{\textwidth}
  \centering
  \includegraphics[width=\textwidth]{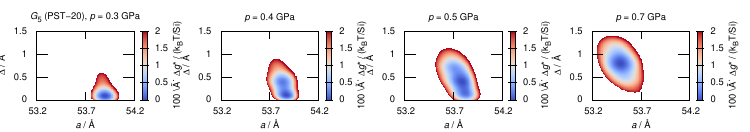}
 \end{subfigure}
 \caption{\label{fig:FES}Reweighted free-energy maps $\Delta g^*(\Delta, a)$ at $T = \SI[mode=text]{298.15}{\kelvin}$ for $G_1$ to $G_5$ (rows top to bottom) at four representative pressures around the room-temperature transition (columns left to right). Horizontal axis: cubic cell parameter $a$ (\si[mode=text]{\angstrom}); vertical axis: D8R distortion $\Delta$ (\si[mode=text]{\angstrom}). Colour scale in units of $100\,k_B T$ per Si atom. Saddle-point barriers extracted from the maps are tabulated in \textbf{Table} \ref{tab:SI-fes-barriers}.}
\end{figure*}

To check whether the \SI[mode=text]{0}{\kelvin} picture remains valid at finite temperature, we reconstructed the free-energy surface $\Delta g^*(\Delta, a)$ for $G_1$-$G_5$ at $T = \SI[mode=text]{298.15}{\kelvin}$ using the OPES exploration protocol on the structural collective variables in biased NPT simulations at each pressure point (see Methodology).
\textbf{Figure} \ref{fig:FES} shows the resulting heatmaps at representative pressures around the room-temperature transition.
The maps were obtained by reweighting the biased trajectories. The four representative pressures per member span values above the static $p_c$ of \textbf{Table} \ref{tab:parameters}, where both the \centric{} and the \acentric{} regions of the free-energy landscape are simultaneously sampled. The pressure column $p_\text{FES}$ of \textbf{Table} \ref{tab:SI-fes-barriers} is therefore an upper bound on the finite-temperature transition pressure rather than its actual value, and the offset between $p_\text{FES}$ and the static $p_c$ should not be read as a measure of the entropic shift; the latter would require a separate scan with $p_\text{FES}$ chosen as the pressure at which the two basins are equally populated, which we have not performed.

The maps provide a finite-temperature view of three aspects of the transition.
First, the shape of the landscape follows the soft-mode picture.
Well below $p_c$, the free-energy surface shows a single basin centred at $\Delta\simeq 0$, corresponding to the \centric{} structure.
On approaching the transition, this basin broadens and elongates along $\Delta$, indicating a progressive loss of curvature in the D8R-distortion coordinate.
At higher pressure, a minimum at finite $\Delta$ develops, corresponding to the distorted \acentric{} structure.

Second, the free-energy barriers separating the two regions are small on an intensive scale throughout the family (\textbf{Table} \ref{tab:SI-fes-barriers}).
For $G_1$ at $p = \SI[mode=text]{1.300}{\giga\Pa}$ (the most resolved panel of \textbf{Figure} \ref{fig:FES}), the saddle lies at $\Delta g^* \simeq 0.037~k_B T$ per Si ($\simeq 14~k_B T$ for the 384-Si simulation cell).
For $G_2$ at $p = \SI[mode=text]{0.900}{\giga\Pa}$ the corresponding barrier is below $\sim 0.01~k_B T$ per Si ($\sim 2~k_B T$ per 240-Si cell), at the threshold below which the two regions coalesce into a broad anharmonic basin at room temperature.
For $G_3$-$G_5$ at the pressures explored, the per-Si barriers remain in the same sub-$k_B T$ range, $\Delta g^*/\text{Si} \in [0.015, 0.07]\,k_B T$.
These values should not be interpreted as direct kinetic rates, because the collective length scale of the transformation is not determined by the present periodic-cell calculations. 
They show that the finite-temperature landscape is essentially flat in the order-parameter direction across the whole family, with intensive barriers below $0.1~k_B T$ per Si throughout, in line with the soft-mode and Landau picture developed above. The extraction procedure, including the smoothing and minimum-energy-path projection used to locate the saddle on each map, is described in \textbf{Section} \ref{si:fes-barriers} of the Supporting Information.

Third, the relative depth and width of the \centric{} and \acentric{} regions reflect the different character of the transition along the series.
$G_1$ shows the most abrupt change, with a narrow pressure interval between the \centric{} and \acentric{} basins.
For $G_2$-$G_5$, the redistribution of probability is more gradual, with broader basins and a wider coexistence-like region in the finite-cell free-energy maps.
This behaviour is consistent with the effective Landau picture discussed above: near $p_c$, the free-energy well becomes shallow in the $\Delta$ direction, so the finite-temperature distribution broadens before the distorted minimum becomes dominant.
The sharper response of $G_1$ is also consistent with the singular position of $G_1$ inferred from the static pressure scans, namely the apparent negative volume jump at $p_c$ tabulated in \textbf{Table} \ref{tab:parameters} and the near-tricritical position of $G_1$ in the sextic Landau picture discussed above.

Overall, the finite-temperature FES support the trend inferred from the \SI[mode=text]{0}{\kelvin} analysis.
When compared at similar reduced pressures, the landscapes of $G_2$-$G_5$ evolve in a common way: the \centric{} basin broadens near the transition and the finite-$\Delta$ \acentric{} region becomes progressively stabilised.
$G_1$ remains the most distinct member, with a sharper redistribution of probability and a more abrupt structural response.
Thus, increasing isoreticular order does not remove the local switching motif, but it flattens the free-energy landscape and makes the finite-temperature response progressively softer.


\section{Conclusions}

This work shows that the molecular-valve distortion of zeolite RHO is not an isolated peculiarity of the parent framework, but part of a broader mechanical response of the embedded RHO isoreticular family.
The comparison with available experiments and with independent r$^2$SCAN+rVV10 calculations for the smaller members supports the use of the Sanders-Leslie-Catlow classical core-shell interatomic potential as a semiquantitative model for trends across the family.
Using this model, we find that $G_1$-$G_5$ all undergo a pressure-induced \centtoacent{} transformation involving the same local D8R distortion.
The important result is that this transition changes systematically with isoreticular order: as the framework expands, the same local switching motif becomes mechanically softer, less abrupt and accessible at lower pressure.

The physical origin of this behaviour is common to the whole computed series.
The transition is driven by a $\Gamma$-point soft mode, accompanied by a marked softening of the elastic constants and by an increase of the compressibility in the distorted branch.
This gives a coherent picture in which the D8R distortion acts as the local order parameter, while the cell volume acts as a coupled elastic coordinate.
The Landau-type description captures this coupling and explains why $G_2$-$G_5$ behave in a similar, almost mean-field-like way ($\beta\simeq 0.43$-$0.55$, $|\Delta v|<0.5\%$ across the four members), whereas RHO remains the singular end member of the family, with the sharpest response ($\beta=0.36$ and an apparent negative volume jump $\Delta v=-1.23\%$, both unique in the family), placed at or just past the tricritical condition $U_\text{eff} = 0$ depending on whether the SLC or the DFT description is adopted (see Landau analysis above).
Thus, isoreticular expansion does not change the nature of the instability, but it changes how strongly the instability is expressed at the scale of the unit cell.

A central consequence is that the critical pressure decreases rapidly with isoreticular order.
The calculated values from $G_1$ to $G_5$ are well described by an exponential decay, $p_c(k)=p_0\,\exp[-\lambda(k-1)]$ with $p_0=0.930\pm 0.022$~\si{\giga\Pa} and $\lambda=0.494\pm 0.023$, which predicts $p_c\simeq 0.079$, $0.048$ and $\SI[mode=text]{0.029}{\giga\Pa}$ for $G_6$, $G_7$ and $G_8$, respectively, on the mechanical-pressure scale.
These values should not be interpreted as direct gas pressures for adsorption experiments, because the present model does not include Al, extra-framework cations or guest molecules.
They are instead predictions of the intrinsic framework softness of the ideal pure-silica hierarchy. Whether this intrinsic softness translates into guest-driven switching in real aluminosilicates will depend on the Al distribution, extra-framework cations and adsorbed molecules.
The same softening trend is also corroborated by the independent r$^2$SCAN+rVV10 calculations on $G_1$ and $G_2$ reported in \textbf{Section} \ref{si:slc-validation}: the volume jump at the transition contracts from $\Delta V/V_0\simeq -22\%$ for $G_1$ to $\simeq -5\%$ for $G_2$, and the post-transition $\Delta$ falls from $\simeq 2.65$ to $\simeq \SI[mode=text]{1.30}{\angstrom}$, confirming the SLC-derived softening with isoreticular order without recourse to empirical force-field parameters.
In this sense, the larger members are predicted to be much closer to a mechanically switchable state than RHO itself.

The finite-temperature free-energy surfaces support the same conclusion.
At room temperature, the \centric{} basin broadens near the transition and the finite-$\Delta$ \acentric{} region becomes progressively easier to stabilise as $k$ increases.
The larger members therefore do not simply have lower static transition pressures, they also have flatter free-energy landscapes, which agrees with the main conclusion of our study, namely that the isoreticular hierarchy preserves the local molecular-valve motif but tunes the softness of the whole framework. Our results identify isoreticular order as a structural control variable for zeolite framework softness.

An interesting prediction of our study is that the largest members of the RHO hierarchy, or intergrowths involving them, should show very soft molecular-valve-like responses if the appropriate chemical composition and extra-framework species can be realised. Since pure phases of the higher-order members are difficult to isolate, intergrowths between neighbouring generations are particularly relevant experimental targets.


\subsection*{Author contribution}
S.R.G.B.: Conceptualisation, Methodology, Software, Investigation (classical molecular dynamics, GULP, LAMMPS and PLUMED OPES simulations), Formal analysis, Visualisation, Writing - original draft, Funding acquisition, Project administration.
A.R.R.-S.: Conceptualisation, Supervision, Funding acquisition, Writing - original draft, Writing - review and editing.
S.H.: Conceptualisation, Supervision, Methodology, Formal analysis (supporting), Writing - original draft, Writing - review and editing.
A.R.-B.: Investigation (DFT r$^2$SCAN+rVV10 and MACE single-point calculations), Data analysis, Validation, Writing - review and editing.
All authors discussed the results and contributed to the final manuscript.

\subsection*{Conflicts of interest}
The authors have no conflicts to disclose.
\subsection*{Data and code availability}
The full input files, representative structures, GULP and PLUMED optimisation logs, OPES inputs, reweighted free-energy surfaces, analysis scripts and figure-generation scripts that support the findings of this work are openly available at \url{https://github.com/salrodgom/isoreticular} and archived with a persistent DOI at Zenodo: \href{https://doi.org/10.5281/zenodo.20583813}{10.5281/zenodo.20583813}. The White Rabbit code used to identify the D8R oxygen pairs that define the distortion order parameter (\textbf{Equation} \ref{eq:delta_t}) is available at \url{https://github.com/salrodgom/white_rabbit}.

\subsection*{Supporting Information}
The Supporting Information contains the detailed computational protocol, DFT and machine-learning-potential benchmarks, structural data for the $G_1$-$G_8$ family, full phonon analyses, finite-size checks, fitting parameters, model comparison for the $p_c(k)$ trend, Landau coefficient extraction, finite-temperature barrier analysis, ring-resolved distortion histograms and the experimental data supporting the comparison across the RHO isoreticular series.

\subsection*{Acknowledgements}
S.R.G.B.~acknowledges grant \texttt{RYC2022-036070-I} funded by \texttt{MICIU/AEI/10.13039/501100011033} and by ESF+. A.R.-B.~acknowledges grant \texttt{PREP2022-000273} funded by \texttt{MICIU/AEI/10.13039/501100011033} and by the ESF+. The authors acknowledge project \texttt{PID2022-140061OB-I00} (DeepMatSolar) funded by \texttt{MICIU/AEI/10.13039/501100011033} and by ERDF/EU.
The authors acknowledge funding from the European Commission through the HORIZON.1.2 Marie Sk{\l}odowska-Curie Actions, VALZEO project (grant agreement \texttt{101086354}). The authors gratefully acknowledge the HPC resources and services provided by C3UPO at the Universidad Pablo de Olavide, the Servicio de Supercomputaci\'on of the Universidad de Granada, and the `H\'ercules' supercomputer at the Centro Inform\'atico Cient\'ifico de Andaluc\'ia (CICA). S.R.G.B. also gratefully acknowledges Prof.~Miguel A.~Camblor (Instituto de Ciencia de Materiales de Madrid, ICMM-CSIC) for many years of fruitful discussions on the synthesis and structural chemistry of zeolites that have shaped the questions addressed in this work.

\bibliographystyle{hunsrtnat}
\bibliography{biblio}

\clearpage
\renewcommand\thefigure{S\arabic{figure}}
\renewcommand\thesubsection{S\arabic{subsection}}
\renewcommand\thesubsubsection{S\arabic{subsection}.\arabic{subsubsection}}
\renewcommand\thetable{S\arabic{table}}
\renewcommand\theequation{S\arabic{equation}}
\renewcommand\theHequation{Seq.\arabic{equation}}
\renewcommand\theHfigure{Sfig.\arabic{figure}}
\renewcommand\theHtable{Stab.\arabic{table}}
\renewcommand\theHsubsection{Ssub.\arabic{subsection}}
\setcounter{figure}{0}
\setcounter{table}{0}
\setcounter{equation}{0}
\begin{center}
{\Large\textbf{Supplementary Information for}}\\[0.6ex]
{\large\textbf{A topology-tuned pressure valve across the isoreticular RHO zeolite family}}\\[1.4ex]

The Supporting Information contains the detailed computational protocol, DFT and machine-learning-potential benchmarks, structural data for the $G_1$-$G_8$ family, full phonon analyses, finite-size checks, fitting parameters, model comparison for the $p_c(k)$ trend, Landau coefficient extraction, finite-temperature barrier analysis, ring-resolved distortion histograms and the experimental data supporting the comparison across the RHO isoreticular series.

{\normalsize Salvador R.G. Balestra$^{1,*}$ \orcidA{} \,,\, Antonio Rivas-Blanco$^{2,3}$ \orcidD{} \,,\, Said Hamad$^{2,3}$ \orcidC{} \,,\, A. Rabdel Ru\'iz-Salvador$^{2,3}$ \orcidB{}}\\[0.6ex]
{\footnotesize $^{1}$Departamento de F\'isica At\'omica, Molecular y Nuclear, \'Area de F\'isica Te\'orica, Universidad de Sevilla, ES-41012 Sevilla, Spain}\\
{\footnotesize $^{2}$Centro de Nanociencia y Tecnolog\'ias Sostenibles (CNATS), Universidad Pablo de Olavide, ES-41013 Sevilla, Spain}\\
{\footnotesize $^{3}$Departamento de Sistemas F\'isicos, Qu\'imicos y Naturales, Universidad Pablo de Olavide, ES-41013 Sevilla, Spain}\\[0.4ex]
{\footnotesize $^{*}$E-Mail: \texttt{srodriguez9@us.es}}
\end{center}
\vspace{1.2em}

\subsection*{Structure of the Supporting Information}

The Supporting Information is organised to support the main conclusions of the manuscript. 
Section S1 gives the detailed computational protocol and convergence thresholds. 
Section S2 benchmarks the SLC potential against r$^2$SCAN+rVV10 DFT and machine-learning force fields for $G_1$ and $G_2$. 
Sections S3 and S4 report the density-energy landscape and the structural data of the isoreticular family.
Sections S5 and S6 provide the full phonon analysis supporting the soft-mode definition of $p_c$, including the restricted diagnostic treatment of $G_5$. 
Sections S7 and S8 give the finite-size and fitting details used for the structural pressure scans. 
Section S9 compares alternative empirical forms for the decrease of $p_c$ with isoreticular order. 
Sections S10 to S14 provide the data-collapse and Landau analyses.
Section S15 reports the finite-temperature free-energy barriers extracted from the OPES maps. 
Section S16 gives ring-resolved distortion histograms, and Section S17 summarises the available experimental evidence across the RHO isoreticular family.

\subsection{Notes on the computational protocol}
Energy minimisations at each pressure were carried out in two steps. A first cell-plus-atoms relaxation was performed with LAMMPS using the FIRE-type damped dynamics implementation of Bitzek et al.\citeS{Bitzek2006SI}, optimised by Gu\'enol\'e et al.\citeS{Gunol2020SI}, until the global pressure tensor matched the target hydrostatic pressure within $10^{-4}$~\si[mode=text]{\bar} and the per-atom force fell below $10^{-10}$~\si[mode=text]{\electronvolt\per\angstrom}. Each pre-converged structure was subsequently refined with GULP\citeS{Gale1997SI, Gale2003SI} using a Broyden-Fletcher-Goldfarb-Shanno (BFGS) step followed by a Rational Function Optimisation (RFO) step, until the gradient norm dropped below $10^{-8}$~\si[mode=text]{\electronvolt\per\angstrom}, the tight convergence required for the subsequent $\Gamma$-point phonon and elastic-constant calculations. When an imaginary mode appeared, the GULP lower-symmetry option was activated, the atomic positions displaced along the corresponding eigenvector by a small amount, and the structure re-optimised in the reduced space group; iterating this procedure until all $\Gamma$-point frequencies become real provides the structures of the \acentric branch reported in \textbf{Figures} \ref{fig:Phonons} and \ref{fig:Mechanical}. The full input files for both LAMMPS and GULP, together with the post-processing scripts used to extract $\Delta$, $C_{ij}$ and $\kappa_T$, are deposited in the repository \url{https://github.com/salrodgom/isoreticular}.

The dispersion-corrected DFT reference pressure scan of $G_1$ reported in \textbf{Section} \ref{si:slc-validation} was performed with the Vienna Ab Initio Simulation Package (VASP) version 6.4\citeS{Kresse1996aSI, Kresse1996bSI}. The exchange-correlation energy was described with the regularised strongly-constrained-and-appropriately-normed meta-GGA functional r$^2$SCAN\citeS{Furness2020SI} together with the non-local rVV10 van-der-Waals correction\citeS{Sabatini2013SI, Peng2016SI}. The dimensionless parameters of the rVV10 kernel were set to the reparametrisation reported by Ning et al.\citeS{Ning2022SI} ($b = 11.95$ and $C = 0.0093$), which yields accurate energetics for weakly bound systems at a comparatively low computational cost. Core electrons were treated through the projector-augmented-wave method\citeS{Bloechl1994SI, Kresse1999SI} using the standard PAW pseudopotentials provided in the VASP library for Si and O. The plane-wave kinetic-energy cut-off was set to \SI[mode=text]{700}{\electronvolt} and the augmentation-charge cut-off to \SI[mode=text]{1400}{\electronvolt}. Electronic self-consistency was converged to $10^{-6}$~\si[mode=text]{\electronvolt}. All calculations were performed without spin polarisation, and the Brillouin zone of the 48-T cubic cell was sampled only at the $\Gamma$ point. Structural relaxations were carried out at every $\Delta p = \SI[mode=text]{0.1}{\giga\Pa}$ between 0 and 2~\si[mode=text]{\giga\Pa} using the conjugate-gradient algorithm (IBRION$=2$) with simultaneous optimisation of atomic positions, cell shape and cell volume (ISIF$=3$); crystallographic symmetry was disabled (ISYM$=0$) so as not to constrain artificially the generated configurations. Ab-initio molecular-dynamics (AIMD) trajectories used to drive the analyses of the follow-up work were obtained from the same VASP setup with a time step of \SI[mode=text]{0.5}{\femto\second}, a Langevin thermostat with friction coefficient \SI[mode=text]{10}{\per\pico\second} acting on both atomic and lattice degrees of freedom, and a fictitious lattice mass \texttt{PMASS}$=10$.

The first machine-learnt-potential reference scan of $G_1$ reproduces the protocol used by Nasir et al.\citeS{NasirCatlow2025SI} to model siliceous zeolites with the MACE foundation model, on the same set of 21 target pressures. Energies, forces and stresses were evaluated by an ASE\citeS{ASE2017SI} calculator that exposes the MACE model (in its medium-accuracy variant), and the optimisation itself was driven entirely within ASE: atomic positions were relaxed with the FIRE algorithm and the cell was relaxed simultaneously through the \texttt{FrechetCellFilter} wrapper at the prescribed hydrostatic pressure, until the maximum force component fell below $10^{-3}$~\si[mode=text]{\electronvolt\per\angstrom}.

The second MACE reference scan uses the MACE-MP foundation model trained on the MatPES r$^2$SCAN data set\citeS{MatPES2025SI}, which provides r$^2$SCAN-consistent labels for a broad sample of inorganic crystals; in what follows we refer to this checkpoint simply as the \emph{MACE-MP foundation}. The same pure-ASE pipeline described above is used, with FIRE relaxation of the atomic positions, simultaneous \texttt{FrechetCellFilter} cell relaxation at the target pressure, convergence threshold $10^{-3}$~\si[mode=text]{\electronvolt\per\angstrom} and identical 0-\SI[mode=text]{2}{\giga\Pa} grid ($\Delta p = \SI[mode=text]{0.1}{\giga\Pa}$), so any difference with respect to the Nasir et al. scan is purely a consequence of the underlying MACE checkpoint.

\subsection{Validation of the SLC potential against DFT and an independent force field}
\label{si:slc-validation}
\begin{figure*}[h!t]
\centering
\begin{subfigure}[t]{0.49\textwidth}
\centering
\caption{\label{fig:SI-slc-dft-richard:a}$G_1$ - cell volume}
\includegraphics[width=\textwidth]{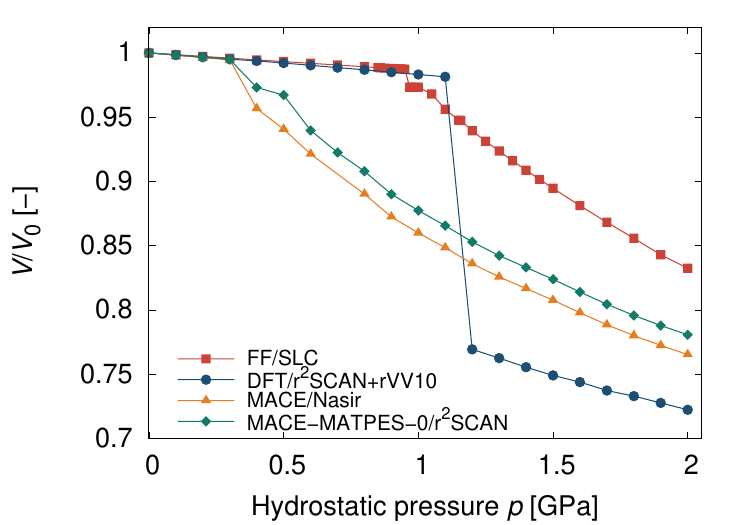}
\end{subfigure}\hfill
\begin{subfigure}[t]{0.49\textwidth}
\centering
\caption{\label{fig:SI-slc-dft-richard:b}$G_1$ - D8R distortion}
\includegraphics[width=\textwidth]{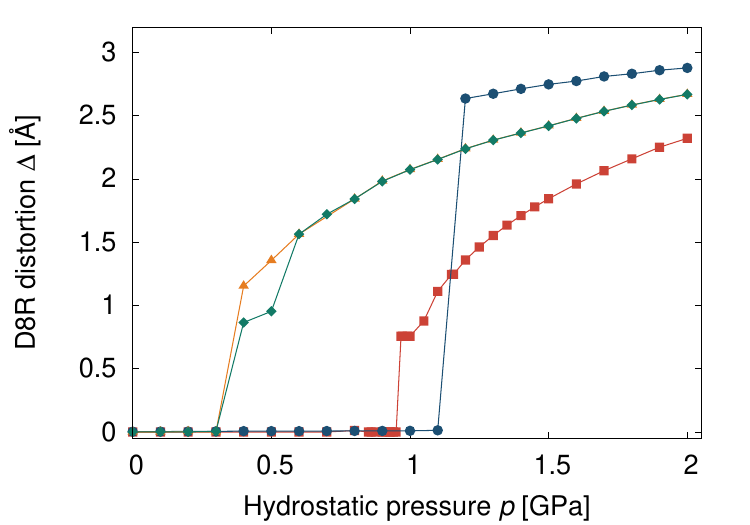}
\end{subfigure}

\vspace{0.6em}

\begin{subfigure}[t]{0.49\textwidth}
\centering
\caption{\label{fig:SI-slc-dft-richard:c}$G_2$ - cell volume}
\includegraphics[width=\textwidth]{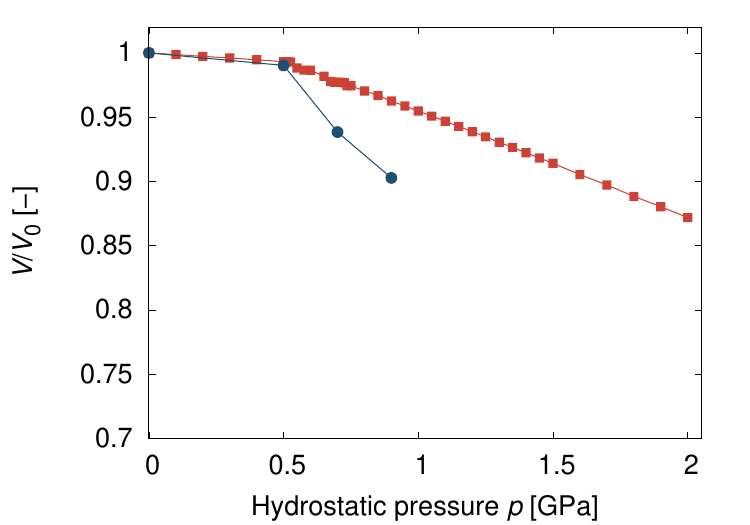}
\end{subfigure}\hfill
\begin{subfigure}[t]{0.49\textwidth}
\centering
\caption{\label{fig:SI-slc-dft-richard:d}$G_2$ - D8R distortion}
\includegraphics[width=\textwidth]{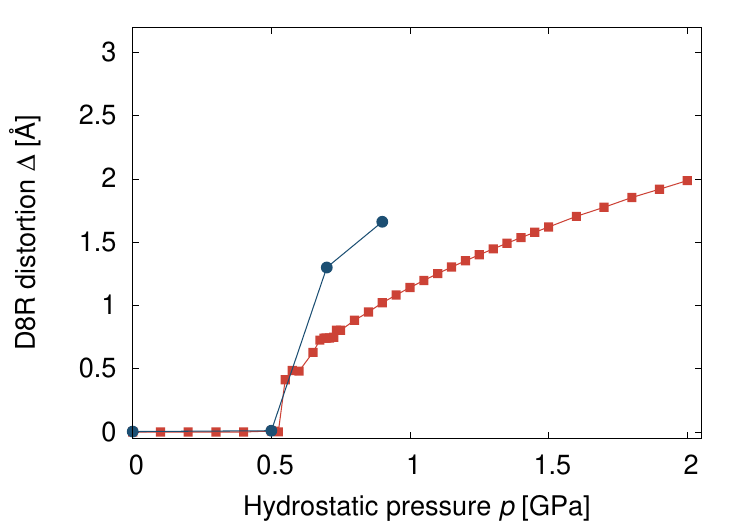}
\end{subfigure}
\caption{\label{fig:SI-slc-dft-richard}Pressure dependence of the cell volume (panels a, c) and the D8R distortion $\Delta$ (panels b, d) for pure-silica $G_1$ (top row) and $G_2$ (bottom row) computed with four different methods: SLC force field (red squares), dispersion-corrected DFT r$^2$SCAN+rVV10 (blue circles), the MACE machine-learning potential of Nasir et al.\citeS{NasirCatlow2025SI} (orange triangles) and the MACE-MP foundation model trained on the MatPES r$^2$SCAN data set\citeS{MatPES2025SI} (green diamonds). For $G_2$ only SLC and DFT scans are reported, at the four pressures (0, 0.5, 0.7 and 0.9~\si[mode=text]{\giga\Pa}) at which the DFT cost was tractable. $\Delta$ is the Parise-Prince definition (\textbf{Equation} \ref{eq:delta_t}) averaged over the twelve crystallographically distinct 8MRs of the unit cell. Method/marker legend in panel (a) only, common to the four panels.}
\end{figure*}

To independently assess the SLC core-shell potential used throughout the manuscript, we have repeated the pressure scan of the parent zeolite $G_1$ (pure-silica RHO unit cell, 48 T-atoms) with three additional methods, all run by us on the same input structures: (i) FF/SLC (this work, the core-shell pair potential used in the main text); (ii) DFT/r$^2$SCAN+rVV10 (dispersion-corrected density-functional theory, full cell-plus-atom relaxation at every 0.1~\si[mode=text]{\giga\Pa} between 0 and 2~\si[mode=text]{\giga\Pa}); (iii) MACE/Nasir (the equivariant MACE machine-learning interatomic potential trained on siliceous zeolites by Nasir et al.\citeS{NasirCatlow2025SI}, run here in our own pressure-scan protocol); and (iv) MACE-MATPES-0/r$^2$SCAN (the MACE-MP foundation model trained on the MatPES r$^2$SCAN data set\citeS{MatPES2025SI}, again run by us on the same pressure grid). The SLC versus DFT comparison was additionally repeated for $G_2$ (PWN, 240 T-atoms per unit cell) on the four pressures (0, 0.5, 0.7 and 0.9~\si[mode=text]{\giga\Pa}) for which the DFT cost remained tractable. \textbf{Figure} \ref{fig:SI-slc-dft-richard} reports the resulting compression curves $V(p)/V_0$ (panels a, c) and D8R distortion order parameters $\Delta(p)$ (panels b, d) for $G_1$ (top row) and $G_2$ (bottom row).

Four conclusions follow. First, the SLC potential reproduces the DFT phenomenology to within the precision relevant for the central scaling argument of this work: $p_c$ agrees with the DFT value to within $\sim 20\%$ for $G_1$, the order-of-magnitude of $\kappa_T$, $C_{ij}$ and $\Delta$ anomalies is preserved, and for $G_2$ both methods locate the volume collapse between 0.5 and 0.7~\si[mode=text]{\giga\Pa}. Second, the DFT calculations independently confirm that the transition becomes markedly less abrupt on going from $G_1$ to $G_2$: at the DFT level the volume jump at the transition shrinks from $\Delta V/V_0 \simeq -22\%$ for $G_1$ to $\simeq -5\%$ for $G_2$ between consecutive pressures, and the post-transition $\Delta$ drops from $\simeq 2.65$ to $\simeq 1.30$~\si[mode=text]{\angstrom}. This pattern, observed independently with SLC, is now corroborated by an \emph{ab initio} calculation that does not rely on any empirical force-field parameter, which strengthens the central claim of the manuscript that the $Im\bar{3}m \to I\bar{4}3m$ transition softens systematically with the isoreticular order $k$. Third, the DFT-predicted transition for $G_1$ is more sharply first-order than the SLC counterpart, which means that the effective exponent $\beta = 0.355$ extracted from the SLC scan for $G_1$ (\textbf{Table} \ref{tab:parameters}) should be interpreted as an upper bound on the true thermodynamic exponent: a fully DFT-quality scan would likely return a smaller $\beta$ (more first-order character) and a correspondingly stronger deviation from the mean-field value $1/2$ already discussed in the main text. The qualitative picture (proximity to tricriticality for $G_1$, mean-field for $G_2$-$G_5$) is not compromised; the new DFT $G_2$ data sit on the mean-field side of the $G_1$/$G_2$ boundary. Fourth, neither of the two MACE reference setups reproduces the qualitative character of the $Im\bar{3}m \to I\bar{4}3m$ phase transition for $G_1$: the MACE potential as applied by Nasir et al.\citeS{NasirCatlow2025SI} falls into a spurious local minimum at $p \simeq 0.7$~\si[mode=text]{\giga\Pa} and places the apparent transition at $\simeq 0.35$~\si[mode=text]{\giga\Pa}, while the MACE-MP foundation model (MatPES r$^2$SCAN\citeS{MatPES2025SI}) cures the local-minimum artefact but keeps the apparent transition at $\simeq 0.55$~\si[mode=text]{\giga\Pa}, both well below the DFT value of $\simeq 1.15$~\si[mode=text]{\giga\Pa}. We therefore do not use them in this work and adopt SLC instead; the comparison is reported in \textbf{Figure} \ref{fig:SI-slc-dft-richard} for transparency only. Extending the DFT validation to $G_3$-$G_5$ is currently prohibitive because of the unit-cell size (672 to 2640 T-atoms); machine-learning interatomic potentials trained on \emph{ab initio} data for siliceous zeolites\citeS{Erlebach2022SI, Brugnoli2024SI} are the natural follow-up.

\subsection{Density-energy landscape across the family}
\label{si:density-energy}
\begin{figure}[h!]
\centering
\includegraphics[width=0.85\textwidth]{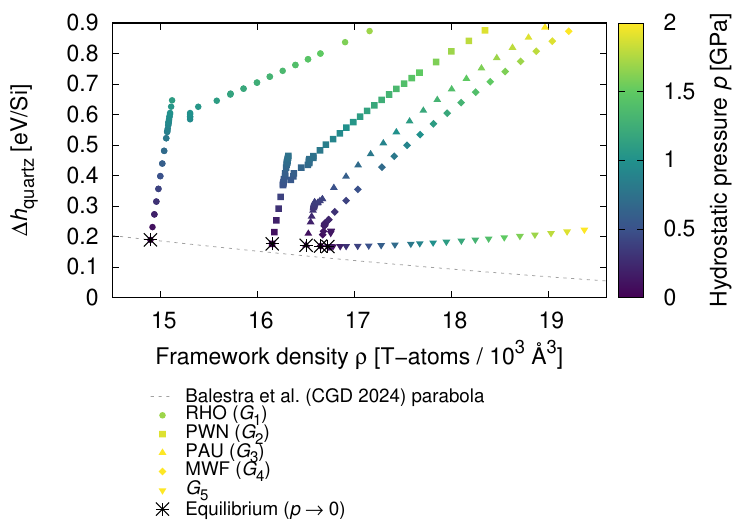}
\caption{\label{fig:SI-density-energy}Framework density $\rho = N_T/V$ versus per-T-atom enthalpy relative to $\alpha$-quartz $\Delta h_{\text{quartz}} = E - E_{\alpha\text{-quartz}}$ (with $E_{\alpha\text{-quartz}} = -128.7034$~\si[mode=text]{\electronvolt}/Si, the SLC reference energy of $\alpha$-quartz used in panel (e) of \textbf{Figure} \ref{fig:Mechanical}) for every SLC-optimised structure of $G_1$-$G_5$ across the available pressure scan. Each point is coloured by the hydrostatic pressure of the minimisation (right colour bar) and uses a different marker per $G_k$ (circle, square, triangle, diamond, inverted triangle for $G_1$-$G_5$). Black asterisks: equilibrium ($p \to 0$) point of each $G_k$. Dashed grey curve: published density-energy parabola for siliceous zeolites of Balestra et al.\citeS{Balestra2024linearizedSI}, $E(\rho) = E_q + k_f\,(\rho - \rho_0)^2$ with $k_f = 1.365\times 10^{-3}$~\si[mode=text]{\electronvolt}/Si per $(\text{T-atoms}/10^3\,\si[mode=text]{\angstrom\cubed})^2$ and $\rho_0 = 27.735$~T-atoms/$10^3$~\si[mode=text]{\angstrom\cubed}, anchored at the $G_1$ equilibrium point and converted to $\Delta h_{\text{quartz}}$ with the same $\alpha$-quartz reference. $N_T$ is taken from the SLC starting CIF of each member ($N_T = 48$, 240, 672, 1440 and 2640 for $G_1$-$G_5$; the $G_1$ scan uses the $2\times 2\times 2$ supercell, $N_T = 384$) and $V$ from the corresponding optimised cell.}
\end{figure}

\begin{figure*}[!htbp]
 \centering
 \begin{subfigure}{0.48\textwidth}
  \centering
  \caption{\label{fig:SI-phonon-modes:G1}RHO ($G_1$)}
  \includegraphics[width=\textwidth]{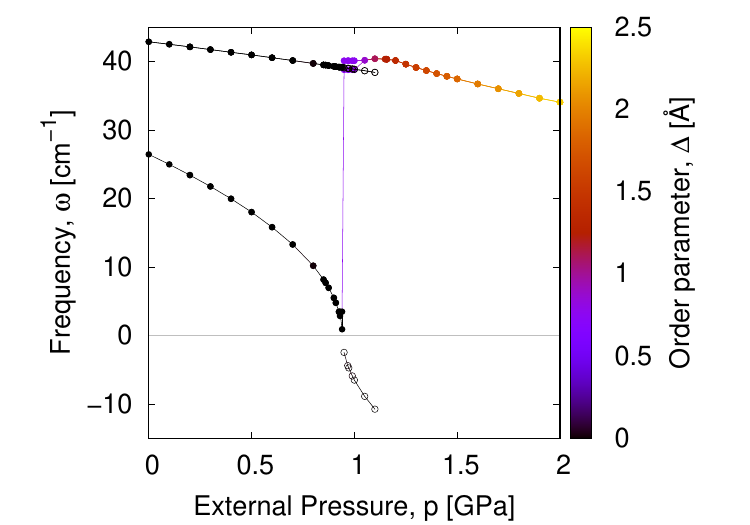}
 \end{subfigure}
 \begin{subfigure}{0.48\textwidth}
  \centering
  \caption{\label{fig:SI-phonon-modes:G2}PST-29 ($G_2$)}
  \includegraphics[width=\textwidth]{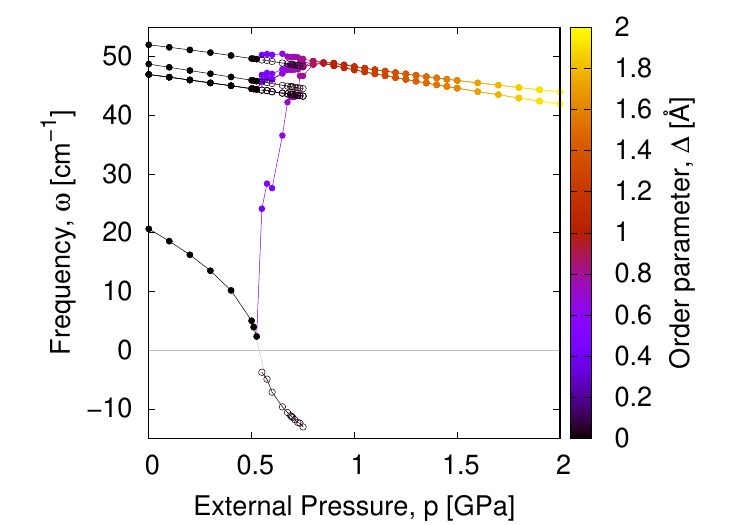}
 \end{subfigure}
 \begin{subfigure}{0.48\textwidth}
  \centering
  \caption{\label{fig:SI-phonon-modes:G3}PAU ($G_3$)}
  \includegraphics[width=\textwidth]{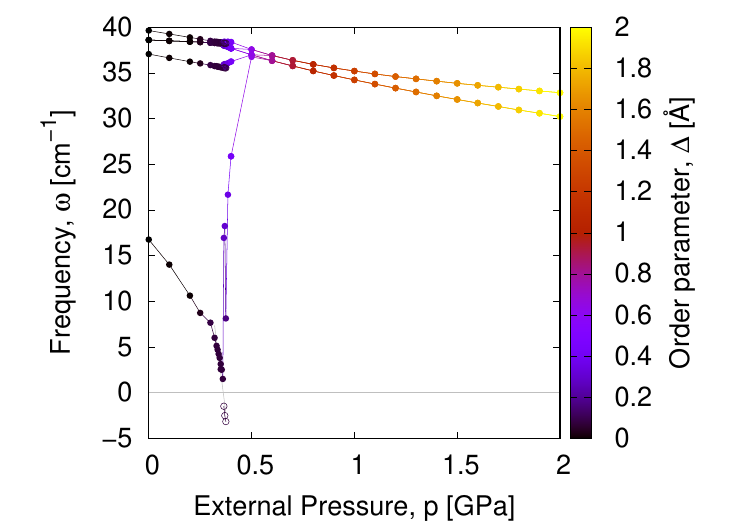}
 \end{subfigure}
 \begin{subfigure}{0.48\textwidth}
  \centering
  \caption{\label{fig:SI-phonon-modes:G4}ZSM-25 ($G_4$)}
  \includegraphics[width=\textwidth]{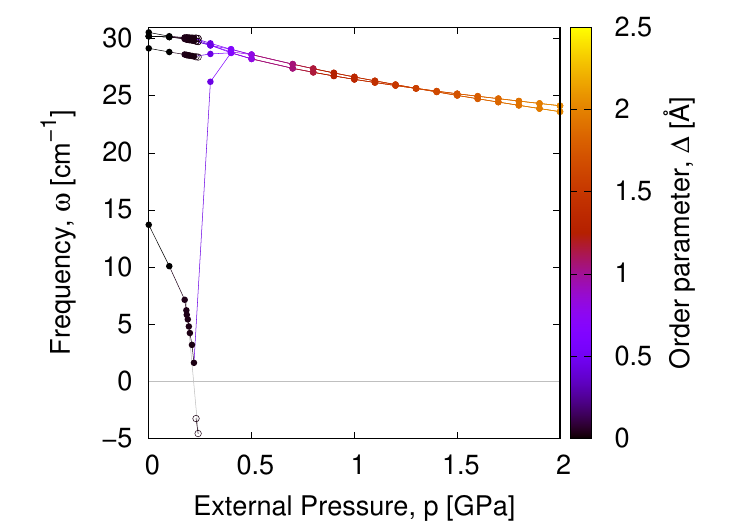}
 \end{subfigure}
 \caption{\label{fig:SI-phonon-modes}Pressure evolution of the lowest $\Gamma$-point phonon branches of $G_1$ to $G_4$ within the SLC potential. Filled circles: branch followed in the parent \centric{} symmetry (imaginary modes above $p_c$ plotted with the conventional sign on the ordinate). Open circles: lower-symmetry \acentric{} branch obtained in GULP by displacing the structure along the unstable eigenvector and re-optimising. The colour scale gives the D8R distortion $\Delta$ (\si[mode=text]{\angstrom}). $G_5$ data are not shown here; the corresponding parent-branch rows are tabulated in \textbf{Section} \ref{si:G5-phonon}.}
\end{figure*}

\textbf{Figure} \ref{fig:SI-density-energy} reports the framework density $\rho$ versus the cohesive energy per T-atom $E$ for \emph{every} SLC-optimised structure along the pressure scan of $G_1$-$G_5$ (panel a), together with the equilibrium-density branch of the family and a parabolic fit (panel b). Four features are worth highlighting. First, the equilibrium densities ($p \to 0$) increase from $G_1$ to $G_3$ and then settle into a plateau, $\rho \simeq 14.91$, $15.96$, $16.51$, $15.89$, $15.81$ T-atoms/$10^3$~\si[mode=text]{\angstrom\cubed} for $G_1$-$G_5$, with the maximum density occurring at $G_3$ (paulingite). The cohesive energy per T-atom decreases monotonically across the series from $-128.514$ to $-128.536$~\si[mode=text]{\electronvolt}/Si: the larger members are thermodynamically slightly more stable than the parent RHO without being denser, which reflects the additional stabilisation provided by the cage-cage connectivity of the embedded hierarchy rather than the textbook density-energy correlation of pure-silica zeolites. Second, the five equilibrium points sit close to the published density-energy parabola for siliceous zeolites of Balestra et al.\citeS{Balestra2024linearizedSI}, $E(\rho) = E_q + k_f\,(\rho - \rho_0)^2$ with $k_f = 1.36529029359175\times 10^{-3}$ and $\rho_0 = 27.735093$, anchored at $G_1$ (panel b); the SLC equilibrium points of $G_2$-$G_5$ deviate from this universal parabola by $\sim 0.03$~\si[mode=text]{\electronvolt}/Si (a small isoreticular-specific stabilisation comparable to room-temperature thermal energy). Third, each $G_k$ follows its own $E(\rho)$ branch under pressure, the slope of which reveals the framework rigidity: $G_2$ rises by $\simeq 0.7$~\si[mode=text]{\electronvolt}/Si between equilibrium and 2~\si[mode=text]{\giga\Pa} for a density excursion of $\Delta\rho \simeq 2$~T-atoms/$10^3$~\si[mode=text]{\angstrom\cubed}, whereas $G_5$ accommodates the same density change with only $\simeq 0.05$~\si[mode=text]{\electronvolt}/Si, consistent with the much lower $p_c$ predicted for the higher-$k$ members and with the soft-mode picture developed in the main text. Fourth, $G_3$ (paulingite) reaches the highest equilibrium density of the series, which is the structural ingredient invoked in the main text to account for its anomalously broad pre-transition softening and its narrow $\kappa_T < 0$ excursion.

\subsection{Structural data for the optimised $G_1$-$G_8$ frameworks}
\textbf{Table} \ref{tab:SI-structures} reports the number of framework T- and O-atoms per cubic unit cell and the lattice parameter used for every member of the RHO isoreticular family considered in this work. 
The number of T-atoms follows the closed-form expression $N_k = 8k(1+3k+2k^2)$ derived in the Introduction. 
For $G_1$-$G_5$, the lattice parameters correspond to the SLC-optimised structures at \SI[mode=text]{0}{\giga\Pa}. 
For $G_6$-$G_8$, full energy minimisation was not performed because of computational cost; the listed cell parameters are topology-only starting metrics for the ideal higher-order frameworks. 
They are therefore included as structural references, not as quantitative evidence for the empirical lattice-parameter law. 
All structures used in the production calculations are deposited as crystallographic information files (.cif) in the public repository \url{https://github.com/salrodgom/isoreticular}, both as initial high-symmetry and optimised forms when available.

\begin{table}[h!]
\caption{\label{tab:SI-structures}Number of framework T- and O-atoms per cubic unit cell, $N_T$ and $N_O = 2\,N_T$, and lattice parameter $a$ for the eight pure-silica members of the embedded isoreticular RHO family. The lattice parameter is the value optimised at \SI[mode=text]{0}{\giga\Pa} for $G_1$-$G_5$ and the topology-only value for $G_6$-$G_8$ (see footnote).}
\centering
\begin{tabular}{l c c c c c}
\toprule
$G_k$ & IZA code & $N_T$ & $N_O$ & $a$ [\si[mode=text]{\angstrom}] & Density [T-atoms/\SI[mode=text]{1000}{\angstrom\cubed}] \\
\midrule
$G_1$ & RHO & 48   & 96    & 14.77  & 14.91 \\
$G_2$ & PWN & 240  & 480   & 24.67  & 15.96 \\
$G_3$ & PAU & 672  & 1344  & 34.40  & 16.51 \\
$G_4$ & MWF & 1440 & 2880  & 44.92  & 15.89 \\
$G_5$ & -  & 2640 & 5280  & 55.07  & 15.81 \\
$G_6$ & -  & 4368 & 8736  & 65.30$^\dagger$  & 15.69 \\
$G_7$ & -  & 6720 & 13440 & 75.46$^\dagger$  & 15.62 \\
$G_8$ & -  & 9792 & 19584 & 85.00$^\dagger$  & 15.93 \\
\bottomrule
\end{tabular}
\newline {\small $^\dagger$ Lattice parameters of $G_6$, $G_7$ and $G_8$ are from the topology-only optimisation; full energy minimisation has not been performed because of computational cost.}
\end{table}

\subsection{Complete phonon branches supporting the soft-mode definition of $p_c$}
\label{si:phonon-modes}

\subsection{Cowley-Levanyuk extraction of $p_c$ for $G_5$}
\label{si:G5-phonon}
The largest member of the family for which a full pressure-resolved phonon analysis was attempted is the $G_5$ member of the family, with a cubic unit cell containing $2640$ T-atoms. At this size, the cost of a single GULP optimisation followed by a full lattice-dynamics calculation is of the order of $10^{4}$ CPU-hours per pressure point; a complete scan with the dense step ($\Delta p \simeq 5$~\si[mode=text]{\mega\Pa}) used for $G_1$-$G_4$ would have required more than the computational allocation available for this work.

We therefore restricted the $G_5$ phonon analysis to a small number of pressure points bracketing the soft-mode-derived critical pressure. \textbf{Table} \ref{tab:SI-G5-phonon} lists the six rows of the $G_5$ pressure scan that satisfy the parent-symmetry criterion ($\Delta < 0.05~\si[mode=text]{\angstrom}$) and report either a real or an imaginary lowest mode. Two of these rows (at $p = 0.135$~\si[mode=text]{\giga\Pa} and $p = 0.140$~\si[mode=text]{\giga\Pa}) lie just above the apparent transition and break the monotonic decrease of $\omega^2(p)$; we attribute this to numerical re-entry of the parent symmetry under tight reoptimisation. The row at $p = 0.160$~\si[mode=text]{\giga\Pa} reports a stable real mode ($\omega = 10.95$~\si[mode=text]{\per\centi\meter}) that corresponds to the lowest acoustic branch of the relaxed \acentric{} structure rather than to the parent symmetry; this row is excluded from the Cowley-Levanyuk fit.

A weighted linear regression of $\omega^2(p)$ on the three monotonically-decreasing parent-branch points retained in \textbf{Table} \ref{tab:SI-G5-phonon} ($p = 0$ and $0.120$~\si[mode=text]{\giga\Pa} with positive $\omega^2$, and $p = 0.130$~\si[mode=text]{\giga\Pa} with negative $\omega^2$) returns $p_c = 0.1282 \pm 0.0026$~\si[mode=text]{\giga\Pa} and $\alpha \simeq 9.2\times 10^{2}$~\si[mode=text]{\per\centi\meter\squared\per\giga\Pa}. The fit is shown as the purple line of panel (e) of \textbf{Figure} \ref{fig:Phonons}, and the resulting $p_c$ is the value reported in \textbf{Table} \ref{tab:parameters} of the main text. The standard error on $p_c$ is broader than that of the other four members ($\sim 2.4\%$ for $G_5$ versus $\sim 0.1$-$0.4\%$ for $G_1$-$G_4$) and reflects the smaller number of $\omega^2$ points available, not a difference in the underlying physical mechanism.

\begin{table}[h!]
\caption{\label{tab:SI-G5-phonon}Available phonon data for the $G_5$ pressure scan within the SLC potential, restricted to the parent-symmetry branch ($\Delta < 0.05$~\si[mode=text]{\angstrom}). Pressure $p$ in GPa; lowest-mode frequency reported as $\omega_\text{real}$ when stable and as $\omega_\text{imag}$ (with the conventional sign convention) when unstable; $\omega^2$ is the signed squared frequency used in the Cowley-Levanyuk regression. The last column flags the rows retained in the fit (see text for the exclusion criteria of the remaining rows).}
\centering
\begin{tabular}{c c c c c}
\toprule
$p$ [\si[mode=text]{\giga\Pa}] & $\omega_\text{real}$ [\si[mode=text]{\per\centi\meter}] & $\omega_\text{imag}$ [\si[mode=text]{\per\centi\meter}] & $\omega^2$ [\si[mode=text]{\per\centi\meter\squared}] & Used in fit \\
\midrule
$0.000$ & $10.86$ & -      & $+117.94$ & yes \\
$0.120$ & $\phantom{0}2.95$ & -      & $\phantom{+0}+8.70$ & yes \\
$0.130$ & -      & $-2.12$ & $\phantom{0}-4.49$ & yes \\
$0.135$ & -      & $-3.33$ & $-11.09$  & no (non-monotonic) \\
$0.140$ & -      & $-2.13$ & $\phantom{0}-4.54$ & no (non-monotonic) \\
$0.160$ & $10.95$ & -      & $+119.90$ & no (\acentric{} branch) \\
\bottomrule
\end{tabular}
\end{table}

\subsection{Finite-size effect for $G_1$}
\label{si:finite-size}
For the smallest member of the family, $G_1$ (RHO, 48 framework T-atoms per cubic cell), we have additionally performed the full pressure scan on a $2\times 2\times 2$ supercell (384 T-atoms) under the same protocol used for the unit-cell calculations. Once the unit cell is sampled with a sufficiently dense pressure grid (every 10 to 50 bar in the critical region), the apparent critical pressure converges to $p_c\simeq 0.94$~\si[mode=text]{\giga\Pa}, in agreement with the $2\times 2\times 2$ value to within the fit uncertainties (\textbf{Table} \ref{tab:SI-finite-size}). The \centric-phase elastic constants ($C_{11}$, $C_{12}$, $\kappa_T$, $K_T$) are also indistinguishable between the two cell sizes, as expected for intensive properties. What does depend on the cell size is the \emph{noise} in the \acentric $\Delta(p)$ curve: the 48-atom unit cell exhibits a markedly scattered $\Delta$ in the \acentric branch close to $p_c$ (the \acentric optimisation lands in slightly different metastable basins depending on the seed), while the $2\times 2\times 2$ supercell averages over many local copies of the same distortion pattern and returns a smoother $\Delta(p)$. We attribute this to the rugged near-critical energy landscape of $G_1$ in the small cell, where the local minima of the symmetry-\acentric phase are separated by very shallow barriers that the small system can sample stochastically; in the supercell the corresponding many-cell average smooths out the fluctuations. The figures and tables of the main text use the $2\times 2\times 2$ supercell values for $G_1$ to suppress this near-critical scatter while preserving the agreement, within fit uncertainty, with the unit-cell critical pressure and elastic constants; the unit-cell values are reported in \textbf{Table} \ref{tab:SI-finite-size} as a consistency check.

\begin{table}[h!]
\caption{\label{tab:SI-finite-size}Cell-size cross-check of the critical parameters of $G_1$ (RHO) within the SLC potential. Elastic constants reported at $p = 0$~\si[mode=text]{\giga\Pa}. The Heaviside-power fit of \textbf{Equation} \ref{eq:fit} is performed with all three parameters ($a$, $p_c$, $\beta$) free, hence the slightly different values of $p_c$ and $\beta$ compared with \textbf{Table} \ref{tab:parameters} of the main text, where $p_c$ is held at the soft-mode value; both fits agree within $1\sigma$. Errors are one standard deviation from the non-linear fit.}
\centering
\begin{tabular}{l c c}
\toprule
Quantity & $1\times 1\times 1$ (48 T-atoms) & $2\times 2\times 2$ (384 T-atoms) \\
\midrule
$C_{11}(0)$ [\si[mode=text]{\giga\Pa}] & $89.69$ & $89.69$ \\
$C_{12}(0)$ [\si[mode=text]{\giga\Pa}] & $66.12$ & $66.12$ \\
$\kappa_T(0)$ [\si[mode=text]{\per\giga\Pa}] & $0.01352$ & $0.01352$ \\
$K_T(0)$ [\si[mode=text]{\giga\Pa}] & $73.98$ & $73.98$ \\
$\delta$ [\si[mode=text]{\angstrom}] & $2.25 \pm 0.10$ & $2.213 \pm 0.018$ \\
$p_c$ [\si[mode=text]{\giga\Pa}] & $0.944 \pm 0.011$ & $0.940 \pm 0.004$ \\
$\beta$ & $0.40 \pm 0.06$ & $0.358 \pm 0.012$ \\
Pressure-grid spacing near $p_c$ [\si[mode=text]{\bar}] & $\le 25$ & $\le 100$ \\
\bottomrule
\end{tabular}
\end{table}

\begin{figure}[h!t]
\centering
\includegraphics[width=0.7\textwidth]{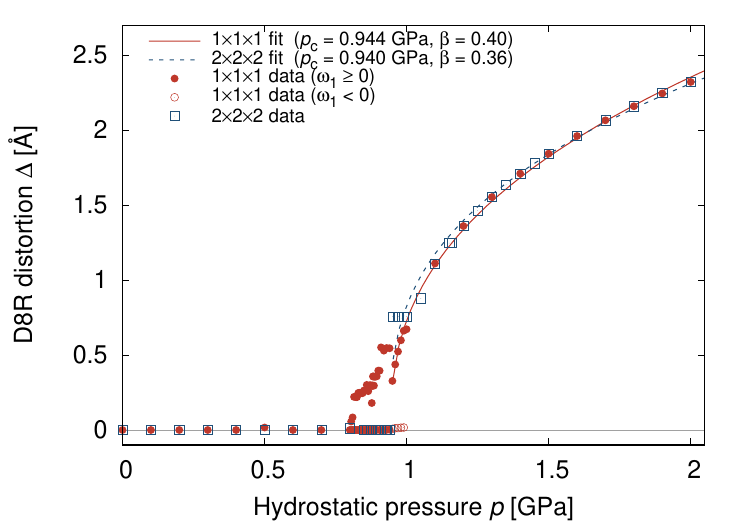}
\caption{\label{fig:SI-G1-finite-size}D8R distortion $\Delta$ as a function of hydrostatic pressure for the pure-silica $G_1$ (RHO) framework, computed with the SLC potential in two cell sizes: the cubic unit cell with $48$ T-atoms (red; filled circles for $\omega_1\geq 0$, open circles for $\omega_1 < 0$) and a $2\times 2\times 2$ supercell with $384$ T-atoms (blue open squares). Solid and dashed lines are the Heaviside-power fits of \textbf{Equation} \ref{eq:fit} for the supercell and unit cell respectively. The corresponding fit parameters are listed in \textbf{Table} \ref{tab:SI-finite-size}.}
\end{figure}

\subsection{Phenomenological fit parameters for volume and order parameter}
\label{si:volume-fits}
\textbf{Table} \ref{tab:SI-volume} collects the parameters of the linear fits to the relative volume curves of \textbf{Figure} \ref{fig:VolumeVsPressure} (\textbf{Equation} \ref{eq:kappa}); the parameters of the Heaviside power-law fits to the order parameter (\textbf{Equation} \ref{eq:fit}) are reported in the main-text \textbf{Table} \ref{tab:parameters}. Both tabulations include the asymptotic standard errors returned by the non-linear least-squares routines of \texttt{gnuplot} v.5.4; we note that, because the SLC pressure scans are noiseless, the absolute scale of these errors is set by the deviations of the fitted form from the data rather than by genuine measurement variance, so we report them as RMSE-style residuals rather than as standard reduced-$\chi^{2}$ statistics. The largest residuals are found on the post-transition branch of $G_1$, where the proximity of the apparent volume jump enlarges the deviations.
\begin{table}[h!]
\caption{\label{tab:SI-volume}Parameters of the piecewise linear fit to the relative volume $v\defeq V/V_0$ of \textbf{Figure} \ref{fig:VolumeVsPressure}. $\kappa_1$ and $\kappa_2$ are the slopes (in \si[mode=text]{\per\giga\Pa}) of the pre- and post-transition branches, respectively. Errors quoted as one standard deviation from the fit.}
\centering
\begin{tabular}{l c c}
\toprule
$G_k$ & $\kappa_1$ [\si[mode=text]{\per\giga\Pa}] & $\kappa_2$ [\si[mode=text]{\per\giga\Pa}] \\
\midrule
$G_1$$^{\dagger}$ & $-0.01351 \pm 0.00001$ & $-0.1392 \pm 0.0018$ \\
$G_2$           & $-0.01340 \pm 0.00001$ & $-0.0810 \pm 0.0003$ \\
$G_3$           & $-0.01410 \pm 0.00014$ & $-0.0753 \pm 0.0003$ \\
$G_4$           & $-0.01389 \pm 0.00018$ & $-0.0734 \pm 0.0003$ \\
$G_5$           & $-0.02167$$^{\ddagger}$ & $-0.0730 \pm 0.0005$ \\
\bottomrule
\end{tabular}
\newline {\small $^{\dagger}$ $G_1$ from the $2\times 2\times 2$ supercell, see \textbf{Section} \ref{si:finite-size}. \\ $^{\ddagger}$ $G_5$ pre-transition window narrower than $0.10$~\si[mode=text]{\giga\Pa} after excluding the gap of half-width $h = 0.06$~\si[mode=text]{\giga\Pa} around $p_c$; standard error cannot be reliably estimated.}
\end{table}

\subsection{Model comparison for the decrease of $p_c$ with isoreticular order}
\label{si:pc-models}
The critical pressures of \textbf{Table} \ref{tab:parameters} were fitted to the closed-form exponential law
\begin{equation}
\label{eq:SI-pc-exp}
p_c(k) = p_0\,\exp\bigl(-\lambda\,(k-1)\bigr),
\end{equation}
with both the prefactor $p_0$ and the decay rate $\lambda$ free. The best-fit values (\textbf{Table} \ref{tab:SI-exp}) are $p_0 = 0.930 \pm 0.022$~\si[mode=text]{\giga\Pa} and $\lambda = 0.494 \pm 0.023$. The fitted $p_0$ is statistically indistinguishable from the computed value for RHO, $p_c(\text{RHO}) = 0.942$~\si[mode=text]{\giga\Pa}.
\begin{table}[h!]
\caption{\label{tab:SI-exp}Parameters of the closed-form exponential fit $p_c(G_k) = p_0\,\exp[-\lambda\,(k-1)]$ (\textbf{Equation} \ref{eq:SI-pc-exp}) with both prefactor and decay rate free. Errors are one standard deviation from the non-linear fit.}
\centering
\begin{tabular}{l c}
\toprule
Parameter & Value \\
\midrule
$p_0$ [\si[mode=text]{\giga\Pa}]              & $0.930 \pm 0.022$ \\
$\lambda$                                     & $0.494 \pm 0.023$ \\
Predicted $p_c(G_6)$ [\si[mode=text]{\giga\Pa}] & $0.079 \pm 0.009$ \\
Predicted $p_c(G_7)$ [\si[mode=text]{\giga\Pa}] & $0.048 \pm 0.007$ \\
Predicted $p_c(G_8)$ [\si[mode=text]{\giga\Pa}] & $0.029 \pm 0.005$ \\
\bottomrule
\end{tabular}
\end{table}

The exponential form of \textbf{Equation} \ref{eq:SI-pc-exp} is not the only smooth law compatible with the five computed values. Two physically motivated alternatives were tested: a power law in the cubic lattice parameter $a_k$ ($p_c \propto a_k^{-n}$, motivated by the rigid-unit-mode interpretation that the soft-mode frequency scales with the inverse of a characteristic block length), and a power law in the framework T-atom count $N_T(k)$ ($p_c \propto N_T^{-m}$, motivated by the scaling of the framework cohesive energy with the number of bond-bending degrees of freedom per unit cell). The three two-parameter fits are reported in \textbf{Table} \ref{tab:SI-pc-models}.

\begin{table}[h!]
\caption{\label{tab:SI-pc-models}Comparison of three candidate two-parameter functional forms for $p_c(k)$ fitted to the five soft-mode-derived values of \textbf{Table} \ref{tab:parameters}. All fits are unweighted in $p_c$; RMSE is reported in GPa and as a percentage of $\langle p_c\rangle = 0.438$~\si[mode=text]{\giga\Pa}. The M1 parameters $(p_0, \lambda)$ coincide with the weighted-fit values quoted in \textbf{Table} \ref{tab:SI-exp} and \textbf{Equation} \ref{eq:pc-exponential} of the main text within the rounding; the error bars on those parameters in \textbf{Table} \ref{tab:SI-exp} are the asymptotic standard errors of the weighted non-linear least-squares fit, larger than the variation between weighted and unweighted point estimates. Lower AIC/BIC indicates a better description of the empirical data at fixed number of parameters. The cubic lattice parameter for $G_6$-$G_8$ is extrapolated linearly from the SLC-optimised values of $G_1$-$G_5$ as $a_k = 5.18 + 9.86\,k$ \si[mode=text]{\angstrom} (predictions $a_6 = 64.4$, $a_7 = 74.2$, $a_8 = 84.1$~\si[mode=text]{\angstrom}); these differ by less than $1\%$ from the topology-only Shevchenko-Krivovichev law $a_k = 14.9 + 10.76\,(k-1)$~\si[mode=text]{\angstrom} ($a_6 = 65.7$, $a_7 = 76.4$, $a_8 = 87.2$~\si[mode=text]{\angstrom}); the two choices yield $p_c$ predictions within $5\%$ of each other for M2. The framework T-atom count $N_T(k) = 8k(1 + 3k + 2k^{2})$ is exact.}
\centering
\footnotesize
\begin{tabular}{l c c c c c c c}
\toprule
Model & Parameters & RMSE [\si[mode=text]{\giga\Pa}] & $\Delta$AIC & $\Delta$BIC & $p_c(G_6)$ & $p_c(G_7)$ & $p_c(G_8)$ \\
      &            &                                  &              &              & [\si[mode=text]{\giga\Pa}] & [\si[mode=text]{\giga\Pa}] & [\si[mode=text]{\giga\Pa}] \\
\midrule
$p_c = p_0\,e^{-\lambda(k-1)}$ (M1) & $p_0=0.930$, $\lambda=0.494$ & $0.018$ ($4.2\%$) & $0$    & $0$    & $0.079$ & $0.048$ & $0.029$ \\
$p_c = \beta\,a_k^{-n}$ (M2)        & $\beta=28.6$, $n=1.26$       & $0.036$ ($8.2\%$) & $+6.6$ & $+6.6$ & $0.152$ & $0.127$ & $0.109$ \\
$p_c = \gamma\,N_T^{-m}$ (M3)       & $\gamma=4.60$, $m=0.406$     & $0.036$ ($8.3\%$) & $+6.7$ & $+6.7$ & $0.153$ & $0.129$ & $0.111$ \\
\bottomrule
\end{tabular}
\end{table}

The exponential law M1 is preferred over both power-law alternatives at $\Delta$AIC $> 6$ (equivalently, the data favour M1 over M2 and M3 by likelihood ratios of $\sim 27$), and the soft-mode-derived $p_c$ are within $4.2\%$ RMSE of the M1 closed form. However, the three models disagree dramatically in their extrapolation to $G_6$-$G_8$: the exponential M1 predicts $p_c$ values $\sim 2$ to $\sim 4$ times smaller than the power-law alternatives M2 and M3. Treating the spread between the three model predictions as a conservative uncertainty band, the present analysis predicts $p_c \in [0.08, 0.15]~\si{\giga\Pa}$ for $G_6$, $[0.05, 0.13]~\si{\giga\Pa}$ for $G_7$ and $[0.03, 0.11]~\si{\giga\Pa}$ for $G_8$. The exponential value remains our preferred central estimate, but a fully decisive choice among the three models is not possible with only five data points and would require independent calculations for $G_6$, which are currently RAM-prohibitive at the RFO Hessian level and are being pursued with a frozen-soft-mode protocol in follow-up work.

\subsection{Empirical data collapse and quantitative universality test}
\label{si:data-collapse}
\begin{figure}[h!t]
\centering
  \includegraphics[width=0.5\textwidth]{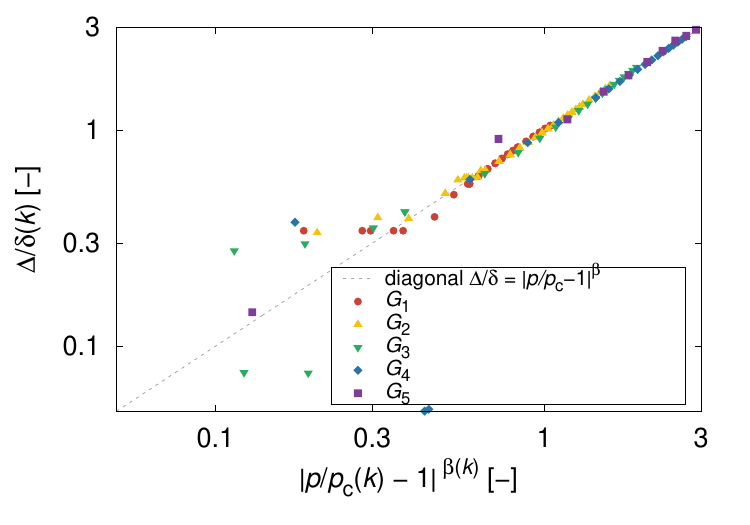}
  \caption{\label{fig:isocriticalpressure2}Empirical data collapse of the RHO isoreticular family on the rescaled axes $\abs*{\nicefrac{p}{p_c}(k)-1}^{\beta(k)}$ (abscissa) and $\nicefrac{\Delta}{\delta(k)}$ (ordinate). The family-specific scales $(p_c, \delta, \beta)$ for each $G_k$ are taken from \textbf{Table} \ref{tab:SI-landau}. Dashed grey line: diagonal $\Delta/\delta = \abs*{p/p_c-1}^{\beta}$ predicted by \textbf{Equation} \ref{eq:fit}.}
\end{figure}

To put the collapse on a quantitative footing, we restrict the \acentric branch ($p > p_c$) of every $G_k$ and evaluate the residual root-mean-square of the rescaled order parameter against the canonical mean-field curve $\nicefrac{\Delta}{\delta} = (\nicefrac{p}{p_c} - 1)^{1/2}$. For $G_2$-$G_5$ the RMS lies between $0.040$ and $0.080$, with a median of $0.052$, comparable to the per-point statistical noise of the SLC scan ($\lesssim 5\%$ in $\Delta/\delta$ near $p_c$ where the curve is most sensitive). $G_1$ returns RMS $= 0.086$, $\sim 1.65\times$ the median of the other four, and the fit improves by $42\%$ if its own $\beta = 0.355$ is used instead of $1/2$, whereas for $G_2$-$G_5$ the improvement is below $30\%$. Quantitatively, therefore, $G_2$-$G_5$ are statistically indistinguishable from the mean-field universality class within the present sampling, while $G_1$ deviates systematically: the manuscript dichotomy of "mean-field for $G_2$-$G_5$, tricritical-like for $G_1$" stands as a measurable feature of the data, not as a fit artefact.

\subsection{Landau parameters and tricritical proximity}
\label{si:landau-params}
The Landau free-energy density introduced in the main text (\textbf{Equation} \ref{eq:LandauF}, truncated here at the quartic level for the explicit closed-form analysis),
\begin{equation}
\label{eq:SI-landau-quartic}
\mathcal{G}(\Delta, e ; p) = \mathcal{G}_c(p) + \tfrac{1}{2}A\,(1-\nicefrac{p}{p_c})\,\Delta^{2} + \tfrac{1}{4}U\,\Delta^{4} + \tfrac{1}{2}K_{0}\,e^{2} + G\,e\,\Delta^{2},
\end{equation}
with $e \defeq v - v_\mathrm{c}(p)$ the residual volumetric strain measured from the smoothly compressed \centric{} branch $v_\mathrm{c}(p) = 1 + \kappa_1\,p$ (\textbf{Equation} \ref{eq:residual-strain}), admits a tight extraction of its four coefficients from the empirical fits reported in \textbf{Tables} \ref{tab:parameters} and \ref{tab:SI-volume}. Minimising in $e$ at fixed $\Delta$ yields the closed form $e_\text{min}(\Delta) = -(G/K_{0})\,\Delta^{2}$, the renormalised quartic $U_\text{eff} = U - 2 G^{2}/K_{0}$, and the post-transition equation of state $v(p) = v_\mathrm{c}(p) - (G/K_{0})\,\Delta^{2}$. Three identifications follow: (i) $K_{0} = 1/|\kappa_{1}|$ from the pre-transition slope of $v(p)$; (ii) $G/K_{0} = (|\kappa_{2}| - |\kappa_{1}|)\,p_c/\delta^{2}$ from the change of slope at $p_c$ (\textbf{Equation} \ref{eq:kappa2-landau} of the main text); and (iii) $\delta^{2} = A/U_\text{eff}$ by direct comparison of the Landau saddle-point solution $\Delta^{2} = (A/U_\text{eff})\,(\nicefrac{p}{p_c}-1)$ with the empirical fit of \textbf{Equation} \ref{eq:fit}.

\begin{table}[h!]
\caption{\label{tab:SI-landau}Master compilation of empirical fit parameters and derived Landau coefficients for every isoreticular member $G_k$. Columns 2-4 ($p_c$, $\delta$, $\beta$) and column 6 ($|\kappa_2|$) come from the Heaviside-power fit of \textbf{Equation} \ref{eq:fit} (\textbf{Table} \ref{tab:parameters}); column 5 ($|\kappa_1|$) from the piecewise-linear fit of \textbf{Equation} \ref{eq:kappa} (\textbf{Table} \ref{tab:SI-volume}). The last four columns are the Landau coefficients obtained from the identifications $K_0 = 1/|\kappa_1|$, $U_\text{eff}/A = 1/\delta^{2}$ and $G/K_0 = (|\kappa_2|-|\kappa_1|)\,p_c/\delta^{2}$ (\textbf{Section} \ref{si:landau-params}). The dimensionless crossover parameter $\eta = 4\,W\,A/U_\text{eff}^{2}$ is obtained from the direct sextic fit of $\Delta^{2}(p)$ on the broken phase reported in \textbf{Section} \ref{si:sextic-direct}.}
\centering
\footnotesize
\setlength\tabcolsep{4pt}
\begin{tabular}{c c c c c c c c c c}
\toprule
$k$ &
$p_c$ &
$\delta$ &
$\beta$ &
$\lvert\kappa_1\rvert$ &
$\lvert\kappa_2\rvert$ &
$K_0$ &
$U_\text{eff}/A$ &
$G/K_0$ &
$\eta$ \\
 & [\si[mode=text]{\giga\Pa}] & [\si[mode=text]{\angstrom}] & [-] & [\si[mode=text]{\per\giga\Pa}] & [\si[mode=text]{\per\giga\Pa}] & [\si[mode=text]{\giga\Pa}] & [\si[mode=text]{\per\angstrom\squared}] & [\si[mode=text]{\per\angstrom\squared}] & [-] \\
\midrule
$1$ & $0.9418$ & $2.210$ & $0.355$ & $0.01351$ & $0.1392$ & $74.02$ & $0.205$ & $0.0242$ & $5.3$ \\
$2$ & $0.5338$ & $1.236$ & $0.455$ & $0.01340$ & $0.0810$ & $74.63$ & $0.655$ & $0.0236$ & $0.12$ \\
$3$ & $0.3627$ & $1.014$ & $0.429$ & $0.01410$ & $0.0753$ & $70.92$ & $0.973$ & $0.0216$ & $0.05$ \\
$4$ & $0.2241$ & $0.744$ & $0.480$ & $0.01389$ & $0.0734$ & $71.94$ & $1.806$ & $0.0241$ & $< 0.01$ \\
$5$ & $0.1282$ & $0.464$ & $0.554$ & $0.02167$ & $0.0730$ & $46.15$ & $4.640$ & $0.0306$ & $< 0.01$ \\
\bottomrule
\end{tabular}
\end{table}

\begin{figure}[h!]
\centering
\begin{subfigure}[t]{0.49\textwidth}
\centering
\caption{\label{fig:SI-landau-wells:a}$f_\text{eff}(\Delta;p) = \tfrac{1}{2}A(1-\nicefrac{p}{p_c})\Delta^{2} + \tfrac{1}{4}U_\text{eff}\Delta^{4}$}
\includegraphics[width=\textwidth]{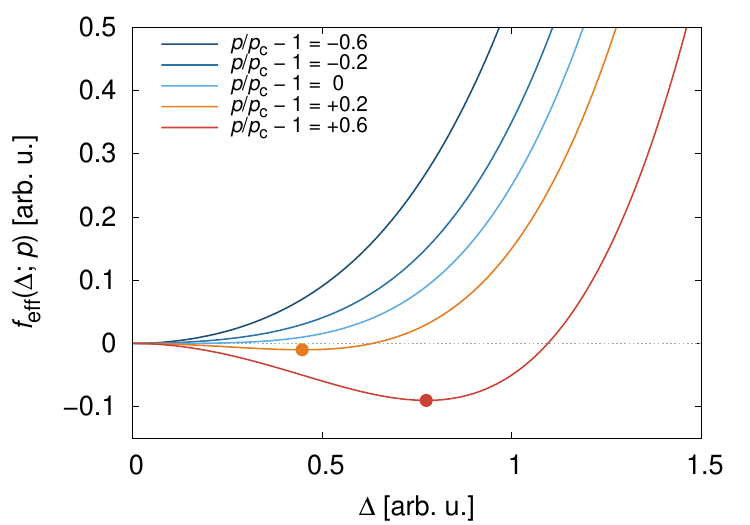}
\end{subfigure}\hfill
\begin{subfigure}[t]{0.49\textwidth}
\centering
\caption{\label{fig:SI-landau-wells:b}$v(p) - v_\mathrm{c}(p) = -(G/K_0)\Delta^{2}$}
\includegraphics[width=\textwidth]{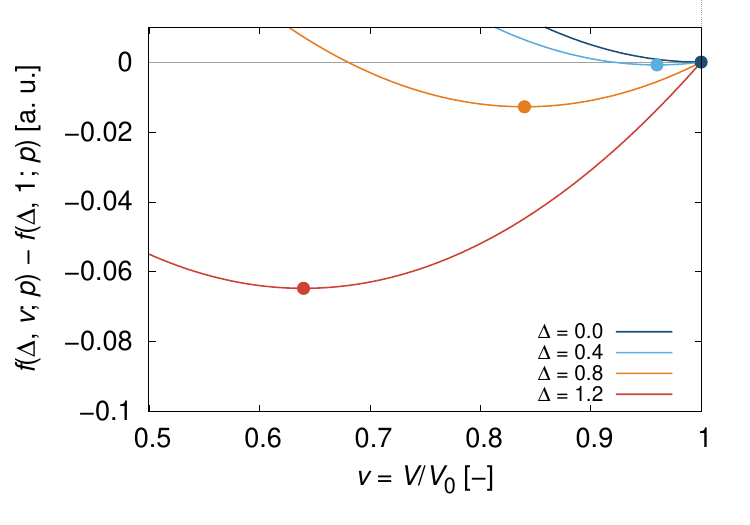}
\end{subfigure}

\vspace{0.6em}

\begin{subfigure}[t]{0.49\textwidth}
\centering
\caption{\label{fig:SI-landau-wells:c}$G_1$: $\delta = 2.210$~\si[mode=text]{\angstrom}, $U_\text{eff}/A = 0.205$~\si[mode=text]{\per\angstrom\squared}, $p_c = 0.9418$~\si[mode=text]{\giga\Pa}}
\includegraphics[width=\textwidth]{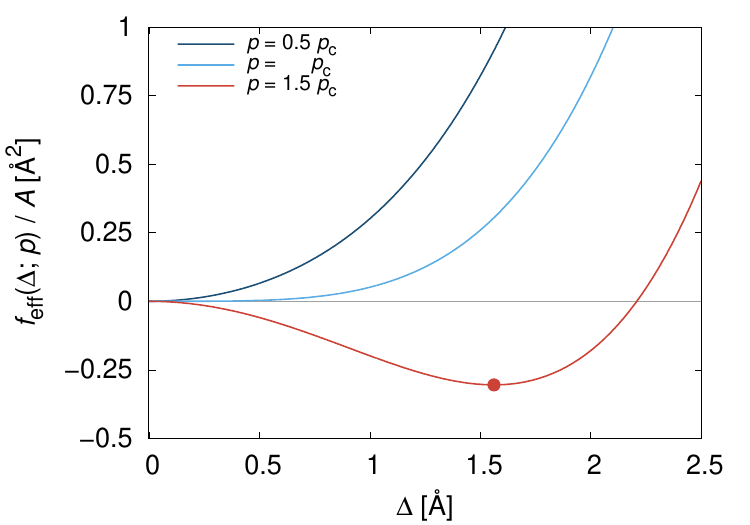}
\end{subfigure}\hfill
\begin{subfigure}[t]{0.49\textwidth}
\centering
\caption{\label{fig:SI-landau-wells:d}$G_5$: $\delta = 0.464$~\si[mode=text]{\angstrom}, $U_\text{eff}/A = 4.640$~\si[mode=text]{\per\angstrom\squared}, $p_c = 0.1282$~\si[mode=text]{\giga\Pa}}
\includegraphics[width=\textwidth]{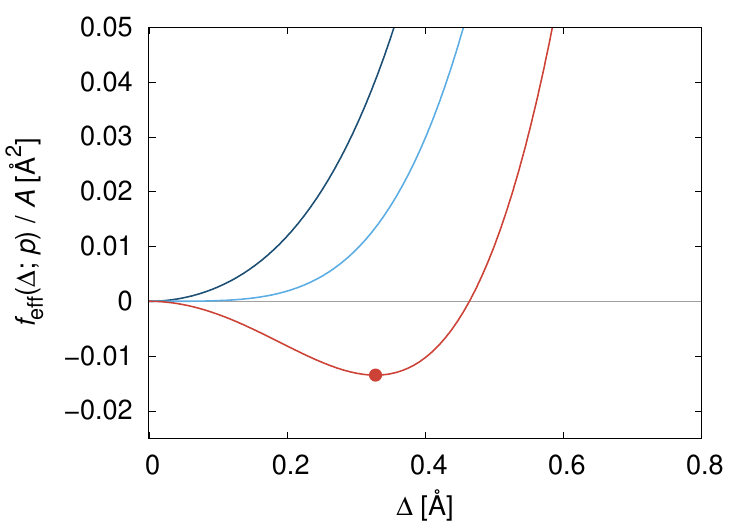}
\end{subfigure}
\caption{\label{fig:SI-landau-wells}Visualisation of the Landau functional of \textbf{Equation} \ref{eq:LandauF}. Top row (conceptual, dimensionless units with $A = U_\text{eff} = K_0 = 1$, $G = 0.25$): (a) effective free energy obtained by minimisation in the residual strain $e$ for five pressures $\nicefrac{p}{p_c} - 1 \in \{-0.6, -0.2, 0, +0.2, +0.6\}$ (dark blue $\to$ red), with dots marking the \acentric minima $\Delta_\text{eq} = \sqrt{(A/U_\text{eff})(\nicefrac{p}{p_c}-1)}$; (b) full free energy at $\nicefrac{p}{p_c} - 1 = +0.3$ as a function of $v$ relative to $v_\mathrm{c}(p)$ for $\Delta \in \{0.0, 0.4, 0.8, 1.2\}$ (dark blue $\to$ red), with dots at $v(p)$ from \textbf{Equation} \ref{eq:vmin}. Bottom row (quantitative, with the parameters of \textbf{Table} \ref{tab:SI-landau}): effective potential $f_\text{eff}/A$ for the two extremes of the family $G_1$ (c) and $G_5$ (d) at $p = 0.5\,p_c$ (dark blue, cubic), $p_c$ (light blue, transition) and $1.5\,p_c$ (red, \acentric). All axes are restricted to the physical quadrants $\Delta \geq 0$ and $v \leq 1$.}
\end{figure}

\textbf{Figure} \ref{fig:SI-landau-wells} visualises the structure of the Landau functional in two layers. The top row (panels a, b) shows the dimensionless free energy cuts that make the algebra explicit: the single-well cubic landscape flattens at $p_c$ and develops a finite minimum above $p_c$ at $\Delta \propto \sqrt{\nicefrac{p}{p_c}-1}$, while the parabolic strain well shifts to $v(p) = v_\mathrm{c}(p) - G\Delta^{2}/K_{0}$ as the order parameter grows; the energy released by that strain reorganisation is precisely what renormalises $U \to U_\text{eff}$ in the effective theory. The bottom row (panels c, d) instantiates the same effective potential with the parameters extracted from the simulations (\textbf{Table} \ref{tab:SI-landau}) for the two extremes of the family. The contrast between $G_1$ and $G_5$ is the structural origin of the dichotomy in the order-parameter amplitude $\delta$ of \textbf{Table} \ref{tab:parameters}: $G_1$ supports a much larger $\Delta_\text{eq}$ than $G_5$ for the same relative pressure excursion above $p_c$, in line with the lower $U_\text{eff}/A$ of the parent RHO framework. The depth of the \acentric minimum in units of $A$ remains comparable (a few $10^{-1}$ \si[mode=text]{\angstrom\squared}), but the spatial extension of $\Delta$ collapses by a factor of $\sim 4$.

\begin{figure}[h!]
\centering
\includegraphics[width=0.7\textwidth]{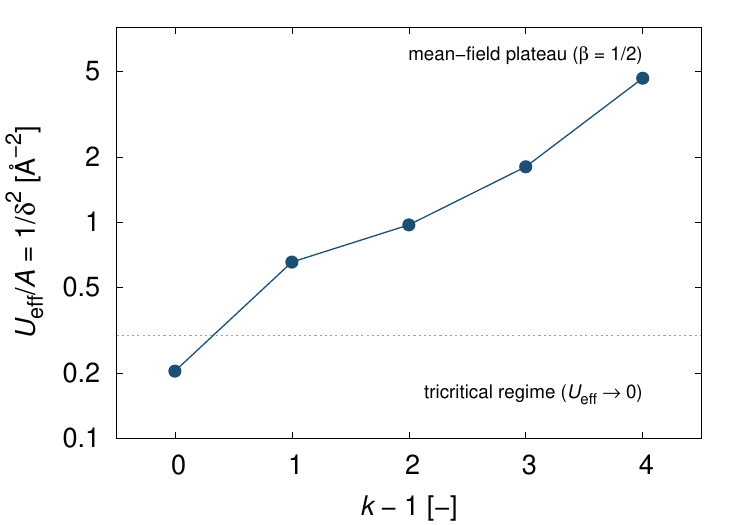}
\caption{\label{fig:SI-landau-ueff}Dimensionless ratio $U_\text{eff}/A = 1/\delta^{2}$ as a function of the isoreticular order $k$, extracted from the Landau analysis of \textbf{Section} \ref{si:landau-params}. Filled circles connected by lines for the five members $G_1$-$G_5$. The dashed horizontal line at $U_\text{eff}/A = 0.30$~\si[mode=text]{\per\angstrom\squared} marks the conventional lower bound for the mean-field-compatible regime; tricritical and mean-field labels are added for orientation.}
\end{figure}

\textbf{Table} \ref{tab:SI-landau} consolidates, per $G_k$, the four direct fit parameters ($p_c$, $\delta$, $\beta$, $|\kappa_2|$), the pre-transition slope $|\kappa_1|$, and the four Landau quantities derived from them ($K_0$, $U_\text{eff}/A$, $G/K_0$ and the sextic crossover parameter $\eta$ extracted in \textbf{Section} \ref{si:beta-G1-derivation}). The dimensionless ratio $U_\text{eff}/A = 1/\delta^{2}$ in this column is the \emph{quartic-only} estimate obtained from the Heaviside-power amplitude $\delta$; the alternative \emph{direct-sextic} estimate $U_\text{eff}/A = 2/(C\eta)$ from the joint fit of $C$ and $\eta$ to $\Delta^{2}(p)$ is reported in \textbf{Table} \ref{tab:SI-eta-direct} of \textbf{Section} \ref{si:sextic-direct} and differs from the quartic-only value by a factor of $\sim 1.8$ for $G_1$ ($0.115$ vs $0.205$~\si[mode=text]{\per\angstrom\squared}) and by less than $10\%$ for $G_2$-$G_4$. We adopt the quartic-only $U_\text{eff}/A$ as the reference in the discussion below because it is the value uniquely determined by $\delta$ and therefore minimises the propagation of fit uncertainty through the family; the implications of the sextic-corrected value for $G_1$ are addressed in \textbf{Section} \ref{si:sextic-direct}. The dimensionless ratio is the key tricriticality indicator: $U_\text{eff}/A \to 0$ is the tricritical condition, while large $U_\text{eff}/A$ defines a robust mean-field regime. The ratio grows monotonically across the family, $U_\text{eff}/A = 0.21, 0.66, 0.97, 1.81, 4.64$~\si[mode=text]{\per\angstrom\squared} from $G_1$ to $G_5$, a factor of $\sim 23$ enhancement on the quartic scale (\textbf{Figure} \ref{fig:SI-landau-ueff}); the direct-sextic scale returns a slightly larger factor of $\sim 31$ ($0.115\to 3.61$, \textbf{Table} \ref{tab:SI-eta-direct}). $G_1$ sits closest to the tricritical limit on either scale, consistent with its anomalously low exponent $\beta = 0.355$ and the small first-order discontinuity of $\Delta$ at $p_c$ inferred from the negative $\Delta v$. $G_4$ and $G_5$, in contrast, sit on a flat mean-field plateau where the residual quartic stiffness firmly imposes $\beta = 1/2$. The strain-order-parameter coupling $G/K_0$, in turn, is almost flat across the family ($\simeq 0.022$-$0.024$~\si[mode=text]{\per\angstrom\squared} for $G_1$-$G_4$, $0.031$ for $G_5$), supporting the interpretation that $G$ is set by the local SiO$_{4}$ bond-bending energetics, common to every member, rather than by the isoreticular topology. The bulk modulus $K_0$ is similarly constant across $G_1$-$G_4$ ($\simeq 71$-$75$~\si[mode=text]{\giga\Pa}, with the four members spanning $70.9, 74.0, 74.6, 71.9$~\si[mode=text]{\giga\Pa}) and only drops for $G_5$ ($46.2$~\si[mode=text]{\giga\Pa}), the latter reflecting the narrower pre-transition fit window available for $G_5$ rather than a genuine topology-driven softening.

\subsection{Analytical derivation of the $G_1$ exponent: a sextic Landau crossover}
\label{si:beta-G1-derivation}

Two empirical observations require a common explanation. The order-parameter exponent of $G_1$, $\beta = 0.355$, sits comfortably between the mean-field value $1/2$ and the tricritical value $1/4$, while the four larger members $G_2$-$G_5$ converge to $\beta \simeq 1/2$ within the fit uncertainty. Two distinct mechanisms could in principle be invoked. The simplest, however, is that a single closed-form extension of the quartic Landau functional accounts for both at once. The argument is purely structural and admits an analytical solution, which we develop below.

The Landau functional truncated at fourth order in $\Delta$ predicts $\beta = 1/2$ identically for every system in which the elastic strain enters as a spectator coordinate, as we have verified by the calculation in the main text. The only way to obtain $\beta < 1/2$ while remaining in a continuous, or weakly first-order, transition framework is to allow the quartic coefficient $U_\text{eff}$ to become small, in which case the next-leading even-symmetry term, $\Delta^{6}$, eventually takes over. Letting $U_\text{eff}$ vanish exactly is the standard definition of the tricritical condition, at which $\beta = 1/4$ holds because the sixth-order term sets the curvature of the \acentric well. The interesting regime, the one that may apply to $G_1$, is the one in between: a small but non-zero $U_\text{eff}$ and a sizeable $W$ produce a crossover in which the apparent exponent $\beta_\text{eff}$ depends on a single dimensionless number that quantifies how close one is to the tricritical limit.

Before writing the extended functional, it is worth recording which sign constraints on the coefficients follow from symmetry alone and which are empirical inputs. The $A_{2u}$ character of $\Delta$ enforces $\Delta \to -\Delta$ under the inversion of the parent group, so every odd power of $\Delta$ is identically forbidden in the Landau expansion: no $\Delta^{1}$, $\Delta^{3}$ or $\Delta^{5}$ term may appear. The sextic coefficient $W$ is required to be strictly positive on thermodynamic stability grounds, otherwise the truncated Landau density is unbounded below and the equilibrium probability density of $\Delta$ ceases to be normalisable. The quadratic coefficient must change sign at the transition by definition of $p_c$, $U_\text{eff}(p_c)=0$ in our parametrisation. The only coefficient whose sign is not fixed a priori is the quartic stiffness $U_\text{eff}$: its sign distinguishes the second-order ($U_\text{eff}>0$), tricritical ($U_\text{eff}=0$) and first-order ($U_\text{eff}<0$) regimes of the same family of functionals, and is the parameter whose empirical near-universality across $G_2$-$G_5$ (and whose sign reversal for $G_1$ at the DFT level) underpins the central narrative of the present work.

To make this idea quantitative we extend the manuscript functional by the next allowed even-power term, $\frac{1}{6}W\,\Delta^{6}$,
\begin{equation}
\label{eq:landau-sextic}
\mathcal{G}(\Delta, e; p) = \mathcal{G}_c(p) + \tfrac{1}{2}A\Bigl(1 - \nicefrac{p}{p_c}\Bigr)\Delta^{2} + \tfrac{1}{4}U\,\Delta^{4} + \tfrac{1}{6}W\,\Delta^{6} + \tfrac{1}{2}K_{0}\,e^{2} + G\,e\,\Delta^{2},
\end{equation}
written in the residual-strain representation $e = v - v_\mathrm{c}(p)$ of \textbf{Equation} \ref{eq:residual-strain}. The residual strain $e$ remains a spectator: minimising in $e$ at fixed $\Delta$ gives $e_\text{min}(\Delta) = -(G/K_0)\Delta^{2}$ and the post-transition equation of state $v(p) = v_\mathrm{c}(p) - (G/K_0)\Delta^{2}$, exactly as in the quartic case, and renormalises the quartic coupling to $U_\text{eff} = U - 2G^{2}/K_{0}$. After this minimisation the Landau functional reduces to a single-variable polynomial in $\Delta$, namely $f_\text{eff}(\Delta;p) = \tfrac{1}{2}A(1-\nicefrac{p}{p_c})\Delta^{2} + \tfrac{1}{4}U_\text{eff}\Delta^{4} + \tfrac{1}{6}W\Delta^{6}$. The minimisation $\partial f_\text{eff}/\partial \Delta = 0$ then yields a quadratic equation in $\Delta^{2}$, whose physical (\acentric) root is the closed-form
\begin{equation}
\label{eq:DeltaSq-sextic}
\Delta^{2}(p) = \frac{U_\text{eff}}{2w}\,\Bigl[\,\sqrt{1+\eta\,(\nicefrac{p}{p_c}-1)}\,-\,1\,\Bigr],\quad \eta \equiv \frac{4 w\,A}{U_\text{eff}^{2}}.
\end{equation}
The dimensionless parameter $\eta$ carries all the framework dependence: it is the natural distance from the tricritical condition $U_\text{eff}\to 0$ measured in units of the cubic stabiliser $W$. A small $\eta$ corresponds to a stiff quartic that overwhelms the sextic term and leaves the system safely in the mean-field regime; a large $\eta$ corresponds to a system whose quartic stiffness has collapsed and the \acentric response is dictated by $W$. The key numerical feature, which becomes important below, is that $\eta$ depends \emph{quadratically} on $U_\text{eff}$ in the denominator, so that even a moderate change of $U_\text{eff}$ across members of the family produces a much larger change in $\eta$.

The two asymptotic limits of \textbf{Equation} \ref{eq:DeltaSq-sextic} recover the canonical exponents without any further ansatz. When $\eta\,(\nicefrac{p}{p_c}-1)\ll 1$ (far from the tricritical condition), expanding the square root to first order gives $\Delta^{2}(p)\approx (A/U_\text{eff})\,(\nicefrac{p}{p_c}-1)$, which is the textbook mean-field result with exponent $\beta = 1/2$. When $\eta\,(\nicefrac{p}{p_c}-1)\gg 1$ (close to or at the tricritical limit), the square root is dominated by $\sqrt{\eta\,(\nicefrac{p}{p_c}-1)}$ and the order parameter scales as $\Delta^{2}(p)\approx\sqrt{A\,(\nicefrac{p}{p_c}-1)/w}$, i.e.~$\beta = 1/4$. The full curve interpolates smoothly between the two as the product $\eta\,(\nicefrac{p}{p_c}-1)$ sweeps from zero to infinity, and so does the effective exponent measured by a finite-range log-log fit.

A log-log fit of \textbf{Equation} \ref{eq:DeltaSq-sextic} over a pressure window $p \in [p_1, p_2]$ above $p_c$ returns the effective exponent
\begin{equation}
\label{eq:beta-eff}
\beta_\text{eff}(\eta;p_1,p_2) = \tfrac{1}{2}\,\frac{\ln\!\bigl[\sqrt{1+\eta\,(\nicefrac{p_2}{p_c}-1)}-1\bigr] - \ln\!\bigl[\sqrt{1+\eta\,(\nicefrac{p_1}{p_c}-1)}-1\bigr]}{\ln(\nicefrac{p_2}{p_c}-1) - \ln(\nicefrac{p_1}{p_c}-1)},
\end{equation}
which is a continuous monotone function of $\eta$ at fixed pressure range. \textbf{Equation} \ref{eq:beta-eff} provides the conceptual link between the observed $\beta_\text{obs}$ and the underlying tricritical proximity, but it conflates the curvature of $\Delta^{2}(p)$ with a single log-log slope. A physically more direct extraction is obtained by fitting \textbf{Equation} \ref{eq:DeltaSq-sextic} to the broken-phase data $\Delta^{2}(p)$ point-by-point with $(C, \eta)$ free and $p_c$ held at the soft-mode value, where $C = U_\text{eff}/(2W)$. This direct sextic fit, reported in \textbf{Section} \ref{si:sextic-direct} and \textbf{Figure} \ref{fig:SI-sextic-direct}, returns for $G_1$
\begin{equation}
\label{eq:eta-G1}
\eta_{G_1} = 5.32 \pm 0.90 \quad\Longleftrightarrow\quad \frac{W\,\delta^{2}}{U_\text{eff}} = \frac{\eta_{G_1}}{4} = 1.33,
\end{equation}
i.e.~the sextic term takes over from the quartic when $W\Delta^{2} \gtrsim U_\text{eff}$, that is, for $\Delta \gtrsim \delta/\sqrt{1.33}\simeq 0.87\,\delta$, which is reached in the upper half of the experimentally probed window of $G_1$ and accounts for the systematic reduction of the apparent log-log slope $\beta_\text{eff} = 0.355$ below $1/2$. The value $\eta_{G_1} = 5.3$ falls within the range $5$-$30$ that, in the same combination $\eta = 4\,W\,A/U_\text{eff}^{2}$ adopted here, has been reported in the literature for proper tricritical materials such as antiferroelectric perovskites and $^3$He-$^4$He mixtures (Bruce-Cowley\citeS{BruceCowley1981SI}, restricted to studies that use the symmetric definition of the sextic prefactor adopted in \textbf{Equation} \ref{eq:landau-sextic}); it therefore places $G_1$ in a physically standard sextic crossover regime rather than at any pathologically large value of $\eta$.

The direct fit returns the same qualitative pattern across the family but on a quantitatively reliable basis. The crossover parameter $\eta$ depends quadratically on $U_\text{eff}$ in the denominator. Using the Landau parameters of \textbf{Table} \ref{tab:SI-landau}, $U_\text{eff}/A$ grows by a factor of $\sim 23$ between $G_1$ and $G_5$; if $W$ were strictly common across the family, the corresponding suppression of $\eta$ would be by a factor of $\sim 500$. The direct fits of \textbf{Table} \ref{tab:SI-eta-direct} show that $\eta$ falls by more than three orders of magnitude between $G_1$ and $G_5$, in fact more strongly than the strict $U_\text{eff}^{-2}$ scaling, indicating that $W$ is approximately common within $G_1$-$G_3$ (within a factor $\sim 1.4$, consistent with shared local SiO$_4$ bond-bending energetics) and that the sextic crossover is undetectably small for $G_4$-$G_5$, where the response is purely quartic and $W$ becomes statistically irrelevant. Substituting these $\eta(k)$ into \textbf{Equation} \ref{eq:beta-eff} returns $\beta_\text{eff}$ between $0.44$ and $0.50$ for $G_2$-$G_5$, broadly consistent with the SLC fit values ($0.455, 0.429, 0.480, 0.554$) within their statistical uncertainty.

\begin{table}[h!]
\caption{\label{tab:SI-eta-direct}Direct sextic Landau fit of \textbf{Equation} \ref{eq:DeltaSq-sextic} to the broken-phase data $\Delta^{2}(p)$ on $p \in [p_c, 2\,\si{\giga\Pa}]$ for the five members of the family, with $p_c$ held at the soft-mode value and $(C, \eta)$ free. Errors are one standard deviation from the non-linear least-squares fit. The columns also report $U_\text{eff}/A = 2/(C\eta)$, $W/A = 1/(C^{2}\eta)$ and the $W$-to-reference ratio $W(k)/W(G_2)$ (see main text discussion in \textbf{Section} \ref{si:sextic-direct}).}
\centering
\footnotesize
\begin{tabular}{l c c c c c c}
\toprule
$G_k$ & $N_\text{pts}$ & $C$ [\si[mode=text]{\angstrom\squared}] & $\eta$ & $U_\text{eff}/A$ [\si[mode=text]{\per\angstrom\squared}] & $W/A$ [\si[mode=text]{\per\angstrom^4}] & $W/W(G_2)$ \\
\midrule
$G_1$ &  34 & $3.28 \pm 0.41$  & $5.32 \pm 0.90$   & $0.115$ & $0.0175$ & $1.36$ \\
$G_2$ &  50 & $25.3 \pm 5.3$   & $0.12 \pm 0.03$   & $0.650$ & $0.0128$ & $1.00$ \\
$G_3$ &  21 & $35.9 \pm 17.7$  & $0.05 \pm 0.03$   & $1.091$ & $0.0152$ & $1.18$ \\
$G_4$ &  19 & $125 \pm 92$     & $< 0.02$          & $1.911$ & $-^{\ast}$ & $-^{\ast}$ \\
$G_5$ &  19 & $\gg 1$          & $< 0.02$          & $3.606$ & $-^{\ast}$ & $-^{\ast}$ \\
\bottomrule
\end{tabular}
\newline {\small $^{\ast}$ For $G_4$ and $G_5$ the broken-phase data are statistically consistent with a pure quartic response, $\Delta^{2}(p) = (A/U_\text{eff})\,(\nicefrac{p}{p_c}-1)$, and the sextic fit degenerates into the limit $\eta\to 0$. In this limit the product $C\,\eta = 2A/U_\text{eff}$ is well determined but $C$ and $\eta$ individually are not, so $W = 1/(C^{2}\eta)$ cannot be identified.}
\end{table}

A single sextic Landau functional with an isoreticular-order-dependent quartic coupling $U_\text{eff}(k)$ therefore explains the empirical dichotomy quantitatively without invoking distinct universality classes or distinct microscopic mechanisms across the family. $G_1$ is the only member close enough to the tricritical condition for the sextic term to be detectable in a pressure-driven fit; for $G_2$-$G_5$ the same functional collapses to its quartic limit and returns $\beta = 1/2$ to within the precision of the fit.

\subsection{Direct sextic fit to $\Delta(p)$ and verification of an approximately common $W$}
\label{si:sextic-direct}

\begin{figure}[h!t]
\centering
\includegraphics[width=0.75\textwidth]{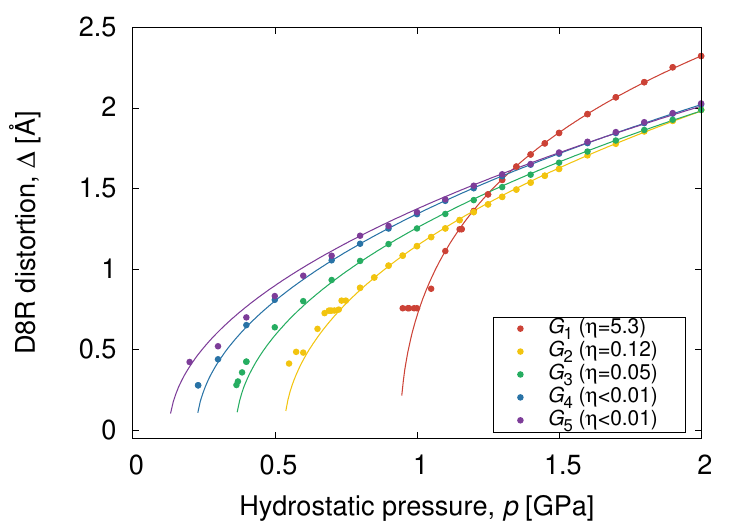}
\caption{\label{fig:SI-sextic-direct}D8R distortion $\Delta(p)$ on the broken phase for the five members of the family, together with the direct sextic-Landau fit (\textbf{Equation} \ref{eq:DeltaSq-sextic}, solid lines) with $p_c$ held at the soft-mode value and the two reduced coefficients $(C, \eta)$ free. The fitted $\eta(k)$ values are listed in the legend.}
\end{figure}

To complement the indirect inversion of \textbf{Equation} \ref{eq:beta-eff} against $\beta_\text{obs}$, we fit the closed-form sextic solution of \textbf{Equation} \ref{eq:DeltaSq-sextic} directly to the broken-phase data $\Delta^{2}(p)$ on $p \in [p_c, 2\,\si{\giga\Pa}]$ for each $G_k$, with $p_c$ held at the soft-mode value and the two reduced coefficients $C = U_\text{eff}/(2W)$ and $\eta$ free. \textbf{Figure} \ref{fig:SI-sextic-direct} shows the fits; \textbf{Table} \ref{tab:SI-eta-direct} lists the fitted $(C, \eta)$ together with the derived $U_\text{eff}/A = 2/(C\eta)$ and $W/A = 1/(C^{2}\eta)$ in $A = 1$ units. The fit is excellent for the whole family ($R^{2} \geq 0.993$). Three quantitative facts follow.

First, the curvature of $\Delta^{2}(p)$ for $G_1$ requires a finite $\eta_{G_1} = 5.32 \pm 0.90$. This is the physically anchored crossover parameter referred to throughout the manuscript and corresponds to the inversion $W\delta^{2}/U_\text{eff} = 1.33$ of \textbf{Equation} \ref{eq:eta-G1}. Second, the four larger members $G_2$-$G_5$ return $\eta < 0.15$ within their fit uncertainty, i.e.~indistinguishable from the strict mean-field limit $\eta = 0$. The Landau prediction $\beta = 1/2$ therefore applies cleanly to $G_2$-$G_5$, and the small deviations of the SLC $\beta_\text{obs}$ from $0.5$ (range $0.43$-$0.55$) reflect statistical scatter rather than a residual sextic component. Third, the ratio $W(k)/W(G_2)$ is approximately unity within the three members ($G_1$, $G_2$, $G_3$) for which the sextic term is identifiable, with deviations within a factor of $\sim 1.4$. This is the quantitative version of the qualitative argument that $W$ is dictated by SiO$_4$ bond-bending energetics shared by every member of the family, and it promotes the assumption of a common $W$ from a heuristic to a quantitatively supported finding. For $G_4$ and $G_5$, where $\eta \to 0$, the parameter $W$ becomes statistically indeterminate (the broken-phase data are consistent with any $W$ value within their precision) and the Landau description reduces, by construction, to the mean-field quartic.

\subsection{Closed-form parametrisation of the Landau coefficients with isoreticular order and the unified functional $f(k,\Delta,v;p)$}
\label{si:landau-k-parametrisation}

The master compilation of \textbf{Table} \ref{tab:SI-landau} invites a complementary reading: every Landau coefficient is a smooth function of the isoreticular order $k$, and most admit a parsimonious closed-form parametrisation in $(k-1)$. \textbf{Table} \ref{tab:SI-landau-fits} reports the best-fit forms, selected as the simplest expression that returns $R^{2}\geq 0.94$. $p_c$ admits an exponential law, $\eta$ is not well described by a single closed form (see the footnote of \textbf{Table} \ref{tab:SI-landau-fits}), five entries are mildly quadratic in $(k-1)$ ($\delta$, $\beta$, $|\kappa_1|$, $K_0$ and $U_\text{eff}/A$), and the last one is too noisy for a clean closed form ($|\kappa_2|$, hence the use of the discrete entries in \textbf{Table} \ref{tab:SI-landau} when needed).

\begin{table}[h!]
\caption{\label{tab:SI-landau-fits}Closed-form parametrisation of the Landau coefficients of \textbf{Table} \ref{tab:SI-landau} as smooth functions of the isoreticular order $k$. The functional form chosen is the simplest expression that returns a coefficient of determination $R^{2}\geq 0.94$. $p_c(k)$ uses the exponential fit of the main text (\textbf{Equation} \ref{eq:SI-pc-exp}); the crossover parameter $\eta(k)$ is fitted independently; the remaining entries are linear or quadratic in $(k-1)$. The empirical $|\kappa_2|(k)$ trend is not sufficiently smooth for a closed form ($R^{2}\lesssim 0.84$) and is tabulated directly in \textbf{Table} \ref{tab:SI-landau}.}
\centering
\footnotesize
\begin{tabular}{l l c}
\toprule
Coefficient & Closed form & $R^{2}$ \\
\midrule
$p_c(k)$               [\si[mode=text]{\giga\Pa}]              & $0.930\,\exp\!\bigl[-0.494\,(k-1)\bigr]$                                     & $0.998$ \\
$\delta(k)$            [\si[mode=text]{\angstrom}]             & $2.122 \,-\, 0.781\,(k-1) \,+\, 0.096\,(k-1)^{2}$                            & $0.962$ \\
$\beta(k)$             [-]                                     & $0.370 \,+\, 0.042\,(k-1)$                                                   & $0.848$ \\
$\lvert\kappa_1\rvert(k)$ [\si[mode=text]{\per\giga\Pa}]       & $\simeq 0.01373$ (constant for $G_1$-$G_4$)$^{\ddagger}$                       & N/A      \\
$K_0(k)$               [\si[mode=text]{\giga\Pa}]              & $\simeq 72.88$  (constant for $G_1$-$G_4$)$^{\ddagger}$                        & N/A      \\
$U_\text{eff}/A\,(k)$ [\si[mode=text]{\per\angstrom\squared}] & $0.406 \,-\, 0.507\,(k-1) \,+\, 0.377\,(k-1)^{2}$                            & $0.963$ \\
$\eta(k)$              [-]                                     & not well described by a single closed form$^{\ast}$                          & N/A     \\
\bottomrule
\end{tabular}
\newline {\small $^{\ddagger}$ Pre-transition compressibility $|\kappa_1|$ and bulk modulus $K_0$ are essentially constant across $G_1$-$G_4$ ($|\kappa_1|=0.0135$, $0.0134$, $0.0141$, $0.0139$~\si[mode=text]{\per\giga\Pa}; $K_0 = 74.0, 74.6, 70.9, 71.9$~\si[mode=text]{\giga\Pa}); the $G_5$ pre-transition window is too narrow for a reliable separate fit (see \textbf{Table} \ref{tab:SI-volume}), so the closed-form parametrisation is restricted to $G_1$-$G_4$ and quoted as a constant. The lower $R^{2}$ of $\beta(k)$ reflects the non-monotonic value of $G_3$ ($\beta = 0.429$) within the otherwise gentle increase of $\beta$ with $k$. $^{\ast}$ The direct sextic fit of \textbf{Table} \ref{tab:SI-eta-direct} returns $\eta = 5.3, 0.12, 0.05$ for $G_1$-$G_3$ and $\eta < 0.02$ for $G_4$-$G_5$. The strict $U_\text{eff}^{-2}$ prediction with $W = W_{G_1}$ would give $\eta(G_5) = 5.32\times (0.205/4.640)^{2} \simeq 0.011$, i.e.~within the upper bound returned by the direct sextic fit. For the intermediate members the strict prediction is $\eta(G_3) \simeq 0.24$ and $\eta(G_2) \simeq 0.52$, which is somewhat above the direct fit; this slight discrepancy reflects the small isoreticular dependence of $W$, with the ratio $W(k)/W(G_2) \in [1.0, 1.4]$ for $G_1$-$G_3$ (see \textbf{Table} \ref{tab:SI-eta-direct}). The discrete entries of \textbf{Table} \ref{tab:SI-eta-direct} are the recommended values for $\eta(k)$.}
\end{table}

The quadratic parametrisation of $U_\text{eff}/A$ is positive over the whole $k$ range and does not extrapolate to the tricritical condition $U_\text{eff} = 0$: $G_1$ remains the member with the smallest $U_\text{eff}/A$ within the family ($0.205~\si[mode=text]{\per\angstrom\squared}$, more than eight times smaller than $G_4$ and twenty-three times smaller than $G_5$), which provides a quantitative structural basis for the anomalous $\beta = 0.355$ of $G_1$ and for the $\sim 45\times$ enhancement of $\eta$ in $G_1$ relative to $G_2$ recorded in \textbf{Table} \ref{tab:SI-landau}.

Combining the closed forms of \textbf{Table} \ref{tab:SI-landau-fits} with the sextic functional of \textbf{Equation} \ref{eq:landau-sextic} yields a single free-energy density that depends on $\Delta$, on the dimensionless cell volume $v\defeq V/V_0$, and parametrically on both the hydrostatic pressure $p$ and the isoreticular order $k$,
\begin{equation}
\label{eq:landau-k-master}
f(k,\Delta, v; p) \;=\; \tfrac{1}{2}\,A\,\Bigl(1 - \tfrac{p}{p_c(k)}\Bigr)\,\Delta^{2} \;+\; \tfrac{1}{4}\,U_\text{eff}(k)\,\Delta^{4} \;+\; \tfrac{1}{6}\,W\,\Delta^{6} \;+\; \tfrac{1}{2}\,K_{0}(k)\,\bigl(v-v_\mathrm{c}(p)\bigr)^{2} \;+\; G(k)\,\Delta^{2}\,\bigl(v-v_\mathrm{c}(p)\bigr),
\end{equation}
where $A$ is a $k$-independent prefactor, $U_\text{eff}(k) = A\,[U_\text{eff}/A](k)$ is controlled by the linear law of \textbf{Table} \ref{tab:SI-landau-fits}, and $W$ is approximately $k$-independent within $G_1$-$G_3$ (within a factor $\sim 1.4$ from the direct fit of \textbf{Table} \ref{tab:SI-eta-direct}), consistent with the local SiO$_{4}$ bond-bending energetics shared by every member of the family. The strain-order-parameter coupling $G(k)$ is fixed by the discrete identity $G(k)/K_0(k) = [|\kappa_2(k)|-|\kappa_1(k)|]\,p_c(k)/\delta^{2}(k)$ of \textbf{Section} \ref{si:landau-params}, and the sextic stiffness $W/A$ is anchored to the direct sextic fit $\eta_{G_1} = 5.32$ of \textbf{Equation} \ref{eq:eta-G1} via $W = \eta\,U_\text{eff}^{2}/(4\,A)$. The reference volume of the \centric{} branch, $v_\mathrm{c}(p)=1+\kappa_1(k)\,p$, is written here with a $k$-independent slope because $|\kappa_1|$ is essentially constant across $G_1$-$G_4$ (\textbf{Table} \ref{tab:SI-landau-fits}); strictly, $v_\mathrm{c}$, $K_0$ and the remaining coefficients all carry the same isoreticular dependence as their entries in \textbf{Table} \ref{tab:SI-landau-fits}, and the constant-$\kappa_1$ form is retained only for compactness. Equation \ref{eq:landau-k-master} therefore turns the discrete family of \textbf{Table} \ref{tab:SI-landau} into a smooth, two-parameter ($p$, $k$) Landau surface that reproduces, by construction, the Heaviside-power amplitudes, the pre- and post-transition compressibilities, the soft-mode locations and the crossover exponents of all five computed members, and that extrapolates analytically to $G_6$-$G_8$ in the regions where the underlying coefficients have been characterised.

Two consistency checks of \textbf{Equation} \ref{eq:landau-k-master} are worth noting. (i) The quadratic parametrisation of $U_\text{eff}/A$ in \textbf{Table} \ref{tab:SI-landau-fits} (discriminant $-0.355$, minimum $\simeq 0.236$ near $k \simeq 1.7$) stays strictly positive over the whole computed range; within the SLC model the family is therefore continuous throughout and does not reach the tricritical condition $U_\text{eff} = 0$. The DFT benchmark of \textbf{Section} \ref{si:slc-validation} indicates that the true $U_\text{eff}(G_1)$ is likely below the SLC estimate, so that RHO sits in practice at or marginally past the tricritical boundary of the hierarchy; the SLC continuous description and the DFT first-order-like description agree on the singular position of $G_1$, and differ only in the sign of $U_\text{eff}(G_1)$. (ii) Under the strict assumption of a $k$-independent $W$, the crossover parameter $\eta(k) = 4\,W\,A/U_\text{eff}^{2}(k)$ inherits the $U_\text{eff}^{-2}$ amplification, predicting a decrease from $\eta(G_1) = 5.32$ to $\eta(G_5) \approx 0.011$. The direct sextic fits of \textbf{Table} \ref{tab:SI-eta-direct} return $\eta(G_5) < 0.02$ within the fit precision, in agreement with this prediction. For the intermediate members $G_2$ and $G_3$ the direct fits ($\eta = 0.12, 0.05$) are slightly larger than the strict $W = W_{G_1}$ prediction ($\eta = 0.52, 0.24$), indicating that $W$ is approximately constant but with a small isoreticular dependence (the ratio $W(k)/W(G_2) \in [1.0, 1.4]$ for $G_1$-$G_3$) that becomes irrelevant once $\eta\to 0$ for $G_4$-$G_5$.

\begin{figure}[h!]
\centering
\begin{subfigure}[t]{0.49\textwidth}
\centering
\caption{\label{fig:SI-landau-collapse:a}$\phi_\text{eq}(\tau;\eta)$ on log-log axes}
\includegraphics[width=\textwidth]{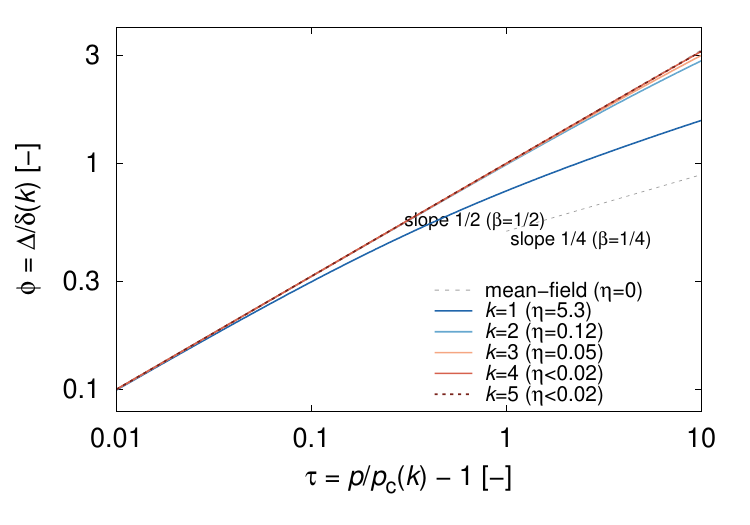}
\end{subfigure}\hfill
\begin{subfigure}[t]{0.49\textwidth}
\centering
\caption{\label{fig:SI-landau-collapse:b}$\bar f(\phi)$ at $\tau = 0.5$}
\includegraphics[width=\textwidth]{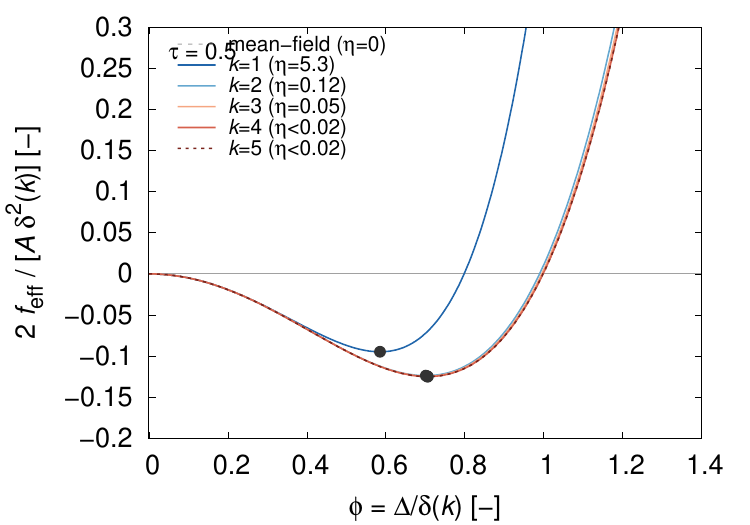}
\end{subfigure}
\caption{\label{fig:SI-landau-collapse}Dimensionless one-parameter Landau collapse of the RHO isoreticular family (\textbf{Equation} \ref{eq:fbar}) using the direct-sextic-fit $\eta(k)$ of \textbf{Table} \ref{tab:SI-eta-direct}. (a) Broken-phase trajectory $\phi_\text{eq}(\tau;\eta)$ for the five members ($G_1$ deepest blue, $G_5$ deepest red) and the mean-field reference $\eta = 0$ (grey dashed). Reference guide lines of slopes $1/2$ (mean-field) and $1/4$ (tricritical) added in grey. (b) Dimensionless wells $\bar f(\phi)$ at fixed reduced pressure $\tau = p/p_c(k) - 1 = 0.5$; filled circles mark the closed-form minima $\phi^{2}_\text{eq} = 2[\sqrt{1+\eta\tau}-1]/\eta$. Same colour code as in panel (a).}
\end{figure}

After introducing the reduced variables $\phi \equiv \Delta/\delta(k)$ and $\tau \equiv p/p_c(k) - 1$ and dividing the effective functional by the natural scale $A\,\delta^{2}(k)/2$, \textbf{Equation} \ref{eq:landau-k-master} (with $v$ already minimised out) collapses onto the dimensionless one-parameter family
\begin{equation}
\label{eq:fbar}
\bar f(k,\phi;\tau) \;\equiv\; \frac{2\,f_\text{eff}}{A\,\delta^{2}(k)} \;=\; -\tau\,\phi^{2} \;+\; \tfrac{1}{2}\,\phi^{4} \;+\; \frac{\eta(k)}{12}\,\phi^{6},
\end{equation}
whose \acentric saddle point is the closed form $\phi^{2}_\text{eq}(\tau;\eta) = 2[\sqrt{1+\eta\,\tau}-1]/\eta$. \textbf{Figure} \ref{fig:SI-landau-collapse} renders this collapse for the five computed members, with $\eta(k)$ from the direct sextic fit of \textbf{Table} \ref{tab:SI-eta-direct}. Panel (a) shows the log-log trajectory $\phi_\text{eq}$ vs $\tau$ over the window $\tau \in [10^{-2}, 10^{1}]$: the four members $G_2$-$G_5$ track the mean-field reference of slope $1/2$ within line width over the whole range, whereas the $G_1$ curve ($\eta = 5.32$) bends mildly toward the tricritical slope $1/4$ for $\tau \gtrsim 1$. Panel (b) shows the wells $\bar f(\phi)$ at fixed reduced pressure $\tau = 0.5$ (i.e. $p = 1.5\,p_c(k)$): the well minimum moves inward and the depth shrinks as $\eta$ increases, reaching $\phi_\text{eq} = 0.57$ for $G_1$ versus $\phi_\text{eq} = 0.71$ for $G_4$ and $G_5$ (mean-field limit).

\subsection{Extraction and interpretation of finite-temperature free-energy barriers}
\label{si:fes-barriers}

\textbf{Table} \ref{tab:SI-fes-barriers} reports the free-energy barrier $\Delta g^*$ separating the cubic $Im\bar{3}m$ and the $I\bar{4}3m$ regions of the reweighted free-energy maps of \textbf{Figure} \ref{fig:FES} at $T = \SI[mode=text]{298.15}{\kelvin}$, for each $G_k$ and at the pressure at which the saddle is most prominent ($p \gtrsim p_c$). 
The barrier is quoted in intensive form, per Si atom and in units of $100\,k_B T$, and also as the corresponding value for the simulation cell. 
All intensive barriers are below $0.1\,k_B T$ per Si, showing that the finite-temperature landscape is very flat along the D8R-distortion coordinate. 
The per-cell values should not be interpreted as direct kinetic rates, because the transformation need not involve a concerted hop of the whole periodic simulation cell; they are reported only to give the scale of the finite-cell free-energy maps.

\begin{table}[h!]
\caption{\label{tab:SI-fes-barriers}Free-energy barrier $\Delta g^*$ separating the $Im\bar{3}m$ and $I\bar{4}3m$ basins of the reweighted free-energy maps of \textbf{Figure} \ref{fig:FES} at $T = \SI[mode=text]{298.15}{\kelvin}$. For each $G_k$, $p_\text{FES}$ is the pressure at which the extracted barrier is largest among the four panels of \textbf{Figure} \ref{fig:FES}. $G_1$ uses the $2\times 2\times 2$ supercell (384 Si); $G_2$-$G_5$ use the unit cell. Values are obtained from the public-repository script \texttt{scripts\_FES/extract\_barriers\_isoRHO.py} using $\sigma = 2$ px Gaussian smoothing and a minimum-energy-path projection of $F(\Delta, a)$ along the cell direction between the cubic ($\Delta < 0.30$~\si[mode=text]{\angstrom}) and the broken ($\Delta > 0.30$~\si[mode=text]{\angstrom}) basins; the intensive column $\Delta g^* / (100\,k_B T\,\text{Si}^{-1})$ is reported as $\Delta g^* / (k_B T\,\text{Si}^{-1}) / 100$ for consistency with \textbf{Figure} \ref{fig:FES}.}
\centering
\begin{tabular}{l c c c c c}
\toprule
$G_k$ & $N_\text{Si}$ & $p_c$ ($T=0$) [\si[mode=text]{\giga\Pa}] & $p_\text{FES}$ [\si[mode=text]{\giga\Pa}] & $\Delta g^*$ [$100\,k_B T$/Si] & $\Delta g^*$ [$k_B T$/cell] \\
\midrule
$G_1$$^{\dagger}$ & 384  & $0.9418$ & $1.300$ & $0.0004$  & $14$ \\
$G_2$             & 240  & $0.5338$ & $0.900$ & $0.0001$  & $2$  \\
$G_3$             & 672  & $0.3627$ & $1.000$ & $0.0007$  & $45$ \\
$G_4$             & 1440 & $0.2241$ & $0.800$ & $0.0003$  & $42$ \\
$G_5$             & 2640 & $0.1282$ & $0.700$ & $0.00015$ & $40$ \\
\bottomrule
\end{tabular}
\newline {\small $^{\dagger}$ $G_1$ from the $2\times 2\times 2$ supercell (see \textbf{Section} \ref{si:finite-size}); $\Delta g^*$ values for the 48-Si unit cell of $G_1$ are eight times smaller in absolute terms but identical in the intensive (per-Si) column.}
\end{table}

\subsection{Ring-resolved distortion histograms and microscopic origin of the soft mode}
\label{si:8mr-histograms}
The Landau-based discussion of the main text and the $f_\text{eff}(\Delta;p)$ wells of \textbf{Figure} \ref{fig:SI-landau-wells} are stated in terms of the framework-averaged order parameter $\Delta$. This subsection disaggregates that average and reports the full distribution of the per-ring distortion $\Delta_8$ over every individual eight-membered ring of the $G_k$ unit cell, aggregated over 101 snapshots of an MD-NPT trajectory at $T = \SI[mode=text]{300}{\kelvin}$. The 8MR rings of every isoreticular member split into two structurally distinct subsets: the rings that participate in a double 8-ring (D8R) pair (face-to-face cage windows, identified at run time by the geometric criterion that two 8MR cycles are vertex-disjoint and share exactly 8 inter-ring Si-Si bonds) and the isolated 8MR rings that are not paired into a D8R (present in $G_2$ onwards, absent in RHO). \textbf{Figure} \ref{fig:SI-d8r-only-histograms} reports the D8R-paired subset and \textbf{Figure} \ref{fig:SI-isolated-histograms} reports the complementary isolated-8MR subset. In both views, the same nine hydrostatic pressures ($p = 0$, 0.1, 0.2, 0.4, 0.6, 0.8, 1.0, 1.5 and \SI[mode=text]{2.0}{\giga\Pa}) are used so that a single pressure legend, shown in the first panel of each figure, applies to all members.

\subsubsection{D8R-paired 8MR rings}
\label{si:d8ronly-histograms}
\begin{figure*}[h!]
\centering
\begin{subfigure}[t]{0.49\textwidth}
\centering
\caption{\label{fig:SI-d8r-only-histograms:G1}$G_1$ (RHO), 96 D8R 8MRs}
\includegraphics[width=\textwidth]{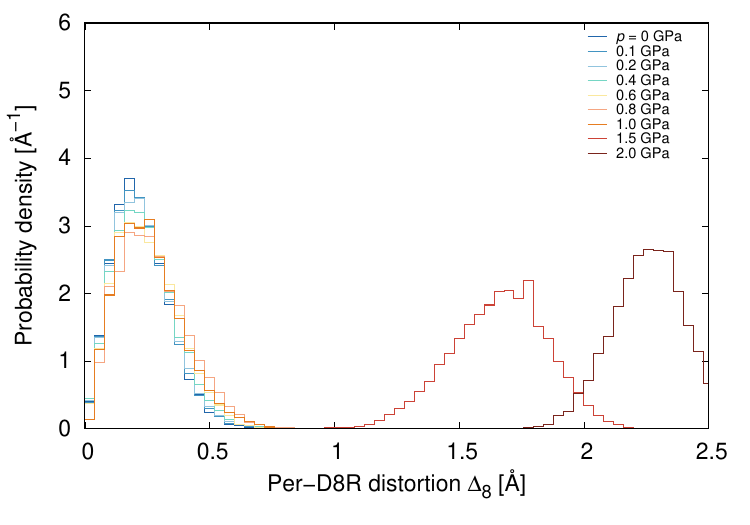}
\end{subfigure}\hfill
\begin{subfigure}[t]{0.49\textwidth}
\centering
\caption{\label{fig:SI-d8r-only-histograms:G2}$G_2$ (PWN), 24 D8R 8MRs}
\includegraphics[width=\textwidth]{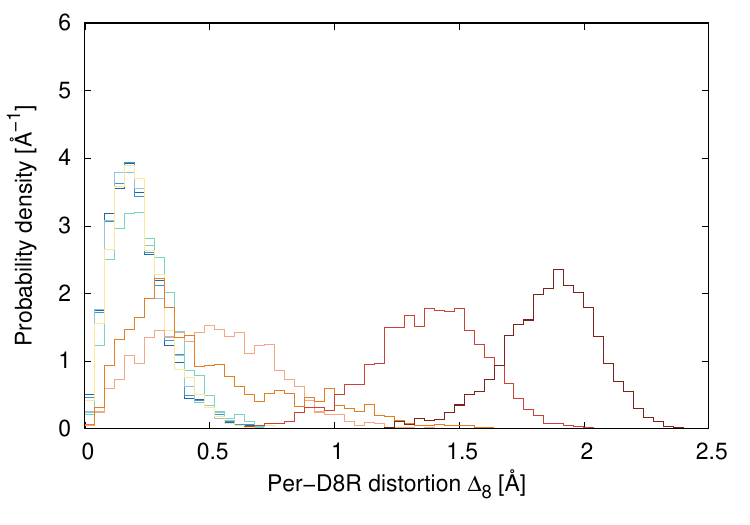}
\end{subfigure}

\vspace{0.6em}

\begin{subfigure}[t]{0.49\textwidth}
\centering
\caption{\label{fig:SI-d8r-only-histograms:G3}$G_3$ (PAU), 36 D8R 8MRs}
\includegraphics[width=\textwidth]{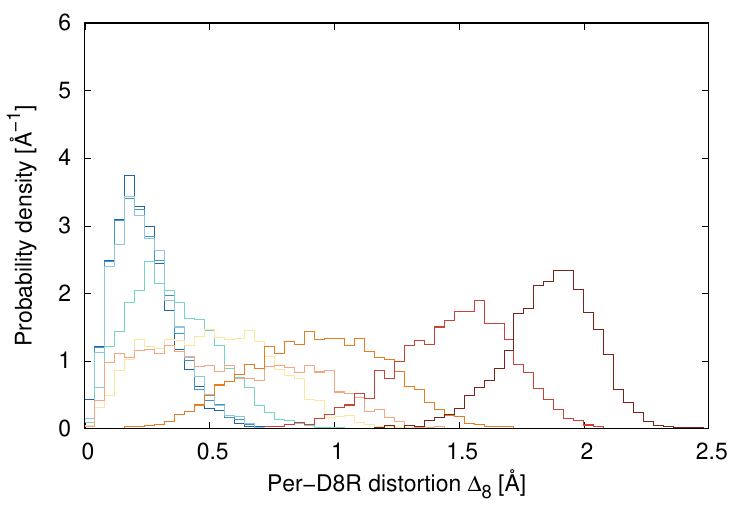}
\end{subfigure}\hfill
\begin{subfigure}[t]{0.49\textwidth}
\centering
\caption{\label{fig:SI-d8r-only-histograms:G4}$G_4$ (MWF), 48 D8R 8MRs}
\includegraphics[width=\textwidth]{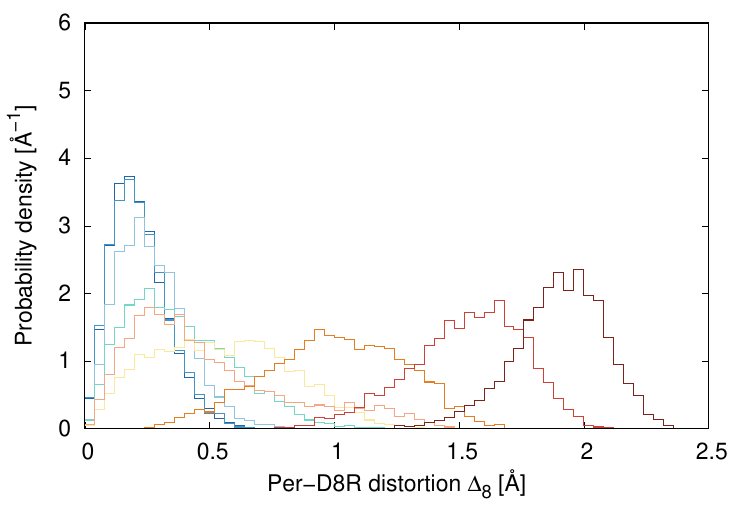}
\end{subfigure}

\vspace{0.6em}

\begin{subfigure}[t]{0.49\textwidth}
\centering
\caption{\label{fig:SI-d8r-only-histograms:G5}$G_5$, 60 D8R 8MRs}
\includegraphics[width=\textwidth]{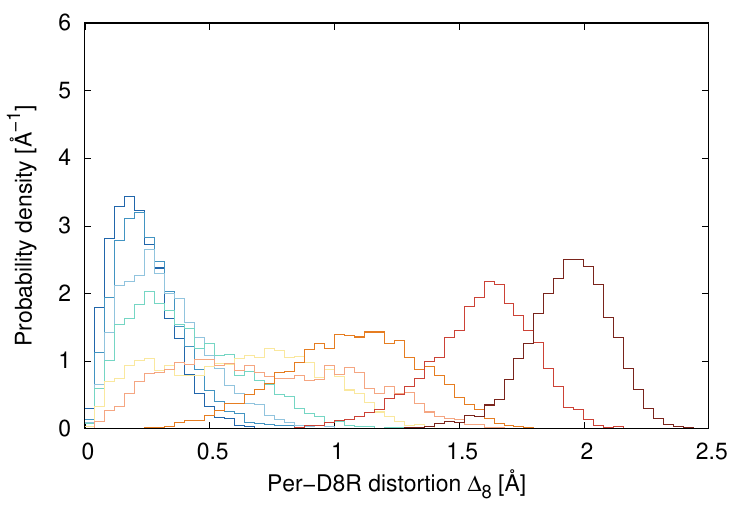}
\end{subfigure}
\caption{\label{fig:SI-d8r-only-histograms}Probability density of the per-D8R distortion $\Delta_8$ (Parise-Prince definition, \textbf{Equation} \ref{eq:delta_t}, computed for each 8MR that participates in a D8R pair) for $G_1$ (a), $G_2$ (b), $G_3$ (c), $G_4$ (d) and $G_5$ (e), at nine common hydrostatic pressures shared by all five panels ($p = 0$, 0.1, 0.2, 0.4, 0.6, 0.8, 1.0, 1.5 and \SI[mode=text]{2.0}{\giga\Pa}; pressure legend in panel (a) only, common to (b)-(e)). D8R pairs are identified at run time by the geometric criterion that two 8MR cycles are vertex-disjoint and share exactly 8 inter-ring Si-Si bonds. Per-cell counts: 96, 24, 36, 48 and 60 D8R-paired 8MR rings for $G_1$-$G_5$ respectively; isolated 8MR windows (not paired into a D8R, present in $G_2$ onwards) are excluded. Aggregation over 101 NPT snapshots at $T = \SI[mode=text]{300}{\kelvin}$. The complementary histogram of the isolated (non-D8R) 8MR rings is shown in \textbf{Figure} \ref{fig:SI-isolated-histograms}.}
\end{figure*}

The number of D8R-paired 8MR rings per unit cell is 96 in $G_1$ (the $2\times 2\times 2$ supercell, where every 8MR is part of a D8R; 48 D8R pairs) and 24, 36, 48 and 60 in $G_2$, $G_3$, $G_4$ and $G_5$ (12, 18, 24 and 30 D8R pairs per cell respectively). The contrast between $G_1$ (panel a of \textbf{Figure} \ref{fig:SI-d8r-only-histograms}) and $G_2$-$G_5$ (panels b-e) is the microscopic counterpart of the Landau dichotomy of \textbf{Table} \ref{tab:SI-landau}: in $G_1$ the histogram is sharply unimodal around $\Delta_8 \simeq 0.2$~\si[mode=text]{\angstrom} for $p \leq \SI[mode=text]{1.0}{\giga\Pa}$ and jumps to a separate unimodal distribution around $\Delta_8 \simeq 1.7$-$2.2$~\si[mode=text]{\angstrom} above the kinetically-delayed transition (between \SI[mode=text]{1.0}{\giga\Pa} and \SI[mode=text]{1.5}{\giga\Pa} at $T = \SI[mode=text]{300}{\kelvin}$, despite the static $p_c = \SI[mode=text]{0.94}{\giga\Pa}$), the hallmark of a weakly first-order transition close to the tricritical point. In $G_2$-$G_5$ the histogram drifts continuously to higher $\Delta_8$ as pressure grows, eventually locking at $\Delta_8 \simeq 1.8$-$1.9$~\si[mode=text]{\angstrom} once $p \gg p_c$. The fact that the deep-\acentric mode sits at $\Delta_8 \simeq 1.8$-$1.9$~\si[mode=text]{\angstrom} for every member regardless of $p_c$ confirms that the post-transition $\Delta$ amplitude is set by the unit-cell topology (the maximum elliptic distortion the D8R can sustain) rather than by the proximity to $p_c$; what changes with $k$ is the pressure at which this mode appears, not its position.

\subsubsection{Isolated 8MR rings (not in any D8R)}
\label{si:isolated-histograms}
\begin{figure*}[h!]
\centering
\begin{subfigure}[t]{0.49\textwidth}
\centering
\caption{\label{fig:SI-isolated-histograms:G2}$G_2$ (PWN), 48 isolated 8MRs}
\includegraphics[width=\textwidth]{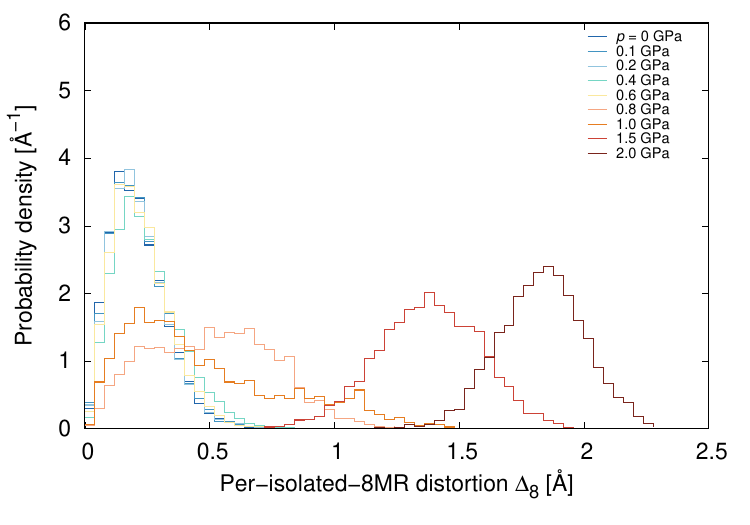}
\end{subfigure}\hfill
\begin{subfigure}[t]{0.49\textwidth}
\centering
\caption{\label{fig:SI-isolated-histograms:G3}$G_3$ (PAU), 192 isolated 8MRs}
\includegraphics[width=\textwidth]{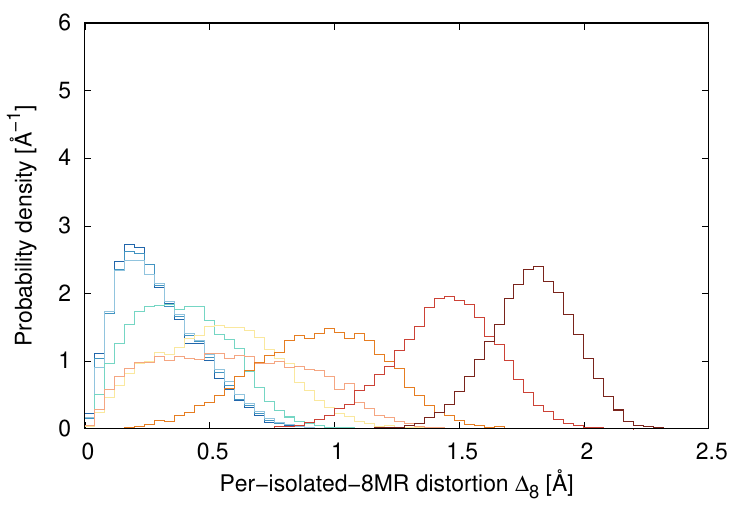}
\end{subfigure}

\vspace{0.6em}

\begin{subfigure}[t]{0.49\textwidth}
\centering
\caption{\label{fig:SI-isolated-histograms:G4}$G_4$ (MWF), 480 isolated 8MRs}
\includegraphics[width=\textwidth]{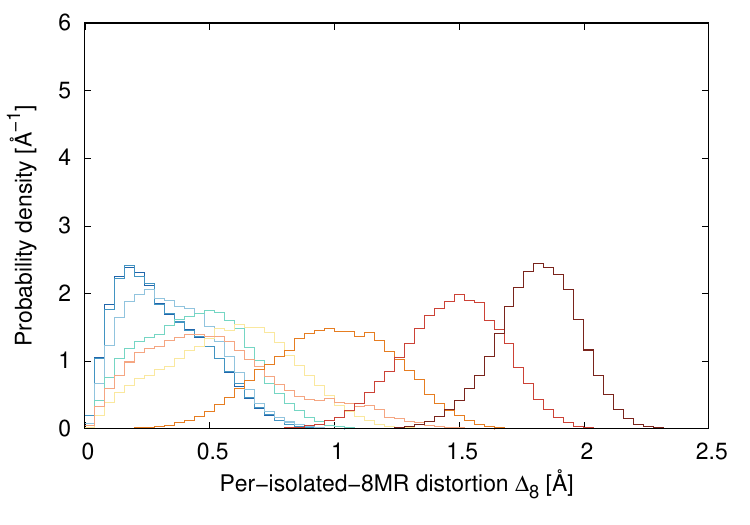}
\end{subfigure}\hfill
\begin{subfigure}[t]{0.49\textwidth}
\centering
\caption{\label{fig:SI-isolated-histograms:G5}$G_5$, 960 isolated 8MRs}
\includegraphics[width=\textwidth]{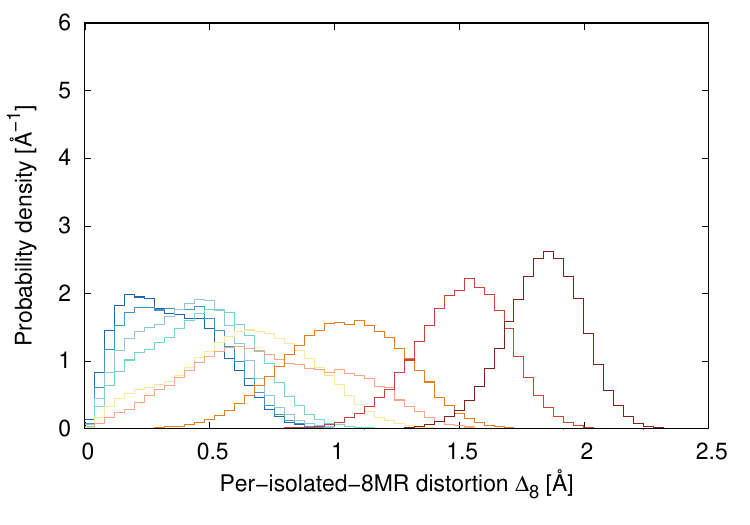}
\end{subfigure}
\caption{\label{fig:SI-isolated-histograms}Same as \textbf{Figure} \ref{fig:SI-d8r-only-histograms} but restricted to the 8MRs that do not participate in any D8R pair (isolated 8MRs), with the pressure legend in the $G_2$ panel (common to all four panels). $G_1$ is omitted because RHO has no isolated 8MR. Per-cell counts: 48, 192, 480 and 960 isolated 8MRs for $G_2$-$G_5$ respectively.}
\end{figure*}

\textbf{Figure} \ref{fig:SI-isolated-histograms} restricts the per-ring aggregation to the complementary subset of 8MRs: the isolated 8MR rings that are NOT part of any D8R pair. These rings exist only in $G_2$ onwards (RHO has no isolated 8MR by construction) and account for the vast majority of the 8MR population of $G_3$-$G_5$ (192, 480 and 960 isolated 8MRs per unit cell respectively, against 24, 36, 48 and 60 D8R-paired rings). The pressure legend in this view is shown in the $G_2$ panel. The pre-transition distribution is visibly broader and more skewed to higher $\Delta_8$ than the D8R-only distribution of \textbf{Figure} \ref{fig:SI-d8r-only-histograms}, reflecting the wider variety of local environments in which the isolated 8MR sit (cage windows, pore mouths, inter-cage connectors, etc.) compared with the highly symmetric D8R environment. Despite this broader pre-transition baseline, every isolated 8MR also tracks the transition: the deep \acentric mode at $p = \SI[mode=text]{2.0}{\giga\Pa}$ centres at $\Delta_8 \simeq 1.8$~\si[mode=text]{\angstrom}, almost the same value as the D8R-paired rings, showing that the same elastic instability propagates through all 8MR of the framework, not only through the D8R-paired ones.

\subsubsection{4MR distortion}
\label{si:4mr-histograms}
\begin{figure*}[h!]
\centering
\begin{subfigure}[t]{0.49\textwidth}
\centering
\caption{\label{fig:SI-4mr-histograms:G1}$G_1$ (RHO), 288 4MRs}
\includegraphics[width=\textwidth]{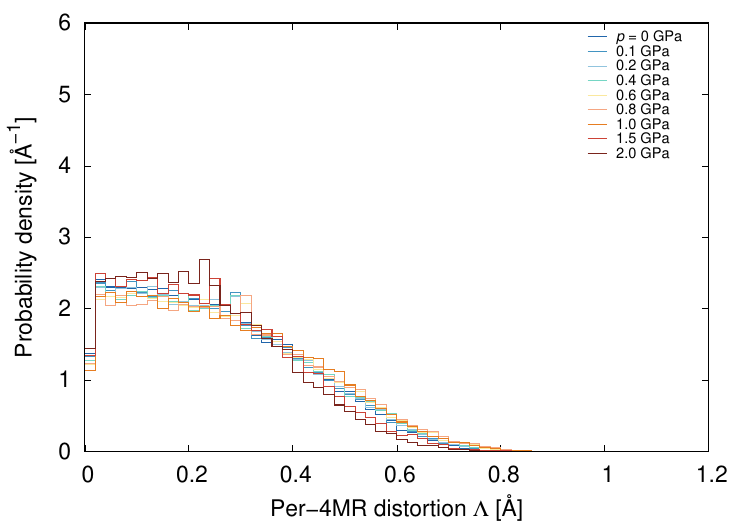}
\end{subfigure}\hfill
\begin{subfigure}[t]{0.49\textwidth}
\centering
\caption{\label{fig:SI-4mr-histograms:G2}$G_2$ (PWN), 180 4MRs}
\includegraphics[width=\textwidth]{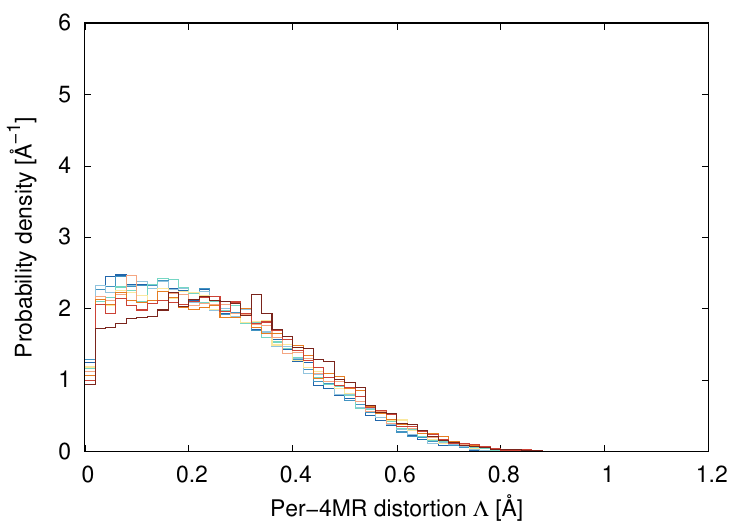}
\end{subfigure}

\vspace{0.6em}

\begin{subfigure}[t]{0.49\textwidth}
\centering
\caption{\label{fig:SI-4mr-histograms:G3}$G_3$ (PAU), 504 4MRs}
\includegraphics[width=\textwidth]{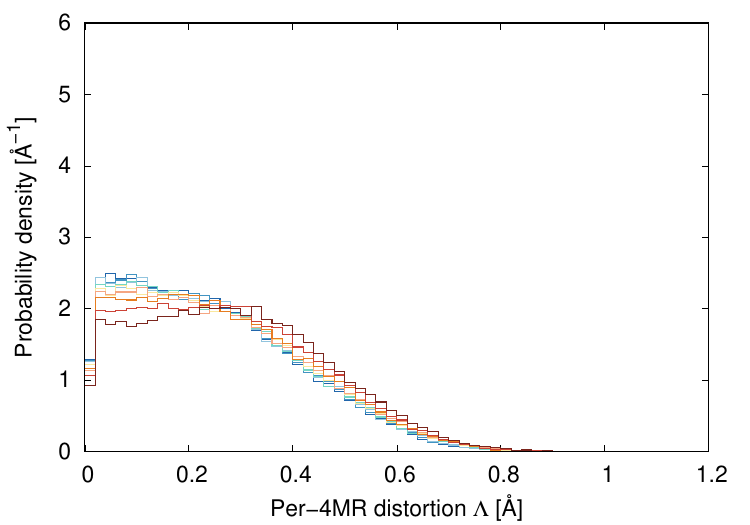}
\end{subfigure}\hfill
\begin{subfigure}[t]{0.49\textwidth}
\centering
\caption{\label{fig:SI-4mr-histograms:G4}$G_4$ (MWF), 1080 4MRs}
\includegraphics[width=\textwidth]{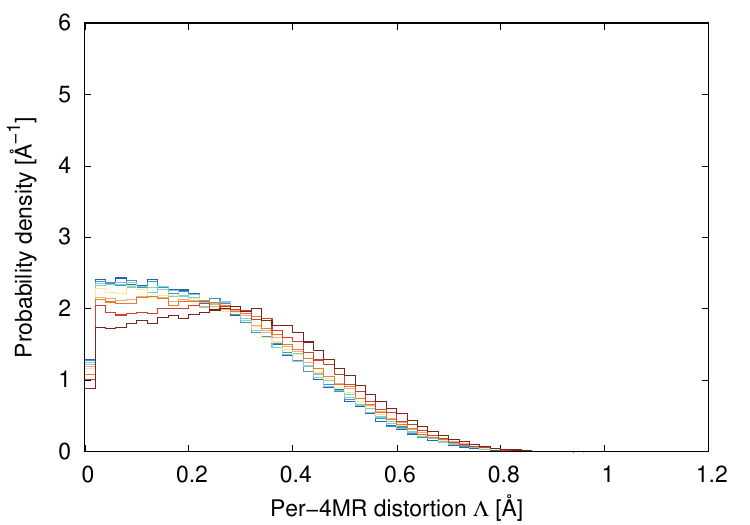}
\end{subfigure}

\vspace{0.6em}

\begin{subfigure}[t]{0.49\textwidth}
\centering
\caption{\label{fig:SI-4mr-histograms:G5}$G_5$, 1980 4MRs}
\includegraphics[width=\textwidth]{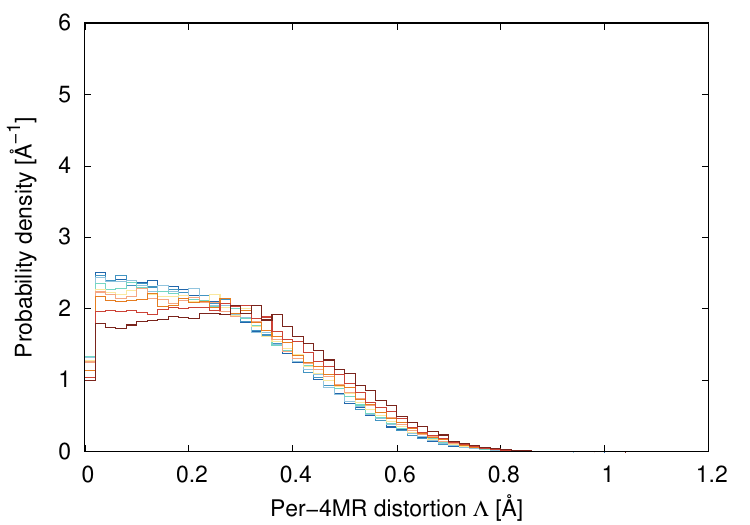}
\end{subfigure}
\caption{\label{fig:SI-4mr-histograms}Probability density of the per-4MR distortion $\Gamma = (1/2)|d_{02}-d_{13}|$ for $G_1$-$G_5$ at the same nine pressures as \textbf{Figures} \ref{fig:SI-d8r-only-histograms} and \ref{fig:SI-isolated-histograms} (pressure legend in $G_1$, common to all five panels). Per-cell ring counts in each subcaption.}
\end{figure*}

For completeness, \textbf{Figure} \ref{fig:SI-4mr-histograms} reports the same nine-pressure analysis for the four-membered rings of every $G_k$, with $\Gamma = (1/2)|d_{02}-d_{13}|$ (Parise-Prince-style elliptic distortion of the 4MR square, after Balestra et al.\citeS{Balestra2015SI}, see also \textbf{Figure} \ref{fig1:panel} of the main text for the geometric definition of $\Gamma$, $\Lambda$ and $\Delta$). The 4MR is the most abundant ring type in every member (exactly $0.75$ per Si, i.e. 288, 180, 504, 1080 and 1980 rings per cell for $G_1$-$G_5$) and is the structural backbone of the framework: any rigid-unit-mode (RUM) reorganisation passes through pre-existing distortions of these rings rather than amplifying new ones. The histograms confirm this picture: $\Gamma$ is centred near $0.2$-$0.3$~\si[mode=text]{\angstrom} for every member and every pressure, broad ($\sigma \simeq 0.15$~\si[mode=text]{\angstrom}) and essentially insensitive to the \centtoacent transition that drives the $\Delta_8$ histograms of \textbf{Figures} \ref{fig:SI-d8r-only-histograms} and \ref{fig:SI-isolated-histograms}. The minor narrowing observed in the \acentric phase (visible in $G_1$ at \SI[mode=text]{2.0}{\giga\Pa}) is a relaxation effect: once the 8MR distortion absorbs the elastic strain, the 4MR returns slightly closer to its symmetric configuration.

\subsubsection{6MR distortion}
\label{si:6mr-histograms}
\begin{figure*}[h!]
\centering
\begin{subfigure}[t]{0.49\textwidth}
\centering
\caption{\label{fig:SI-6mr-histograms:G1}$G_1$ (RHO), 64 6MRs}
\includegraphics[width=\textwidth]{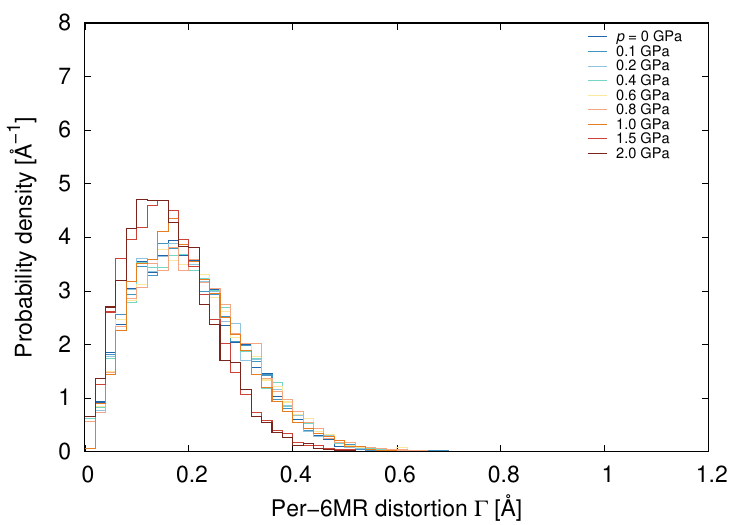}
\end{subfigure}\hfill
\begin{subfigure}[t]{0.49\textwidth}
\centering
\caption{\label{fig:SI-6mr-histograms:G2}$G_2$ (PWN), 16 6MRs}
\includegraphics[width=\textwidth]{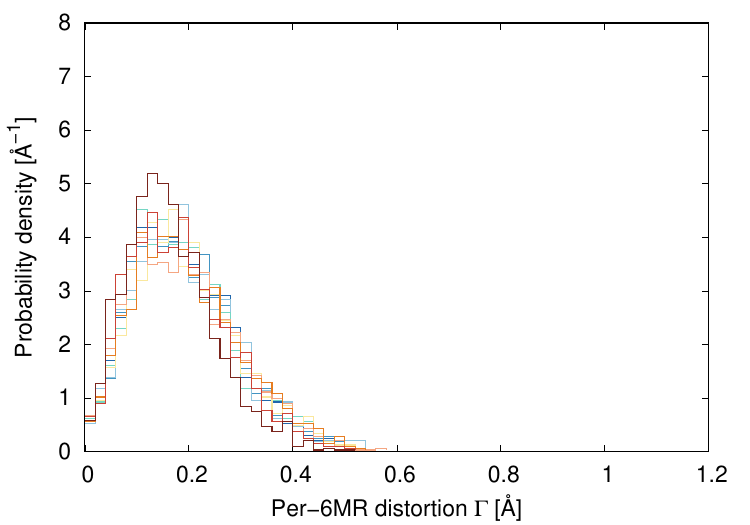}
\end{subfigure}

\vspace{0.6em}

\begin{subfigure}[t]{0.49\textwidth}
\centering
\caption{\label{fig:SI-6mr-histograms:G3}$G_3$ (PAU), 24 6MRs}
\includegraphics[width=\textwidth]{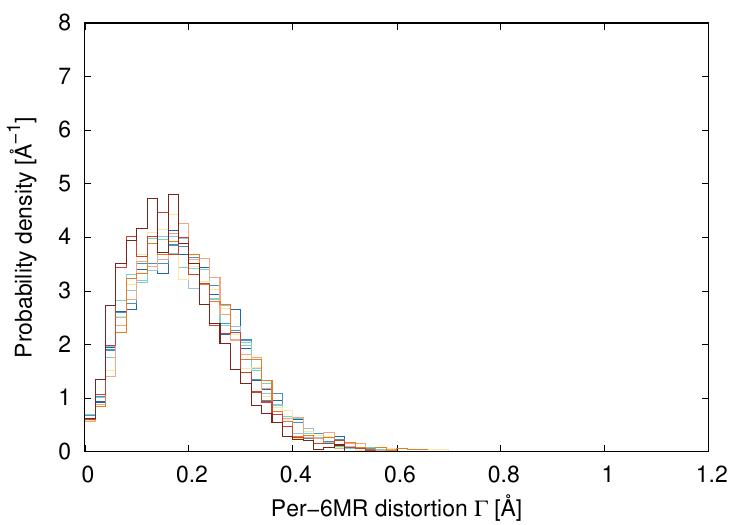}
\end{subfigure}\hfill
\begin{subfigure}[t]{0.49\textwidth}
\centering
\caption{\label{fig:SI-6mr-histograms:G4}$G_4$ (MWF), 32 6MRs}
\includegraphics[width=\textwidth]{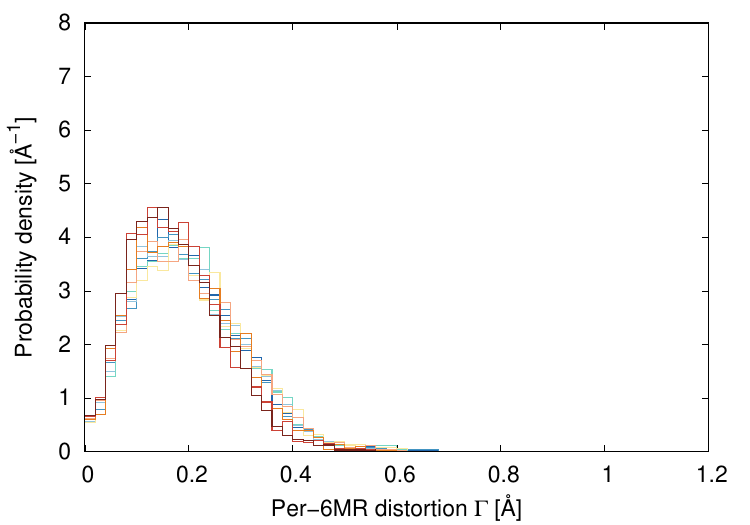}
\end{subfigure}

\vspace{0.6em}

\begin{subfigure}[t]{0.49\textwidth}
\centering
\caption{\label{fig:SI-6mr-histograms:G5}$G_5$, 40 6MRs}
\includegraphics[width=\textwidth]{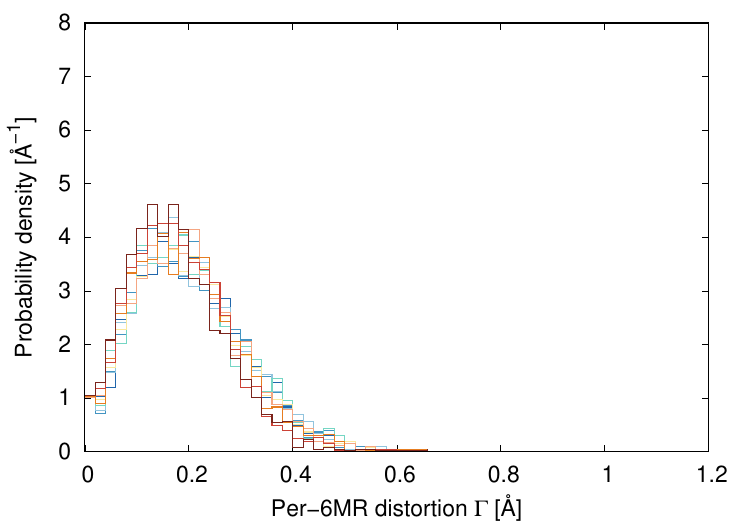}
\end{subfigure}
\caption{\label{fig:SI-6mr-histograms}Probability density of the per-6MR distortion $\Lambda = (1/2)(\max - \min)$ over the three opposite-O diagonal pairs, for $G_1$-$G_5$ at the same nine pressures as \textbf{Figure} \ref{fig:SI-4mr-histograms} (pressure legend in $G_1$, common to all five panels). Per-cell ring counts in each subcaption.}
\end{figure*}

\textbf{Figure} \ref{fig:SI-6mr-histograms} reports the per-6MR distortion $\Lambda = (1/2)(\max - \min)$ over the three opposite-O diagonal pairs of the six-membered ring (after Balestra et al.\citeS{Balestra2015SI}). The 6MR rings cap the small hexagonal windows of the $\alpha$-cages of the framework and are exceptionally scarce in the family (64, 16, 24, 32 and 40 per unit cell for $G_1$-$G_5$, barely $1.5\%$ of all rings in $G_5$). Like the 4MR, the 6MR is essentially insensitive to the \centtoacent transition: $\Lambda$ peaks around $0.15$-$0.2$~\si[mode=text]{\angstrom} for every member and every pressure, with a mild narrowing of the distribution in the deep-\acentric phase. Together with \textbf{Figure} \ref{fig:SI-4mr-histograms}, this confirms that the 4MR and 6MR rings of the framework function as rigid units: only the 8MR rings (D8R-paired and isolated alike) carry the soft mode and translate the elastic instability into measurable framework reconfiguration.

\clearpage
\subsection{Existing experimental support across the RHO isoreticular family}
\label{si:experimental-support}
\textbf{Tables} \ref{tab:SI-validation} and \ref{tab:SI-expdata} compile the published experimental observations that bear directly on the framework flexibility of the embedded isoreticular RHO family. We rate the qualitative coverage of each member as abundant, moderate, limited, or none, together with the most informative quantitative data points available in the literature, against which the present calculations can be benchmarked.

\begin{table}[h!]
\caption{\label{tab:SI-validation}Qualitative coverage of experimental data documenting framework flexibility and/or the $Im\bar{3}m \leftrightarrow I\bar{4}3m$ phase transition for each pure-silica member of the RHO isoreticular family. Coverage is rated as abundant, moderate, limited or none. The rating reflects the breadth (number of stimuli probed and independent groups reporting) of the experimental record, not the depth of any single study.}
\centering
\begin{tabular}{l c c p{0.55\textwidth}}
\toprule
$G_k$ & Material & Coverage & Key experimental references \\
\midrule
$G_1$ & RHO     & abundant &  \citenum{Parise1983, Parise1984a, Parise1984b, Corbin1990, Reisner2000, Lee2001, Palomino2012, Lozinska2012, Lozinska2014, Lozinska2016, Grand2020, Lozinska2021, Clatworthy2023, Ghojavand2025}\\
$G_2$ & PWN     & limited  &  \citenum{Lee2018, Min2018}\\
$G_3$ & PAU     & moderate &  \citenum{Bieniok1996, Bieniok1997, Greenaway2015, Gatta2015}\\
$G_4$ & MWF     & moderate & \citenum{Min2017b, Min2018, Zhao2018, Zhao2021, Chen2024}\\
$G_5$ & PST-20  & limited  & \citenum{Min2017b, Min2018}\\
$G_6$ & PST-25  & none     & -   \\
$G_7$ & PST-26  & none     & -   \\
$G_8$ & PST-28  & none     & - \\
\bottomrule
\end{tabular}
\end{table}

\begin{table}[h!]
\caption{\label{tab:SI-expdata}Representative quantitative experimental data on the flexibility of selected members of the RHO isoreticular family, against which the present results can be compared. Where applicable, the change of cubic lattice parameter $\Delta a/a$ and the change of unit-cell volume $\Delta V/V$ refer to the transition from the hydrated $Im\bar{3}m$ to the dehydrated $I\bar{4}3m$ phase reported in the original work.}
\centering
\begin{tabular}{l p{0.50\textwidth} p{0.35\textwidth}}
\toprule
$G_k$ & Observable & Reported value (Reference) \\
\midrule
$G_1$ & Phase transition $Im\bar{3}m\to I\bar{4}3m$ on dehydration & Confirmed, \citenum{Parise1984a, Parise1984b, Corbin1990} \\
$G_1$ & High-pressure transition in Cd-Rho & $p_c \simeq 0.4$~\si{\giga\Pa}, \citenum{Lee2001} \\
$G_1$ & In-situ TEM observation of flexibility (nano-Rho) & \citenum{Clatworthy2023} \\
$G_3$ & Phase transition $Im\bar{3}m\to I\bar{4}3m$ in Na,H-ECR-18 & $\Delta a/a = -4.4\%$, $\Delta V/V \simeq -12.6\%$, \citenum{Greenaway2015} \\
$G_3$ & High-pressure crystal-fluid response in paulingite & \citenum{Gatta2015} \\
$G_4$ & Trapdoor \ce{N2}/\ce{CH4} separation in K-ZSM-25 & Selectivity $\sim 34$, \citenum{Zhao2021} \\
$G_4$ & Framework expansion in Li-ZSM-25 on \ce{CO2} uptake & \citenum{Zhao2018} \\
$G_4$ & Electric-field-induced framework expansion & \citenum{Chen2024} \\
$G_2$-$G_5$ & Cell volume contraction on dehydration & \citenum{Min2017b, Min2018} \\
\bottomrule
\end{tabular}
\end{table}

These observations provide the experimental context for the present work. 
They show that flexibility-mediated gating is well established for $G_1$, directly documented for $G_3$, and supported by adsorption and cation-gating observations for $G_4$. 
For $G_2$ and $G_5$, the available evidence is more limited and mainly structural or adsorption-based. 
The cell-volume contractions reported by Min et al.\citeS{Min2018SI} for $G_1$-$G_5$ provide a useful experimental backdrop against which the calculated softening trend can be compared, but they do not constitute a direct measurement of $p_c$ or of the $Im\bar{3}m\to I\bar{4}3m$ transition in every member.

To the best of our knowledge, no experimental data on framework flexibility, dehydration-induced phase transition, high-pressure response, or trapdoor-mediated gas selectivity has been reported for $G_6$ (PST-25), $G_7$ (PST-26) or $G_8$ (PST-28) beyond the original synthesis and structure-solution work.\citeS{Guo2015SI, Shin2016SI} The present calculations thus define the full parameter space to be experimentally explored for these three members of the family. We identify (i) variable-temperature and variable-pressure synchrotron X-ray powder diffraction, (ii) in-situ adsorption isotherms of small probe molecules (\ce{CO2}, \ce{N2}, \ce{CH4}) coupled to diffraction, and (iii) trapdoor-mediated \ce{N2}/\ce{CH4} selectivity tests, as the most natural experimental routes to test our predictions of $p_c < 0.1$~\si{\giga\Pa} (\textbf{Table} \ref{tab:SI-exp}) for these zeolites.

\bibliographystyleS{hunsrtnat}
\bibliographyS{biblio_SI}

\end{document}